\documentclass[12pt]{article}
\usepackage{amsthm,amsmath,amsfonts,amssymb}
\usepackage{enumerate}
\usepackage{natbib}
\usepackage{rotating}
\usepackage[colorlinks,citecolor=blue,urlcolor=blue]{hyperref}
\usepackage{array}

\input{tcilatex}
\newtheorem{thrm}{Theorem}
\newtheorem{lmm}{Lemma}
\theoremstyle{definition}
\newtheorem{assumption}{Assumption}
\newtheorem{formulation}{Formulation}
\newcommand{\PreserveBackslash}[1]{\let\temp=\\#1\let\\=\temp}
\newcolumntype{C}[1]{>{\PreserveBackslash\centering}p{#1}}

\begin{document}

\title{Locally robust semiparametric estimation of sample selection models
without exclusion restrictions}
\author{Zhewen Pan$^{1}$ ~~ Yifan Zhang$^2$\thanks{%
Corresponding to: Yifan Zhang, Business School, Sun Yat-sen University,
Shenzhen 518107, China. Email: zhangyf278@mail.sysu.edu.cn.} \\
$^1$\textit{\small School of Economics, Zhejiang University of Finance \&
Economics, Hangzhou 310018, China} \\
$^2$\textit{\small Business School, Sun Yat-sen University, Shenzhen 518107,
China}}
\maketitle

\textbf{Abstract}{\small : Existing identification and estimation methods
for semiparametric sample selection models rely heavily on exclusion
restrictions. However, it is difficult in practice to find a credible
excluded variable that has a correlation with selection but no correlation
with the outcome. In this paper, we establish a new identification result
for a semiparametric sample selection model without the exclusion
restriction. The key identifying assumptions are nonlinearity on the
selection equation and linearity on the outcome equation. The difference in
the functional form plays the role of an excluded variable and provides
identification power. According to the identification result, we propose to
estimate the model by a partially linear regression with a nonparametrically
generated regressor. To accommodate modern machine learning methods in
generating the regressor, we construct an orthogonalized moment by adding
the first-step influence function and develop a locally robust estimator by
solving the cross-fitted orthogonalized moment condition. We prove
root-n-consistency and asymptotic normality of the proposed estimator under
mild regularity conditions. A Monte Carlo simulation shows the satisfactory
performance of the estimator in finite samples, and an application to wage
regression illustrates its usefulness in the absence of exclusion
restrictions. }



\newpage

\section{Introduction}

\label{sec:intro}Sample selection refers to the situation when we attempt to
infer about population parameters from a special subpopulation. Potential
causes of sample selection include nonrandom sampling, self-selectivity,
nonresponse on survey questions, attrition from social programs, and so
forth. In observational studies, sample selection is so frequently
encountered that it seems to be the rule rather than the exception %
\citep[p.253]{lee2010micro}. Ordinary regression using only the selected
sample may incur biased estimation and misleading inference for the
population parameters of interest. A modeling strategy for correcting the
sample selection bias is to specify an equation describing the selection
mechanism in addition to the main outcome equation 
\citep{gronau1974wage,
heckman1974shadow, lewis1974comments}: 
\begin{subequations}
\label{paraSSM}
\begin{eqnarray}
D &=&1\left\{ X^{\prime }\gamma \geq \varepsilon \right\} , \\
Y^{\ast } &=&X^{\prime }\beta +U, \\
Y &=&Y^{\ast }\cdot D,
\end{eqnarray}
\end{subequations}
where $D$ is the binary selection variable, $Y^{\ast }$ is the latent
outcome variable, and $Y$ is the observed outcome variable.

The sample selection model (\ref{paraSSM}) is conventionally identified via
parameterizing the distribution of the error term $\left( \varepsilon
,U\right) $ and then estimated by the maximum likelihood or \cite%
{heckman1979sample}'s two-step methods. However, the parametric
likelihood-based estimators have been found possessing evident bias when the
error distribution is misspecified %
\citep[e.g.,][]{arabmazar1982investigation}. This finding motivates
econometricians to relax parametric distributional assumptions and consider
semiparametric identification and estimation of the sample selection model.
An important development is gained by \cite{chamberlain1986asymptotic}, who
shows that with a nonparametric specification on the error distribution, $%
\beta $ in model (\ref{paraSSM}) is not semiparametrically identified if no
extra restriction is imposed. To regain identification of $\beta $, \cite%
{chamberlain1986asymptotic} proposes a sufficient identifying assumption
termed exclusion restriction, which requires some component of $\beta $ to
be zero. The exclusion restriction has been widely accepted by the sample
selection literature and underlies the consistency of a considerable amount
of semiparametric estimators for $\beta $ 
\citep[e.g.,][]{gallant1987semi, chen1998efficient,
powell2001semiparametric, chen2003semiparametric, newey2009twostep,
escanciano2015simple, liu2022sample}.

In empirical studies, it is usually difficult to find a credible excluded
variable that has a correlation with selection but no correlation with the
outcome 
\citep[e.g.,][]{krueger2001effect, blundell2007changes,
chen2024estimation}. In particular, the exclusion restriction is hardly
justified by economic theories \citep{van2007economic,
jones2015economics} and may not hold exactly 
\citep{conley2012plausibly,
nevo2012identification, van2018beyond}. If we attempt to waive the exclusion
restriction, nonetheless, we will return to %
\citeauthor{chamberlain1986asymptotic}'s dilemma that $\beta $ in model (\ref%
{paraSSM}) is not (point) identified unless we impose parametric
distributional assumptions on the model error. Another remedy for this
dilemma is to settle for partial identification. For instance, \cite%
{honore2020selection} construct the identified set for $\beta $ in model (%
\ref{paraSSM}) in the absence of the exclusion restriction or parametric
distributional assumptions.

This paper provides an alternative approach to addressing the sample
selection problem, by establishing a result of point identification but
without imposing the exclusion restriction or parametric distributional
assumptions. To resolve \citeauthor{chamberlain1986asymptotic}'s dilemma, we
instead impose a mild functional form restriction on the selection equation,
which will be proven sufficient for identification of coefficients in the
outcome equation. Specifically, we consider a distribution-free sample
selection model with a nonparametric selection mechanism: 
\begin{subequations}
\label{semiSSM}
\begin{eqnarray}
D &=&1\left\{ h\left( X\right) \geq \varepsilon \right\} , \\
Y^{\ast } &=&X^{\prime }\beta +U, \\
Y &=&Y^{\ast }\cdot D,
\end{eqnarray}
\end{subequations}
where the functional form restriction on $h$ will be given in Section \ref%
{sec:ident}. Loosely speaking, the function $h$ is assumed to be nonlinear
when $X$ includes at least two continuous covariates or be nonmonotone when $%
X$ includes only one or even no continuous covariate. This functional form
restriction is largely innocuous, as nonlinear or nonmonotone functions are
more prevalent in the real world than their opposite counterparts. Moreover,
the nonparametric nature of $h$ has an additional advantage of reducing the
risk of functional form misspecification over the linear setting (\ref%
{paraSSM}a). Although the nonparametricity may reduce interpretability of
the selection equation, it would not cause trouble in the context of sample
selection models where the marginal effects of covariates on the latent
outcome are of primary interest.

To illustrate the difference between our identification strategy and those
based on the exclusion restriction or parametric distributional assumptions,
we write the population regression function of model (\ref{semiSSM}) for the
selected sample: 
\begin{equation}
E\left[ Y\left\vert X,D=1\right. \right] =X^{\prime }\beta +\lambda \left(
h\left( X\right) \right) ,  \label{RRF1}
\end{equation}%
where $\lambda \left( t\right) =E\left[ U\left\vert \varepsilon \leq
t\right. \right] $ provided that $X\perp \left( \varepsilon ,U\right) $. The
key identifying assumption is that $\left( X^{\prime },\lambda \left(
h\left( X\right) \right) \right) $ must have full rank. In the case of $%
h\left( X\right) =X^{\prime }\gamma $, a specific nonlinear functional form
of $\lambda $ as a result of the parametric distributional assumption on $%
\left( \varepsilon ,U\right) $ will suffice for the full rank condition. The
identification power depends on the extent of nonlinearity of $\lambda $.
For instance, the inverse Mill's ratio function implied by normal
distribution is nearly piecewise linear; therefore, if the range of $%
X^{\prime }\gamma $ is small, then $\lambda \left( X^{\prime }\gamma \right) 
$ and $X$ will be highly correlated and the identification power will be low %
\citep{leung1996choice}. When $\lambda $ is nonparametrically specified
after relaxing the parametric distributional assumption, it does not
preclude the special case of $\lambda \left( t\right) =t$ in which $\left(
X^{\prime },\lambda \left( X^{\prime }\gamma \right) \right) $ suffers from
perfect multicollinearity. The exclusion restriction exploits a priori
knowledge of zero value in some component of $\beta $. This ensures
identification of the remaining coefficients because, after excluding some
covariate, the full rank condition will still be met even when $\lambda
\left( t\right) =t$. By contrast, we relax the linear specification on $h$
and then rely on nonlinear or nonmonotone variations of $h$ for point
identification of $\beta $. The functional form restriction that
differentiates $h$ from linear or generalized linear functions validates the
full rank condition of $\left( X^{\prime },\lambda \left( h\left( X\right)
\right) \right) $, while leaving $\lambda $ unspecified. Intuitively, the
nonlinear part $h\left( X\right) -X^{\prime }\gamma $, which is nonzero for
any $\gamma $, serves as an excluded variable and provides identification
power for $\beta $.

On the basis of identification, $\beta $ in model (\ref{semiSSM}) can be
estimated semiparametrically via a partially linear regression that
incorporates a generated regressor within the nonparametric function. The
generated regressor is the propensity score of selection rather than $%
h\left( X\right) $ as in the regression function (\ref{RRF1}), because the
function $h$ and the distribution function of $\varepsilon $ cannot be
separately identified when they are both specified nonparametrically %
\citep[Section 2.1]{horowitz2009semiparametric}. After estimating the
propensity score in the first step and plugging it into the nonparametric
function, we can then employ the partialling-out method %
\citep{robinson1988root} to construct a consistent estimator for $\beta $.
This naive estimator may behave well when the propensity score is estimated
by conventional nonparametric techniques that satisfy Donsker-type
properties, such as kernel and sieve estimation. However, if we use the
newly developed nonparametric techniques, the so-called machine learning
(ML) methods, in the first-step estimation, the naive estimator for $\beta $
will possess a heavy bias and fail to be $\sqrt{n}$-consistent in general %
\citep{chernozhukov2018double}, where $n$ is the sample size. To address
this problem, we derive explicit form of the first-step influence function
and accordingly construct an orthogonal moment function that is insensitive
to the first-step estimation bias \citep{chernozhukov2022locally}. By
combining with cross-fitting, we propose a locally robust semiparametric
estimator for $\beta $, and establish its $\sqrt{n}$-consistency and
asymptotic normality under a first-step convergence rate condition which can
hold for most machine learners under regularity conditions.

The paper is organized as follows. The rest of this section briefly reviews
the related literature. Section \ref{sec:ident} gives a rigorous treatment
of point identification of $\beta $ in model (\ref{semiSSM}). Particularly,
we discuss the functional form restriction on the selection equation (\ref%
{semiSSM}a), which is the key identifying assumption of our method. Section %
\ref{sec:estim} proposes an orthogonal moment condition and accordingly a
locally robust estimator, and provides sufficient conditions for the
asymptotic property of the estimator. Section \ref{sec:simul} investigates
the finite-sample property of the estimator through Monte Carlo simulations.
Section \ref{sec:appli} presents an empirical application to the wage data
of \cite{honore2020selection} and Section \ref{sec:concl} concludes. The
technical proofs and derivations are gathered in the Appendix.

\subsection{Related Literature}

There is a small but growing literature on relaxing the exclusion
restriction in the semiparametric sample selection model. In a seminal
paper, \cite{lee2009training} develops bounds for the coefficients in a
sample selection model that leaves the selection mechanism totally
unspecified. As \citeauthor{lee2009training}'s bounds are often too wide to
be informative, \cite{honore2020selection} impose a linear structure on the
selection mechanism, namely, consider model (\ref{paraSSM}), and show that
the additional structure can significantly reduce the bounds for the
parameters of interest. \cite{honore2022sample} extend the analysis in \cite%
{honore2020selection} by allowing for heteroskedasticity and parameter
heterogeneity. Our considered model (\ref{semiSSM}) is essentially the same
as \citeauthor{lee2009training}'s model in the sense that the selection
equation (\ref{semiSSM}a) is completely nonparametric in nature. The
difference lies only in the functional form restriction we impose on the
selection equation, which strongly reduces \citeauthor{lee2009training}'s
bounds to singletons. While the identified set derived by \cite%
{honore2020selection} can be small enough, this gain in identification power
stems from a linear structure that is quite restrictive and susceptible to
misspecification. In comparison, our nonlinearity or nonmonotonicity
restriction provides more identification power and, importantly, is arguably
more realistic.

The second solution to the lack of the exclusion restriction is the
so-called identification at infinity suggested by \cite%
{chamberlain1986asymptotic} and \cite{heckman1990varieties}. This approach
to identification is grounded in the observation that in model (\ref{paraSSM}%
), the sample selection problem disappears for individuals whose values of $%
X^{\prime }\gamma $ go to infinity, because these individuals face an
arbitrarily large probability of selection and a zero-valued selectivity
correction term. Accordingly, the coefficients in the outcome equation will
be point identified if a component of $X$, often called a special regressor,
exhibits infinite support. \cite{andrews1998semiparametric} propose a
semiparametric estimator for the intercept coefficient in equation (\ref%
{paraSSM}b) based on identification at infinity. \cite{lewbel2007endogenous}
generalizes this result by demonstrating that identification can be achieved
as long as the special regressor has a large, not necessarily infinite,
support that encompasses that of the error $\varepsilon $. Motivated by the
possible inaccessibility of such a special regressor, \cite{d2013another}
suggest another means of identification at infinity, under the condition
that selection becomes independent of the covariates when the outcome takes
arbitrarily large values. \cite{d2018extremal} build on the identification
result of \cite{d2013another} and develop an extremal quantile regression
estimator for the semiparametric sample selection model without the
exclusion restriction or large support regressor. Although identification at
infinity can lead to point identification, it is typically featured as
irregular identification \citep{khan2010irregular}. As a result, the derived
estimators will converge at rates slower than $1\left/ \sqrt{n}\right. $. An
alternative solution is to find additional conditions under which the
regression coefficients subject to sample selection are point identified and 
$\sqrt{n}$-estimable. For instance, \cite{chen2010semiparametric} establish
identification of $\beta $ in both models (\ref{paraSSM}) and (\ref{semiSSM}%
) under a joint symmetric distribution assumption on the errors $\varepsilon 
$ and $U$, and propose $\sqrt{n}$-consistent estimators for $\beta $. We
consider our paper as a useful supplement to this particular solution. We
show that $\beta $ in model (\ref{semiSSM}) can be identified under a
natural functional form restriction on the selection equation, thus
obviating the need for relatively more binding assumptions such as joint
symmetry.

The approach to identification based on functional form can be traced back
to \cite{heckman1979sample}, who exploits nonlinearity of the selectivity
correction function to achieve identification and $\sqrt{n}$-consistent
estimation of the parametric sample selection model. Recently, \cite%
{escanciano2016identification} extend \citeauthor{heckman1979sample}'s
approach to a general semiparametric model and establishes identification of
the linear coefficients by exploiting nonlinearity elsewhere in the model.
Nevertheless, when applied to model (\ref{semiSSM}), their identification
result implies that $\beta $ can be identified only up to scale and only
when $X$ includes at least two continuous covariates. In this paper, we
adapt the result of \cite{escanciano2016identification} to the
semiparametric sample selection model by establishing complete
identification of $\beta $, irrespective of the number of continuous
covariates and even in the absence of continuous covariates. Our result
formalizes \citet[p.13]{ahn1993semiparametric}'s allegation that
nonlinearity of the function $h$ in equation (\ref{semiSSM}a) will suffice
for identification of $\beta $. Similar identification results based on
functional form restrictions can be developed for a variety of
semiparametric models that lack exclusion restrictions, such as the
endogenous regressor binary choice model without instruments %
\citep{dong2010endogenous}, the censored regression model subject to
nonparametric sample selection \citep{pan2022semiparametric}, and the
marginal treatment effect model without instruments \citep{pan2024marginal}.
Our paper differs from this line of literature in that we allow the
first-step nonparametric estimate to be generated by modern machine
learners. Therefore, this paper also connects to the literature on
double/debiased machine learning (DML) or locally robust estimation.

ML methods perform well by employing regularization to reduce variance and
trading off regularization bias with overfitting in practice. However, ML
estimators live in highly complex spaces which fail to meet Donsker-type
conditions. Consequently, the classical asymptotic theories for
semiparametric estimators \citep[e.g.,][]{newey1994large} are inapplicable
to naive estimators that are obtained by simply plugging first-step ML
estimates into moment functions for the parameters of interest. Actually, 
\cite{chernozhukov2018double} discover that the naive estimators generally
have slower than $1\left/ \sqrt{n}\right. $ rates of convergence due to the
regularization bias in the first step. To obtain $\sqrt{n}$-consistent
estimators and simple procedures for inference, \cite{chernozhukov2018double}
propose DML estimation methods by combining orthogonalized moment functions
with cross-fitting procedures. Orthogonalization implies that moment
functions are robust to local perturbations in the first-step estimates, and
cross-fitting provides an efficient form of data splitting which removes the
bias induced by overfitting while avoiding Donsker-type conditions. \cite%
{chernozhukov2022locally} further propose a general method of
orthogonalizing moment functions by adding the corresponding first-step
influence functions. However, it is not so straightforward to derive
first-step influence functions for sample seletion models because of the
presence of the propensity score of selection as a generated regressor. \cite%
{bia2024double} devise an elaborate cross-fitting procedure to circumvent
the derivation of first-step influence functions when addressing sample
selection in evaluating effects of discretely distributed treatments.
Alternatively, we explicitly construct the first-step influence function and
thus the orthogonalized moment function by employing \cite%
{hahn2013asymptotic}'s derivation of first-step influence functions for
three-step estimators, and propose a DML/locally robust estimator for $\beta 
$ in the semiparametric sample selection model (\ref{semiSSM}). Our
construction contributes to the emerging literature on adapting DML to
models with generated regressors 
\citep[e.g.,][]{sasaki2023estimation,
escanciano2023automatic}.

\section{Identification}

\label{sec:ident}Recall that the function $h$ in the selection equation (\ref%
{semiSSM}a) cannot be identified in our setting where $h$ and the
distribution function of $\varepsilon $ are both specified
nonparametrically. To facilitate the identification of model (\ref{semiSSM}%
), we normalize equation (\ref{semiSSM}a) into a reduced form as%
\begin{equation}
D=1\left\{ \pi \left( X\right) \geq V\right\} ,  \label{sel_eq}
\end{equation}%
where $\pi \left( x\right) \equiv F_{\left. \varepsilon \right\vert X}\left(
\left. h\left( x\right) \right\vert x\right) $, $V\equiv F_{\left.
\varepsilon \right\vert X}\left( \left. \varepsilon \right\vert X\right) $,
and $F_{\left. \varepsilon \right\vert X}\left( \left. \cdot \right\vert
x\right) $ is the conditional distribution function of $\varepsilon $ given $%
X=x$. We term equation (\ref{sel_eq}) as \textquotedblleft reduced
form\textquotedblright\ because the normalized error term $V$ is by
definition statistically independent of $X$ and uniformly distributed over
the unit interval $\left[ 0,1\right] $. Moreover, by the properties of $V$,
we have%
\begin{equation*}
E\left[ \left. D\right\vert X=x\right] =\Pr \left( \left. V\leq \pi \left(
x\right) \right\vert X=x\right) =\Pr \left( V\leq \pi \left( x\right)
\right) =\pi \left( x\right) .
\end{equation*}%
Namely, $\pi \left( x\right) $ is the propensity score of selection given $%
X=x$, which is an identified quantity for any $x$ in the support of $X$.
Under the normalized or reduced-form selection equation (\ref{sel_eq}), the
functional form restriction can be imposed on the identified function $\pi $
rather than on the unidentifiable function $h$ in (\ref{semiSSM}a).

To formalize the identifying assumptions, we need several additional
notations. First, we denote $g\left( v\right) \equiv E\left[ U\left\vert
V\leq v\right. \right] $ as the selectivity correction function. Second, we
let $X$ be partitioned as $\left( X^{C},X^{D}\right) $. In general, $X^{C}$
and $X^{D}$ consist of covariates that are continuously and discretely
valued, respectively. In the special case of no continuous covariate, $X^{C}$
can be any one discrete covariate that satisfies the following Assumption %
\ref{ass:ff}.(ii), and $X^{D}$ consists of the remaining discrete
covariates. Third, we denote $X_{k}$, $X_{k}^{C}$, and $X_{k}^{D}$ as the $k$%
-th coordinates of $X$, $X^{C}$, and $X^{D}$, respectively. And we denote $%
x^{C}$ as a generic element in the support of $X^{C}$; likewise for $x^{D}$, 
$x_{k}^{C}$, and $x_{k}^{D}$. The functional form restriction will entail
setting a benchmark value of $X^{D}$ within its support, under which $\pi $
will be required to be nonlinear or nonmonotone in $X^{C}$. Without loss of
generality, we suppose that the vector of zeros is in the support of $X^{D}$
and is the benchmark value. Then, we denote $\pi _{0}\left( x^{C}\right)
=\pi \left( x^{C},0\right) $, where the discrete covariates are equal to
zero. If $\pi _{0}$ is differentiable, we further denote $\partial _{k}\pi
_{0}\left( x^{C}\right) =\left. \partial \pi _{0}\left( x^{C}\right) \right/
\partial x_{k}^{C}$\ as its partial derivative with respect to the $k$-th
argument. Moreover, for any $x_{k}^{D}\neq 0$, we denote $x^{Dk}$ as the $%
\dim \left( X^{D}\right) \times 1$ vector with the $k$-th coordinate being
equal to $x_{k}^{D}$ and all the other coordinates being equal to zero.

\begin{assumption}[Model]
\label{ass:model}Assume that (i) $\left\{ \left( Y_{i},D_{i},X_{i}\right)
\right\} _{i=1}^{n}$ is a random sample of observations from the
distribution of $\left( Y,D,X\right) $ that satisfies model (\ref{semiSSM});
and (ii) $E\left[ U\left\vert V,X\right. \right] =E\left[ U\left\vert
V\right. \right] $ with probability one, where $V\equiv F_{\left.
\varepsilon \right\vert X}\left( \left. \varepsilon \right\vert X\right) $
is the normalized error term in the selection equation.
\end{assumption}

\begin{assumption}[Functional Form]
\label{ass:ff}Assume that (i) when $\dim \left( X^{C}\right) \geq 2$, $\pi
_{0}\left( x^{C}\right) $ and $g\left( v\right) $ are differentiable
functions, and there exist two vectors $x^{C}$, $\tilde{x}^{C}$ in the
support of $X^{C}$ and two elements $k,j$ in set $\left\{ 1,2,\cdots ,\dim
\left( X^{C}\right) \right\} $ such that $\partial _{k}\pi _{0}\left(
x^{C}\right) \neq 0$, $\partial _{j}\pi _{0}\left( x^{C}\right) \neq 0$, $%
\partial _{k}\pi _{0}\left( \tilde{x}^{C}\right) \neq 0$, $\partial _{j}\pi
_{0}\left( \tilde{x}^{C}\right) \neq 0$, and $\left. \partial _{k}\pi
_{0}\left( x^{C}\right) \right/ \partial _{j}\pi _{0}\left( x^{C}\right)
\neq \left. \partial _{k}\pi _{0}\left( \tilde{x}^{C}\right) \right/
\partial _{j}\pi _{0}\left( \tilde{x}^{C}\right) $; or (ii) when $\dim
\left( X^{C}\right) =1$, there exist two constants $x^{C}$, $\tilde{x}^{C}$
in the support of $X^{C}$ such that $\pi _{0}\left( x^{C}\right) =\pi
_{0}\left( \tilde{x}^{C}\right) $.
\end{assumption}

\begin{assumption}[Support]
\label{ass:support}For each $k\in \left\{ 1,2,\cdots ,\dim \left(
X^{D}\right) \right\} $, assume for some $x_{k}^{D}\neq 0$ in the support of 
$X_{k}^{D}$ that there exists $x^{C}\left( k\right) $ in the support of $%
X^{C}$ such that $\pi \left( x^{C}\left( k\right) ,x^{Dk}\right) $ is in the
support of $\pi _{0}\left( X^{C}\right) $.
\end{assumption}

Assumption \ref{ass:model} describes the model and data. The conditional
mean independence assumption \ref{ass:model}.(ii) is implied by and much
weaker than the full independence assumption $X\perp \left( \varepsilon
,U\right) $ that is commonly imposed in the sample selection literature. In
particular, throughout our analysis, we do not impose any constraints on the
statistical relationship between $X$ and the structural selection error $%
\varepsilon $. Instead, we exploit the independence between $X$ and the
reduced-form selection error $V$, given which Assumption \ref{ass:model}%
.(ii) is essentially equivalent to the independence of the covariance of $U$
and $V$ from $X$. Under Assumption \ref{ass:model}.(ii), the population
regression function of model (\ref{semiSSM}) on the selected sample is%
\begin{equation}
E\left[ Y\left\vert X,D=1\right. \right] =X^{\prime }\beta +E\left[
U\left\vert X,V\leq \pi \left( X\right) \right. \right] =X^{\prime }\beta
+g\left( \pi \left( X\right) \right) .  \label{RF2}
\end{equation}%
Our aim is to identify $\beta $ from the regression function (\ref{RF2}).

Assumption \ref{ass:ff} imposes a functional form restriction that
distinguishes the propensity score from a linear function, given a benchmark
value of $X^{D}$. When $\dim \left( X^{C}\right) \geq 2$, where $X^{C}$ must
consist of continuous covariates, Assumption \ref{ass:ff}.(i) requires $\pi
_{0}$ to exhibit some nonlinear variations through a derivative formulation.
Specifically, Assumption \ref{ass:ff}.(i) will not hold if $\pi _{0}\left(
x^{C}\right) =f\left( x^{C\prime }\eta \right) $ for a smooth function $f$,
because in this case $\left. \partial _{k}\pi _{0}\left( x^{C}\right)
\right/ \partial _{j}\pi _{0}\left( x^{C}\right) =\left. \eta _{k}\right/
\eta _{j}$ for any $x^{C}$. Otherwise, however, it is difficult to construct
examples that violate Assumption \ref{ass:ff}.(i). When $\dim \left(
X^{C}\right) =1$, $X^{C}$ may be continuous or discrete by definition.
Assumption \ref{ass:ff}.(ii) requires the univariate function $\pi _{0}$ to
be not one-to-one, but imposes no smoothness or continuity restriction on $%
\pi _{0}$. This assumption will hold if the probability of selection is
unaffected by some change in $X^{C}$. A similar local irrelevance assumption
is imposed in nonseparable models to attain point identification %
\citep[e.g.,][]{torgovitsky2015identification, d2021testing}. In the special
case of $X^{C}$ being a binary covariate, Assumption \ref{ass:ff}.(ii)
implies the full irrelevance of $X^{C}$ to the selection probability, which
is a condition suggested by \cite{chamberlain1986asymptotic} for
identification of the linear selection model (\ref{paraSSM}).

The combination of Assumptions \ref{ass:model} and \ref{ass:ff} enables
identification of the linear coefficients of $X^{C}$ based on the functional
form difference of $\pi _{0}$. In order to identify the linear coefficients
of $X^{D}$, we further impose a support overlapping condition as Assumption %
\ref{ass:support}. This condition holds if the support of $\pi _{0}\left(
X^{C}\right) $ is overlapped with that of $\pi \left( X^{C},x^{D}\right) $
for any $x^{D}\neq 0$, or if we can find an $x_{k}^{D}\neq 0$ for each $k$
such that the support of $\pi _{0}\left( X^{C}\right) $ is overlapped with
that of $\pi \left( X^{C},x^{Dk}\right) $. Under these assumptions, we can
establish identification of the linear coefficients of interest via a
construction method.

\begin{thrm}
\label{theorem:iden} If Assumptions \ref{ass:model}-\ref{ass:support} hold,
then $\beta $ in the semiparametric sample selection model (\ref{semiSSM})
is identified.
\end{thrm}

The proof of Theorem \ref{theorem:iden} follows the method of \cite%
{pan2024marginal}. Nonetheless, for completeness, we also provide the proof
in Appendix \ref{appendix:th1}. It is worth mentioning that our
identification strategy is characterized with overidentification in the
sense that generally more than one pair of points in the support of $X^{C}$
will satisfy Assumption \ref{ass:ff}. Moreover, there may be more than one
benchmark value of $X^{D}$ that satisfies Assumptions \ref{ass:ff} and \ref%
{ass:support}. Consequently, the identification can also be represented as
an average over all pairs of points satisfying Assumption \ref{ass:ff} and
over all benchmark values satisfying Assumptions \ref{ass:ff} and \ref%
{ass:support}.

\section{Estimation}

\label{sec:estim}On the basis of identification, $\beta $ in the
semiparametric sample selection model (\ref{semiSSM}) can be estimated via
the partially linear regression (\ref{RF2}) that incorporates a generated
regressor $\pi \left( X\right) $ within the nonparametric selectivity
correction function $g$. To this end, the unknown function $\pi $ needs to
be estimated in the first step to generate $\pi \left( X\right) $. Since $%
\pi $ is specified nonparametrically, albeit subject to a nonbinding
functional form restriction, we suggest to use modern ML methods in
estimating $\pi $, which have shown desirable performance in the setting of
highly complex $\pi $ or high-dimensional $X$. Given the estimated $\pi $,
we may employ standard partially linear regression techniques to estimate $%
\beta $ by plugging in the generated regressor. One popular choice could be
the kernel-weighted pairwise difference estimator proposed by \cite%
{ahn1993semiparametric}:%
\begin{equation}
\hat{\beta}^{AP}=\arg \min_{b}\sum_{i=1}^{n-1}\sum_{j=i+1}^{n}\frac{1}{h}%
k\left( \frac{\hat{\pi}\left( X_{i}\right) -\hat{\pi}\left( X_{j}\right) }{h}%
\right) D_{i}D_{j}\left[ \left( Y_{i}-Y_{j}\right) -\left(
X_{i}-X_{j}\right) ^{\prime }b\right] ^{2},  \label{AP}
\end{equation}%
where $k\left( \cdot \right) $ and $h$ are the kernel function and
bandwidth, respectively. However, as discussed before, $\hat{\beta}^{AP}$
with plug-in machine learned $\hat{\pi}$ will generally fail to be $\sqrt{n}$%
-consistent due to the regularization bias of $\hat{\pi}$. Moreover, it is
very difficult to deduce the first-step influence function of $\hat{\pi}$
for $\hat{\beta}^{AP}$, which blocks the construction of a locally robust
estimator that asymptotically eliminates the influence of $\hat{\pi}$. This
difficulty arises because the generated regressor enters the definition of $%
\hat{\beta}^{AP}$ in the kernel weights, which implies that the generated
regressor will serve as a conditioning variable in the population moment
function of (\ref{AP}). Therefore, we consider constructing a locally robust
estimator for $\beta $ from the partialling-out estimation %
\citep{robinson1988root}. The theoretical connection between the
partialling-out estimation and the kernel-weighted pairwise difference
estimation is exposited in, e.g., \citet[p.307]{pagan1999nonparametric}.

For notational convenience, we denote $P\equiv \pi \left( X\right) $, $%
P_{i}\equiv \pi \left( X_{i}\right) $, $\hat{P}\equiv \hat{\pi}\left(
X\right) $, and $\hat{P}_{i}\equiv \hat{\pi}\left( X_{i}\right) $. And we
denote $\mu _{Z}\left( P\right) \equiv E\left[ Z\left\vert P\right. \right] $%
, with $Z$ being $Y$ or $X$.\ We start with the partially linear regression
for the whole sample:%
\begin{equation}
E\left[ Y\left\vert X\right. \right] =E\left[ \left( X^{\prime }\beta
+U\right) D\left\vert X\right. \right] =PX^{\prime }\beta +\eta \left(
P\right) ,  \label{RF3}
\end{equation}%
where $\eta \left( p\right) \equiv E\left[ U\cdot 1\left\{ V\leq p\right\} %
\right] =p\cdot g\left( p\right) $. We consider the whole-sample regression (%
\ref{RF3}) rather than the selected-sample regression (\ref{RF2}) in the
estimation mainly for two reasons. First, the whole-sample regression can
utilize the information of covariates in the unselected sample, which may
help increase the efficiency of estimation. Second, the probability space
and $\sigma $-field pertaining to whole-sample conditional expectations keep
consistent with those pertaining to the first-step propensity score, which
facilitates the derivation of the first-step influence function. To set out
the partialling-out estimation, we write the conditional expectation of $Y$
given the generated regressor based on (\ref{RF3}):%
\begin{equation}
\mu _{Y}\left( P\right) =E\left[ Y\left\vert P\right. \right] =E\left[ E%
\left[ Y\left\vert X\right. \right] \left\vert P\right. \right] =P\mu
_{X}\left( P\right) ^{\prime }\beta +\eta \left( P\right) ,  \label{RF4}
\end{equation}%
where the second equality follows from the law of iterated expectations.
Subtracting equation (\ref{RF4}) from equation (\ref{RF3}) cancels out the
unknown function $\eta $ and yields%
\begin{equation}
E\left[ Y-\mu _{Y}\left( P\right) \left\vert X\right. \right] =P\left( X-\mu
_{X}\left( P\right) \right) ^{\prime }\beta ,  \label{RF5}
\end{equation}%
according to which the \citeauthor{robinson1988root}-type partialling-out
estimator can be defined as%
\begin{eqnarray*}
\hat{\beta}^{R} &=&\arg \min_{b}\sum_{i=1}^{n}\left[ Y_{i}-\hat{\mu}%
_{Y}\left( \hat{P}_{i}\right) -\hat{P}_{i}\left( X_{i}-\hat{\mu}_{X}\left( 
\hat{P}_{i}\right) \right) ^{\prime }b\right] ^{2} \\
&=&\left[ \sum_{i=1}^{n}\hat{P}_{i}^{2}\left( X_{i}-\hat{\mu}_{X}\left( \hat{%
P}_{i}\right) \right) \left( X_{i}-\hat{\mu}_{X}\left( \hat{P}_{i}\right)
\right) ^{\prime }\right] ^{-1}\sum_{i=1}^{n}\hat{P}_{i}\left( X_{i}-\hat{\mu%
}_{X}\left( \hat{P}_{i}\right) \right) \left( Y_{i}-\hat{\mu}_{Y}\left( \hat{%
P}_{i}\right) \right) ,
\end{eqnarray*}%
where $\hat{\mu}_{Y}$ and $\hat{\mu}_{X}$ are nonparametric estimates of $%
\mu _{Y}$ and $\mu _{X}$, respectively.

In this section, we propose a locally robust estimator for $\beta $ on the
basis of $\hat{\beta}^{R}$. We derive the first-step influence function and
construct the orthogonalized moment in Subsection \ref{subsec1}, develop the
locally robust estimator in Subsection \ref{subsec2}, and investigate its
asymptotic property in Subsection \ref{subsec3}.

\subsection{Orthogonalization}

\label{subsec1}The population moment condition corresponding to $\hat{\beta}%
^{R}$ is%
\begin{equation}
E\left[ P\left( X-\mu _{X}\left( P\right) \right) \left( Y-\mu _{Y}\left(
P\right) -P\left( X-\mu _{X}\left( P\right) \right) ^{\prime }\beta \right) %
\right] =0.  \label{R_moment}
\end{equation}%
It is now well known that this \citeauthor{robinson1988root}-type moment
funciton is Neyman orthogonal with respect to $\mu _{X}$ and $\mu _{Y}$ %
\citep[e.g.,][Theorem 4.1]{chernozhukov2018double}, and hence that the
second-step estimates $\hat{\mu}_{X}$ and $\hat{\mu}_{Y}$ have no
first-order effect on the asymptotic properties of $\hat{\beta}^{R}$. In
consequence, we only need to consider the influence function of the
first-step generated regressor $\hat{P}=\hat{\pi}\left( X\right) $ in order
to construct an orthogonal moment. Note that $P$ enters the moment function
not only in a direct manner, but also in an indirect manner as an argument
of the unknown functions $\mu _{X}$ and $\mu _{Y}$. The direct effect of $%
\hat{\pi}$ is readily derived according to the approach of \cite%
{newey1994asymptotic}. In contrast, the indirect effect of $\hat{\pi}$ is
more complicated since $P$ plays a dual role in the nonparametric
regressions $\mu _{X}\left( P\right) =E\left[ X\left\vert P\right. \right] $
and $\mu _{Y}\left( P\right) =E\left[ Y\left\vert P\right. \right] $, that
is, that of conditioning variable and that of argument. Fortunately, in the
present case where the influence functions of second-step regressions are
equal to zero, it is sufficient to only account for the first-step generated
regressor as an argument \cite[Remark 3]{hahn2013asymptotic}, which
simplifies the derivation of the first-step influence function. For example,
the indirect effect of $\hat{\pi}$ as an argument of $\mu _{Y}$ on the
moment (\ref{R_moment}) is {\small 
\begin{equation*}
E\left[ \left. \frac{\partial P\left( X-\mu _{X}\left( P\right) \right)
\left( Y-\mu _{Y}\left( P\right) -P\left( X-\mu _{X}\left( P\right) \right)
^{\prime }\beta \right) }{\partial \mu _{Y}}\cdot \frac{d\mu _{Y}\left( \pi
\left( X\right) \right) }{d\pi }\right\vert X\right] =-P\left( X-\mu
_{X}\left( P\right) \right) \mu _{Y}^{\left( 1\right) }\left( P\right) ,
\end{equation*}%
} where the term in the bracket is the naive derivative of the moment
function with respect to $\pi $. The complete derivation of the first-step
influence function is provided in Appendix \ref{appendix:derivation}. Then
we can construct the orthogonal moment function by adding the first-step
influence function to the identifying moment function in (\ref{R_moment}) as%
\begin{equation*}
\psi \left( W,\pi ,\mu ,\alpha ,\beta \right) =r\left( W,\pi ,\mu ,\beta
\right) +\alpha \left( X\right) \cdot \left( D-P\right) ,
\end{equation*}%
where $W=\left( Y,D,X\right) $, $\mu =\left( \mu _{X},\mu _{Y}\right) $, and%
\begin{eqnarray}
r\left( W,\pi ,\mu ,\beta \right) &=&P\left( X-\mu _{X}\left( P\right)
\right) \left[ Y-\mu _{Y}\left( P\right) -D\left( X-\mu _{X}\left( P\right)
\right) ^{\prime }\beta \right] ,  \notag \\
\alpha \left( X\right) &=&-P\left( X-\mu _{X}\left( P\right) \right) \left[
\mu _{Y}^{\left( 1\right) }\left( P\right) -P\mu _{X}^{\left( 1\right)
}\left( P\right) ^{\prime }\beta \right] ,  \label{alpha}
\end{eqnarray}%
with $\mu _{Y}^{\left( 1\right) }$ and $\mu _{X}^{\left( 1\right) }$ being
the derivative functions of $\mu _{Y}$ and $\mu _{X}$, respectively.
Therefore, the orthogonalized moment condition for $\beta $ derived from the %
\citeauthor{robinson1988root}-type moment (\ref{R_moment}) is%
\begin{equation}
E\left[ \psi \left( W,\pi ,\mu ,\alpha ,\beta \right) \right] =0.
\label{O_moment}
\end{equation}

As revealed by \cite{chernozhukov2022locally}, the moment condition (\ref%
{O_moment}) has two key orthogonality properties. First, varying the
additional nuisance function $\alpha $ away from its true value has no
effect, globally, on $E\left[ \psi \left( W,\pi ,\mu ,\alpha ,\beta \right) %
\right] $, which is a direct consequence of the definition of $\psi $.

\begin{lmm}
\label{lemma1}Denote $\tilde{\alpha}$ and $\tilde{\mu}$ as generic functions
in the sets of possible values of $\alpha $ and $\mu $, respectively, and $%
\tilde{\beta}$ as a generic vector in the set of possible values of $\beta $%
. Then we have%
\begin{equation*}
E\left[ \psi \left( W,\pi ,\tilde{\mu},\tilde{\alpha},\tilde{\beta}\right) %
\right] =E\left[ \psi \left( W,\pi ,\tilde{\mu},\alpha ,\tilde{\beta}\right) %
\right]
\end{equation*}%
for any $\tilde{\alpha}$, $\tilde{\mu}$, and $\tilde{\beta}$.
\end{lmm}

\noindent Second, varying the first-step nuisance function $\pi $ away from
its true value has no effect, locally, on $E\left[ \psi \left( W,\pi ,\mu
,\alpha ,\beta \right) \right] $, which is an expected consequence of
bringing in the first-step influence function.

\begin{lmm}
\label{lemma2}(i) Suppose that $\mu _{Y}$ and $\mu _{X}$ are differentiable,
and that the condition for interchangeability of the derivative and
expectation holds. Then we have%
\begin{equation*}
\left. \frac{\partial }{\partial t}E\left[ \psi \left( W,\pi +t\delta ,\mu
,\alpha ,\beta \right) \right] \right\vert _{t=0}=0
\end{equation*}%
for any $\delta $ as a possible direction of deviation of $\pi $ from the
true value, where $t$ is a scalar representing the size of the deviation and
the derivative is evaluated at $t=0$.

(ii) Suppose that $\mu _{Y}$ and $\mu _{X}$ are second-order continuously
differentiable, with their first and second derivatives bounded. Suppose
that $E\left[ \left\Vert X\right\Vert ^{s}\right] <\infty $ for a constant $%
s>1$. Let $q=2s\left/ \left( s-1\right) \right. >2$. Then for any $\tilde{\pi%
}$ in a small neighborhood of $\pi $, we have%
\begin{equation*}
\left\Vert E\left[ \psi \left( W,\tilde{\pi},\mu ,\alpha ,\beta \right) %
\right] \right\Vert \leq C\left\Vert \tilde{\pi}-\pi \right\Vert _{q,F}^{2}
\end{equation*}%
for a constant $C$, where $\left\Vert \tilde{\pi}-\pi \right\Vert
_{q,F}=\left( E\left[ \left\vert \tilde{\pi}\left( X\right) -\pi \left(
X\right) \right\vert ^{q}\right] \right) ^{1/q}$ denotes the $L_{q}\left(
F\right) $-norm, with $F$ representing the distribution law of $X$.
\end{lmm}

Lemma \ref{lemma2}.(i) is a directional derivative characterization of
orthogonality with respect to the generated regressor, while Lemma \ref%
{lemma2}.(ii) bounds the departure from zero of the expected moment function
as just $\pi $ varies. The bound implies that the orthogonal moment shrinks
to zero at a squared rate of convergence of the first-step estimation, which
will be useful for establishing $\sqrt{n}$-consistency of the estimation of $%
\beta $. Lemmas \ref{lemma1} and \ref{lemma2} follow from direct inspection
of conditions in Theorems 2 and 3 of \cite{chernozhukov2022locally}. For
completeness, we give an explicit derivation of these orthogonality
properties in Appendix \ref{appendix:lemmas}. The following lemma extends
the rate result in Lemma \ref{lemma2}.(ii) to the case of varying $\pi $ and 
$\mu $ simultaneously.

\begin{lmm}
\label{lemma3}Suppose that the conditions in Lemma \ref{lemma2}.(ii) hold.
Then for any $\tilde{\pi}$ and $\tilde{\mu}$ in small neighborhoods of $\pi $
and $\mu $, respectively, we have%
\begin{eqnarray*}
\left\Vert E\left[ \psi \left( W,\tilde{\pi},\tilde{\mu},\alpha ,\beta
\right) \right] \right\Vert &\leq &C_{1}\left\Vert \tilde{\pi}-\pi
\right\Vert _{q,F}^{2}+C_{2}\left\Vert \tilde{\mu}-\mu \right\Vert _{q,F}^{2}
\\
&&+C_{3}\left\Vert \tilde{\pi}-\pi \right\Vert _{q,F}\left\Vert \tilde{\mu}%
-\mu \right\Vert _{q,F} \\
&&+C_{4}\left\Vert \tilde{\pi}-\pi \right\Vert _{q,F}\left\Vert \tilde{\mu}%
^{\left( 1\right) }-\mu ^{\left( 1\right) }\right\Vert _{q,F}
\end{eqnarray*}%
for constants $C_{j}$ ($j=1,2,3,4$), where%
\begin{eqnarray*}
\left\Vert \tilde{\mu}-\mu \right\Vert _{q,F} &=&\left( E\left[ \left\vert 
\tilde{\mu}_{Y}\left( P\right) -\mu _{Y}\left( P\right) \right\vert ^{q}%
\right] \right) ^{1/q}+\left( E\left[ \left\Vert \tilde{\mu}_{X}\left(
P\right) -\mu _{X}\left( P\right) \right\Vert ^{q}\right] \right) ^{1/q}, \\
\left\Vert \tilde{\mu}^{\left( 1\right) }-\mu ^{\left( 1\right) }\right\Vert
_{q,F} &=&\left( E\left[ \left\vert \tilde{\mu}_{Y}^{\left( 1\right) }\left(
P\right) -\mu _{Y}^{\left( 1\right) }\left( P\right) \right\vert ^{q}\right]
\right) ^{1/q}+\left( E\left[ \left\Vert \tilde{\mu}_{X}^{\left( 1\right)
}\left( P\right) -\mu _{X}^{\left( 1\right) }\left( P\right) \right\Vert ^{q}%
\right] \right) ^{1/q}.
\end{eqnarray*}
\end{lmm}

\subsection{Locally Robust Estimation}

\label{subsec2}We follow the DML method that combines orthogonal moment
functions with cross-fitting, to construct debiased sample moments and thus
a locally robust estimator for the target parameter $\beta $. Cross-fitting
means that the moment function $\psi $ for each observation is evaluated at
estimates of $\left( \pi ,\mu ,\alpha \right) $ that only use other
observations. Specifically, partition the sample of observations $\left\{
W_{i}=\left( Y_{i},D_{i},X_{i}\right) \right\} _{i=1}^{n}$ into $L$
subsamples $\left\{ W_{i}\right\} _{i\in I_{\ell }}$ ($\ell =1,2,\cdots ,L$%
), such that $\bigcup_{\ell =1}^{L}I_{\ell }=\left\{ 1,2,\cdots ,n\right\} $
and $I_{\ell }\cap I_{\ell ^{\prime }}=\varnothing $. Denote $\left( \hat{\pi%
}_{\ell },\hat{\mu}_{\ell },\hat{\alpha}_{\ell }\right) $ to be estimates of 
$\left( \pi ,\mu ,\alpha \right) $ that are constructed using all
observations not in $I_{\ell }$, and denote $\hat{P}_{\ell i}=\hat{\pi}%
_{\ell }\left( X_{i}\right) $. Then the debiased sample moment function is%
\begin{equation*}
\hat{\Psi}_{n}\left( \beta \right) =\frac{1}{n}\sum_{\ell =1}^{L}\sum_{i\in
I_{\ell }}\psi \left( W_{i},\hat{\pi}_{\ell },\hat{\mu}_{\ell },\hat{\alpha}%
_{\ell },\beta \right) =\frac{1}{n}\sum_{\ell =1}^{L}\sum_{i\in I_{\ell }}%
\left[ r\left( W_{i},\hat{\pi}_{\ell },\hat{\mu}_{\ell },\beta \right) +\hat{%
\alpha}_{\ell }\left( X_{i}\right) \left( D_{i}-\hat{P}_{\ell i}\right) %
\right] ,
\end{equation*}%
and the locally robust estimator $\hat{\beta}^{LR}$ is defined as a solution
to the moment condition $\hat{\Psi}_{n}\left( \beta \right) =0$, which has
an explicit form:%
\begin{eqnarray*}
\hat{\beta}^{LR} &=&\left[ \frac{1}{n}\sum_{\ell =1}^{L}\sum_{i\in I_{\ell
}}D_{i}\hat{P}_{\ell i}\left( X_{i}-\hat{\mu}_{X\ell }\left( \hat{P}_{\ell
i}\right) \right) \left( X_{i}-\hat{\mu}_{X\ell }\left( \hat{P}_{\ell
i}\right) \right) ^{\prime }\right] ^{-1} \\
&&\cdot \frac{1}{n}\sum_{\ell =1}^{L}\sum_{i\in I_{\ell }}\left[ \hat{P}%
_{\ell i}\left( X_{i}-\hat{\mu}_{X\ell }\left( \hat{P}_{\ell i}\right)
\right) \left( Y_{i}-\hat{\mu}_{Y\ell }\left( \hat{P}_{\ell i}\right)
\right) +\hat{\alpha}_{\ell }\left( X_{i}\right) \left( D_{i}-\hat{P}_{\ell
i}\right) \right] .
\end{eqnarray*}

The first-step estimation $\hat{P}_{\ell i}=\hat{\pi}_{\ell }\left(
X_{i}\right) $ for $P_{i}=\pi \left( X_{i}\right) =E\left[ D_{i}\left\vert
X_{i}\right. \right] $\ can be generated by any of the ML methods, such as
random forests, neural nets, lasso or post-lasso, boosted regression trees,
and various hybrids and ensembles of these methods. The cross-fitting device
eliminates the need for Donsker conditions for $\hat{\pi}_{\ell }$, which is
important because most ML methods are not known to satisfy such conditions.
The only requirement for $\hat{\pi}_{\ell }$ is a convergence rate condition
given by Assumption 4 in the following subsection. We will adopt an
off-the-shelf random forest algorithm in the first step in subsequent
simulation and application.

The second-step estimation $\hat{\mu}_{\ell }=\left( \hat{\mu}_{Y\ell },\hat{%
\mu}_{X\ell }\right) $ for $\mu _{Y}\left( p\right) =E\left[ Y\left\vert
P=p\right. \right] $ and $\mu _{X}\left( p\right) =E\left[ X\left\vert
P=p\right. \right] $ can also be constructed by ML methods on account of the
nonparametric nature of $\mu =\left( \mu _{Y},\mu _{X}\right) $. However,
since the regressions in this step are all univariate, we will adopt the
kernel regression estimation for computational simplicity:%
\begin{equation}
\hat{\mu}_{Z\ell }\left( p\right) =\left. \left[ \sum_{\ell ^{\prime }\neq
\ell }\sum_{j\in I_{\ell ^{\prime }}}\frac{1}{h_{Z}}k\left( \frac{\hat{P}%
_{\ell \ell ^{\prime }j}-p}{h_{Z}}\right) Z_{j}\right] \right/ \left[
\sum_{\ell ^{\prime }\neq \ell }\sum_{j\in I_{\ell ^{\prime }}}\frac{1}{h_{Z}%
}k\left( \frac{\hat{P}_{\ell \ell ^{\prime }j}-p}{h_{Z}}\right) \right] ,
\label{muhat}
\end{equation}%
where $Z$ represents $Y$ or any element of $X$, $k\left( \cdot \right) $ and 
$h_{Z}$ are the kernel function and regression-specific bandwidth, and $\hat{%
P}_{\ell \ell ^{\prime }j}=\hat{\pi}_{\ell \ell ^{\prime }}\left(
Z_{j}\right) $ with $\hat{\pi}_{\ell \ell ^{\prime }}$ being an ML estimate
of $\pi $ using all observations not in $I_{\ell }$ and not in $I_{\ell
^{\prime }}$.

Lastly, the additional nuisance parameter $\alpha $ can be estimated by
plugging proper estimates of the unknown components into (\ref{alpha}):%
\begin{equation}
\hat{\alpha}_{\ell }\left( X_{i}\right) =-\hat{P}_{\ell i}\left( X_{i}-\hat{%
\mu}_{X\ell }\left( \hat{P}_{\ell i}\right) \right) \left[ \hat{\mu}_{Y\ell
}^{\left( 1\right) }\left( \hat{P}_{\ell i}\right) -\hat{P}_{\ell i}\hat{\mu}%
_{X\ell }^{\left( 1\right) }\left( \hat{P}_{\ell i}\right) ^{\prime }\hat{%
\beta}_{\ell }\right] ,  \label{alphahat}
\end{equation}%
where $\hat{\mu}_{X\ell }^{\left( 1\right) }\left( p\right) =\left. d\hat{\mu%
}_{X\ell }\left( p\right) \right/ dp$, $\hat{\mu}_{Y\ell }^{\left( 1\right)
}\left( p\right) =\left. d\hat{\mu}_{Y\ell }\left( p\right) \right/ dp$, and 
$\hat{\beta}_{\ell }$ is an initial estimate of $\beta $ that is formed
using all observations not in $I_{\ell }$. We adopt the cross-fitted
partialling-out estimator as $\hat{\beta}_{\ell }$:%
\begin{eqnarray*}
\hat{\beta}_{\ell } &=&\left[ \sum_{\ell ^{\prime }\neq \ell }\sum_{j\in
I_{\ell ^{\prime }}}\hat{P}_{\ell \ell ^{\prime }j}^{2}\left( X_{j}-\hat{\mu}%
_{X\ell \ell ^{\prime }}\left( \hat{P}_{\ell \ell ^{\prime }j}\right)
\right) \left( X_{j}-\hat{\mu}_{X\ell \ell ^{\prime }}\left( \hat{P}_{\ell
\ell ^{\prime }j}\right) \right) ^{\prime }\right] ^{-1} \\
&&\cdot \sum_{\ell ^{\prime }\neq \ell }\sum_{j\in I_{\ell ^{\prime }}}\hat{P%
}_{\ell \ell ^{\prime }j}\left( X_{j}-\hat{\mu}_{X\ell \ell ^{\prime
}}\left( \hat{P}_{\ell \ell ^{\prime }j}\right) \right) \left( Y_{j}-\hat{\mu%
}_{Y\ell \ell ^{\prime }}\left( \hat{P}_{\ell \ell ^{\prime }j}\right)
\right) .
\end{eqnarray*}%
Under the specification (\ref{alphahat}) of $\hat{\alpha}_{\ell }$, the
locally robust estimator can also be written as%
\begin{eqnarray*}
\hat{\beta}^{LR} &=&\left[ \frac{1}{n}\sum_{\ell =1}^{L}\sum_{i\in I_{\ell
}}D_{i}\hat{P}_{\ell i}\left( X_{i}-\hat{\mu}_{X\ell }\left( \hat{P}_{\ell
i}\right) \right) \left( X_{i}-\hat{\mu}_{X\ell }\left( \hat{P}_{\ell
i}\right) \right) ^{\prime }\right] ^{-1}\frac{1}{n}\sum_{\ell
=1}^{L}\sum_{i\in I_{\ell }}\hat{P}_{\ell i}\left( X_{i}-\hat{\mu}_{X\ell
}\left( \hat{P}_{\ell i}\right) \right) \\
&&\cdot \left[ Y_{i}-\hat{\mu}_{Y\ell }\left( \hat{P}_{\ell i}\right)
-\left( D_{i}-\hat{P}_{\ell i}\right) \left( \hat{\mu}_{Y\ell }^{\left(
1\right) }\left( \hat{P}_{\ell i}\right) -\hat{P}_{\ell i}\hat{\mu}_{X\ell
}^{\left( 1\right) }\left( \hat{P}_{\ell i}\right) ^{\prime }\hat{\beta}%
_{\ell }\right) \right] .
\end{eqnarray*}

\subsection{Asymptotic Property}

\label{subsec3}To investigate the asymptotic property of the proposed
estimator, we make the following assumptions.

\begin{assumption}
\label{ass:pd}The matrix $M=E\left[ P^{2}\left( X-\mu _{X}\left( P\right)
\right) \left( X-\mu _{X}\left( P\right) \right) ^{\prime }\right] $ is
positive definite, and $E\left[ \left\Vert \psi \left( W,\pi ,\mu ,\alpha
,\beta \right) \right\Vert ^{2}\right] <\infty $.
\end{assumption}

\begin{assumption}
\label{ass:rate}For each $\ell =1,\cdots ,L$,%
\begin{eqnarray*}
n^{1/4}\left\Vert \hat{\pi}_{\ell }-\pi \right\Vert _{q,F} &\overset{p}{%
\longrightarrow }&0, \\
n^{1/4}\left\Vert \hat{\mu}_{\ell }-\mu \right\Vert _{q,F} &\overset{p}{%
\longrightarrow }&0, \\
n^{1/4}\left\Vert \hat{\mu}_{\ell }^{\left( 1\right) }-\mu ^{\left( 1\right)
}\right\Vert _{q,F} &\overset{p}{\longrightarrow }&0, \\
n^{1/4}\left\Vert \hat{\alpha}_{\ell }-\alpha \right\Vert _{q,F} &\overset{p}%
{\longrightarrow }&0,
\end{eqnarray*}%
where $\left\Vert \cdot \right\Vert _{q,F}$ denotes the $L_{q}\left(
F\right) $-norm with $q>2$ given in Lemma \ref{lemma2}.(ii).
\end{assumption}

Assumption \ref{ass:pd} is standard and it ensures nonsingularity of the
asymptotic covariance matrix of the estimator. Assumption \ref{ass:rate}
requires an $n^{-1/4}$-rate of convergence for the first-step machine
learner as is familiar from the DML literature. The difference resulting
from the multi-step nature of our estimator lies in two aspects. First, we
need a rate of $L_{q}\left( F\right) $-convergence that implies the usual
mean-square convergence. Second, the construction of $\hat{\mu}_{\ell }$
needs the first-step estimation of $\pi $ as an input, such as in (\ref%
{muhat}), which implicitly imposes additional restrictions on the first-step
estimation. \cite{sperlich2009note} and \cite{mammen2012nonparametric}
derive the convergence rate of the kernel estimation with a generated
regressor, under a bias-variance structure or a complexity restriction on
the generated regressor, respectively. However, such structure or complexity
restriction is tailored to the traditional kernel or sieve estimators and
not known to hold for most ML estimators. In comparison, we employ the
cross-fitting devise in the construction of $\hat{\mu}_{\ell }$ as in (\ref%
{muhat}) and provide an alternative set of sufficient conditions for
Assumption \ref{ass:rate} that only require a faster rate of the first-step
estimation but without requiring any structure or complexity restrictions.

\begin{assumption}
\label{ass:rate1}(i) $\hat{\mu}_{\ell }$ and $\hat{\alpha}_{\ell }$ are
defined in (\ref{muhat}) and (\ref{alphahat}), respectively, and $\hat{\mu}%
_{\ell }^{\left( 1\right) }\left( p\right) =\left. d\hat{\mu}_{\ell }\left(
p\right) \right/ dp$. (ii) For each $\ell =1,\cdots ,L$ and $\ell ^{\prime
}\neq \ell $, we have $n^{1/4}\left\Vert \hat{\pi}_{\ell }-\pi \right\Vert
_{q,F}\overset{p}{\longrightarrow }0$ and $\left. n^{1/4}\left\Vert \hat{\pi}%
_{\ell \ell ^{\prime }}-\pi \right\Vert _{q,F}\right/ h_{Z}^{1+1/q}\overset{p%
}{\longrightarrow }0$, where $h_{Z}$ is the bandwidth specific to $Z$ that
represents $Y$ or any element of $X$. (iii) The probability density function
of $P$, $f_{P}\left( p\right) $, is bounded above and below from zero over $%
p\in \left( 0,1\right) $, and is differentiable with the derivative function
satisfying the Lipschitz-continuity condition $\left\vert f_{P}^{\left(
1\right) }\left( p_{1}\right) -f_{P}^{\left( 1\right) }\left( p_{2}\right)
\right\vert \leq C\left\vert p_{1}-p_{2}\right\vert $ for some $C>0$. (iv)
For $Z$ being $Y$ or any element of $X$, $\mu _{Z}\left( p\right) $ is twice
differentiable with the derivative functions being Lipschitz-continuous, and 
$E\left[ \left. \left\vert Z\right\vert ^{q\left/ \left( q-1\right) \right.
}\right\vert P=p\right] $ is Lipschitz-continuous and bounded over $p\in
\left( 0,1\right) $. (v) The kernel function $k\left( \cdot \right) $ is
symmetric, is twice differentiable with bounded derivatives, has compact
support, and satisfies $\left\vert \left\vert v_{1}\right\vert ^{l}k\left(
v_{1}\right) -\left\vert v_{2}\right\vert ^{l}k\left( v_{2}\right)
\right\vert \leq C\left\vert v_{1}-v_{2}\right\vert $ for some $C>0$ for all 
$0\leq l\leq 3$. (vi) For $Z$ being $Y$ or any element of $X$, the sequence
of bandwidths $h_{Z}=h_{Z}\left( n\right) $ satisfies $n^{1/8}h_{Z}%
\rightarrow 0$, $n^{1/4}h_{Z}^{1-2/q}\rightarrow \infty $, and $\left.
n^{1/2}h_{Z}\right/ \ln n\rightarrow \infty $ as $n$ goes to infinity.
\end{assumption}

\begin{lmm}
\label{lemma4}Assumption \ref{ass:rate1} implies Assumption \ref{ass:rate}.
\end{lmm}

Along with regularity conditions that are standard in the kernel estimation,
Assumption \ref{ass:rate1} requires a faster than $n^{-1/4}$ rate for the
preliminary machine learner $\hat{\pi}_{\ell \ell ^{\prime }}$.
Specifically, Assumption \ref{ass:rate1}.(ii) and (vi) imply $\left\Vert 
\hat{\pi}_{\ell \ell ^{\prime }}-\pi \right\Vert _{q,F}=o_{p}\left(
n^{-3/8}\right) $. Under these assumptions, we can establish $\sqrt{n}$%
-consistency and asymptotic normality of the proposed estimator for the
sample selection model without exclusion restrictions.

\begin{thrm}
\label{theorem:AN}Suppose that the conditions in Lemma \ref{lemma2}.(ii) and
Assumptions \ref{ass:model}, \ref{ass:pd}, and \ref{ass:rate} hold. Then the
locally robust estimator $\hat{\beta}^{LR}$ for $\beta $ is $\sqrt{n}$%
-consistent and asymptotically normal:%
\begin{equation*}
\sqrt{n}\left( \hat{\beta}^{LR}-\beta \right) \longrightarrow N\left(
0,M^{-1}\Sigma M^{-1}\right) ,
\end{equation*}%
where $M$ is defined in Assumption \ref{ass:pd} and $\Sigma =E\left[ \psi
\left( W,\pi ,\mu ,\alpha ,\beta \right) \psi \left( W,\pi ,\mu ,\alpha
,\beta \right) ^{\prime }\right] .$
\end{thrm}

To conduct statistical inference, we need to construct a consistent
estimator for $M^{-1}\Sigma M^{-1}$. A choice at hand is to estimate $M$ and 
$\Sigma $ separately by%
\begin{eqnarray*}
\hat{M}_{n} &=&\frac{1}{n}\sum_{\ell =1}^{L}\sum_{i\in I_{\ell }}D_{i}\hat{P}%
_{\ell i}\left( X_{i}-\hat{\mu}_{X\ell }\left( \hat{P}_{\ell i}\right)
\right) \left( X_{i}-\hat{\mu}_{X\ell }\left( \hat{P}_{\ell i}\right)
\right) ^{\prime }, \\
\hat{\Sigma}_{n} &=&\frac{1}{n}\sum_{\ell =1}^{L}\sum_{i\in I_{\ell }}\psi
\left( W_{i},\hat{\pi}_{\ell },\hat{\mu}_{\ell },\hat{\alpha}_{\ell },\hat{%
\beta}^{LR}\right) \psi \left( W_{i},\hat{\pi}_{\ell },\hat{\mu}_{\ell },%
\hat{\alpha}_{\ell },\hat{\beta}^{LR}\right) ^{\prime }.
\end{eqnarray*}%
The consistency of $\hat{M}_{n}^{-1}$ for $M^{-1}$ has been shown in the
proof of Theorem \ref{theorem:AN}. By the same logic, it is ready to show
the consistency of $\hat{\Sigma}_{n}$ for $\Sigma $. It follows that $\hat{M}%
_{n}^{-1}\hat{\Sigma}_{n}\hat{M}_{n}^{-1}\overset{p}{\longrightarrow }%
M^{-1}\Sigma M^{-1}$.

\section{Simulation}

\label{sec:simul}We examine the finite sample property of the locally robust
estimator $\hat{\beta}^{LR}$ by a Monte Carlo simulation. The simulation
data are generated from the semiparametric sample selection model (\ref%
{semiSSM}) without any exclusion restriction, which means that all
covariates enter the outcome equation (\ref{paraSSM}b) with nonzero
coefficients. The vector of covariates consists of ten dimensions, that is, $%
X=\left( X_{1},\cdots ,X_{10}\right) ^{\prime }$, where every two adjacent
covariates are statistically correlated with each other with a common
correlation coefficient 0.5. To simulate the situation usually encountered
in practice, we let the first two covariates to be continuously distributed
and let the remaining ones to be discretely distributed. Specifically, $%
X_{1} $ follows a standard normal distribution, $X_{2}$ follows a uniform
distribution over the unit interval $\left[ 0,1\right] $, and $X_{k}$'s ($%
k=3,\cdots ,10$) are $\left\{ 0,1\right\} $-valued binary variables with $%
\Pr \left( X_{k}=1\right) =0.5$. The error term $\varepsilon $ is generated
from a standardized distribution with zero mean and unit variance, and $U$
is generated by $U=\rho \varepsilon +\sqrt{1-\rho ^{2}}e$, where $e$ is a
standard normal random variable independent of $\varepsilon $. Hence, $E%
\left[ U\right] =E\left[ \varepsilon \right] =0$ and $Corr\left(
U,\varepsilon \right) =\rho $. The linear coefficients are all set to one,
that is, $\beta =\boldsymbol{1}_{10\times 1}$. The nonlinear function $%
h\left( \cdot \right) $ in the selection equation (\ref{paraSSM}a) is set to
be $h\left( x\right)
=c+x_{1}+x_{1}^{2}-x_{2}-x_{2}^{2}+x_{3}x_{4}-x_{5}x_{6} $, where the
constant $c$ is introduced to control for the censoring rate $\Pr \left(
D=0\right) =\Pr \left( h\left( X\right) <\varepsilon \right) $. In the
benchmark setting, $c$ is set to produce approximately 50\% censoring, $%
\varepsilon $ follows a standard normal distribution such that $\left(
\varepsilon ,U\right) $ follows a bivariate normal distribution with the
correlation coefficient being $\rho $, and $\rho $ is set 0.5. Different
designs are constructed by varying the censoring rate $\Pr \left( D=0\right) 
$, the correlation coefficient $\rho $, the distribution of $\varepsilon $,
as well as the functional form of $h\left( \cdot \right) $. The sample size
is set $n=250,500,1000$, and the simulation replicates $100$ times for each
design.

In every replication $b=1,\cdots ,100$, we compute the proposed estimator $%
\hat{\beta}^{LR,b}$ by using a random forest as the first-step estimate of
the propensity score of selection, using the kernel regression (\ref{muhat})
as the second-step estimate of $\mu =\left( \mu _{X},\mu _{Y}\right) $, and
plugging in the corresponding estimates of unknown components of the
first-step influence function as in (\ref{alphahat}). The random forest is
an ensemble learning technique that estimates the propensity score by
averaging predictions from multiple decision trees \citep{biau2016random}.
While various machine learners are available, the random forest stands out
for the ability to capture complex relationships in data. This makes it well
suited for estimating the regression relationship between the sample
selection indicator and covariates, which is assumed to be nonlinear
(Assumption \ref{ass:ff}). The random forest has gained popularity in recent
years and is frequently applied in the DML literature, such as in \cite%
{chernozhukov2018double}. We employ a cross-validation strategy to tune the
hyperparameters of the random forest, such as the number of trees in the
forest and the minimum number of observations required at each leaf node.
The number of folds in the cross-fitting and cross-validation is set five.
In the second-step kernel estimation, we choose the Gaussian kernel and
rule-of-thumb bandwidths for simplicity. Along with $\hat{\beta}^{LR,b}$, we
compute the \citeauthor{robinson1988root}-type partialling-out estimator $%
\hat{\beta}^{R,b}$ as comparison, which is the naive (i.e.,
non-orthogonalized, non-cross-fitting) version of $\hat{\beta}^{LR,b}$. As
in $\hat{\beta}^{LR,b}$, we compute $\hat{\beta}^{R,b}$ by using the random
forest as the first-step estimate and using the kernel regression as the
second-step estimate, but without cross-fitting. As mentioned before, $\hat{%
\beta}^{R,b}$\ would likely fail to be $\sqrt{n}$-consistent due to the
first-order influence of the first-step learning bias.

For every component $k=1,\cdots ,10$ of each estimator $\hat{\beta}$, we
summarize simulation results by the absolute bias $Bias\left( \hat{\beta}%
_{k}\right) =\left. \sum_{b=1}^{100}\left\vert \hat{\beta}%
_{k}^{b}-1\right\vert \right/ 100$, the standard deviation $SD\left( \hat{%
\beta}_{k}\right) =\left( \left. \sum_{b=1}^{100}\left[ \hat{\beta}%
_{k}^{b}-\left( \left. \sum_{b=1}^{100}\hat{\beta}_{k}^{b}\right/ 100\right) %
\right] ^{2}\right/ 100\right) ^{1/2}$, and the coverage probability of an
asymptotic 95\% confidence interval defined as $Coverage\left( \hat{\beta}%
_{k}\right) =\left. \sum_{b=1}^{100}1\left\{ \left\vert \hat{\beta}%
_{k}^{b}-1\right\vert \leq 1.96\times SE\left( \hat{\beta}_{k}\right)
\right\} \right/ 100$, where $SE\left( \hat{\beta}_{k}\right) =\left[ \left.
\left( \hat{M}_{n}^{-1}\hat{\Sigma}_{n}\hat{M}_{n}^{-1}\right) _{k,k}\right/
n\right] ^{1/2}$. Considering that the primary aim of the locally robust
estimation procedure is to provide a valid method of conducting inference,
we focus mainly on the estimators' performance in terms of the coverage
probability. In Table \ref{table1}, we report the average absolute bias $%
\left. \sum_{k=1}^{10}Bias\left( \hat{\beta}_{k}\right) \right/ 10$, the
average standard deviation $\left. \sum_{k=1}^{10}SD\left( \hat{\beta}%
_{k}\right) \right/ 10$, the average coverage probability $\left.
\sum_{k=1}^{10}Coverage\left( \hat{\beta}_{k}\right) \right/ 10$, the
maximum coverage probability $\max_{k}Coverage\left( \hat{\beta}_{k}\right) $%
, and the minimum coverage probability $\min_{k}Coverage\left( \hat{\beta}%
_{k}\right) $ of $\hat{\beta}^{LR}$ and $\hat{\beta}^{R}$ in Columns (1) and
(2), respectively.

\begin{table}[tbh]
\caption{Simulation results for the benchmark setting}
\label{table1}
\begin{center}
\begin{tabular}{cccccc}
\hline\hline
&  & (1) & (2) & (3) & (4) \\ 
&  & \ \ Locally \ \  & \ Robinson \  & Robinson with & Robinson with \\ 
&  & Robust &  & Orthogonalization & Cross-fitting \\ \hline
\multicolumn{6}{l}{\textbf{Panel A:} $n=250$} \\ 
Average Bias &  & 0.240 & 0.172 & 0.118 & 0.535 \\ 
Average SD &  & 0.257 & 0.170 & 0.140 & 0.435 \\ 
Average Coverage &  & 0.953 & 0.874 & 0.918 & 0.638 \\ 
Max Coverage &  & 0.990 & 0.960 & 0.990 & 0.870 \\ 
Min Coverage &  & 0.860 & 0.670 & 0.810 & 0.070 \\ \hline
\multicolumn{6}{l}{\textbf{Panel B:} $n=500$} \\ 
Average Bias &  & 0.154 & 0.133 & 0.084 & 0.364 \\ 
Average SD &  & 0.151 & 0.117 & 0.098 & 0.265 \\ 
Average Coverage &  & 0.950 & 0.824 & 0.922 & 0.581 \\ 
Max Coverage &  & 1.000 & 0.970 & 0.960 & 0.900 \\ 
Min Coverage &  & 0.820 & 0.580 & 0.830 & 0.010 \\ \hline
\multicolumn{6}{l}{\textbf{Panel C:} $n=1000$} \\ 
Average Bias &  & 0.099 & 0.106 & 0.060 & 0.248 \\ 
Average SD &  & 0.095 & 0.086 & 0.071 & 0.161 \\ 
Average Coverage &  & 0.948 & 0.760 & 0.907 & 0.525 \\ 
Max Coverage &  & 1.000 & 0.930 & 0.960 & 0.930 \\ 
Min Coverage &  & 0.880 & 0.260 & 0.810 & 0.020 \\ \hline\hline
\end{tabular}%
\end{center}
\end{table}

As the sample size $n$ increases, the bias and standard deviation of the
locally robust estimator decrease at a roughly root-$n$ rate as expected.
Moreover, the locally robust estimator performs rather well in terms of the
coverage probability regardless of $n$. In comparison, the Robinson-type
estimator's bias decreases at a rate that is evidently slower than $n^{-1/2}$
as $n$ increases, which is in accordance with the theoretical conjecture
that the Robinson-type estimator with ML first-step cannot achieve $\sqrt{n}$%
-consistency. As a result, its coverage probability deviates from the
nominal level even farther for larger $n$. This means that the naive
Robinson-type estimator for the sample selection model will generally lead
to misleading conclusions in inference and hypothesis testing. The locally
robust approach corrects for the size distortion of hypothesis testing
constructed by the Robinson-type estimator.

Nevertheless, the bias of the locally robust estimator seems a bit larger
than it is supposed to be. In particular, when the sample size is small, the
locally robust estimator is more biased than the naive Robinson-type
estimator. To probe into the reason of this phenomenon, we separately
consider orthogonalization and cross-fitting, the two ingredients of DML.
Namely, we compute the Robinson estimator with orthogonalization (or,
equivalently, the locally robust estimator without cross-fitting) and the
Robinson estimator with cross-fitting (or the locally robust estimator
without orthogonalization). The results of these two auxiliary estimators
are reported in Columns (3) and (4) of Table \ref{table1}.

By comparing Column (3) with Column (2), we can see that adding the
first-step influence function effectively reduces the estimation bias of the
target parameter, which is in agreement with the reduction of the influence
of the first-step learning bias from first order to second order in theory.
Additionally, the orthogonalization increases the convergence rate of the
bias to an order of nearly $n^{-1/2}$ by Column (3). However, from Column
(3) to Column (1), we find that using cross-fitting remarkably enlarges the
bias, which accounts for the large bias of the locally robust estimator. The
negative impact of cross-fitting on estimation can also be found by
comparing Columns (2) and (4). These results partly contradict with the
theory where cross-fitting is the key to relaxing the Donsker condition and
thus admitting the ML first-step. We conjecture that inaccurate
out-of-sample prediction relative to in-sample prediction results in the
discrepancy between theory and finite-sample performance in using
cross-fitting.

To demonstrate this conjecture, we run a simulation for repeated samples
such that every observation can find a copy of itself and thus the ML
first-step actually carry out in-sample prediction even with cross-fitting.
Specifically, we first generate a random sample of size $n/2$ following the
benchmark setting, and then make a copy of it to produce a sample of size $n$%
. The results reported in Table \ref{table2} show that the negative impact
of cross-fitting disappears under repeated samples, which supports the
conjecture.

\begin{table}[tbh]
\caption{Simulation results for repeated samples}
\label{table2}
\begin{center}
\begin{tabular}{cccccc}
\hline\hline
&  & (1) & (2) & (3) & (4) \\ 
&  & \ \ Locally \ \  & \ Robinson \  & Robinson with & Robinson with \\ 
&  & Robust &  & Orthogonalization & Cross-fitting \\ \hline
\multicolumn{6}{l}{\textbf{Panel A:} $n=250$} \\ 
Average Bias &  & 0.180 & 0.208 & 0.179 & 0.251 \\ 
Average SD &  & 0.221 & 0.242 & 0.220 & 0.275 \\ 
Average Coverage &  & 0.940 & 0.930 & 0.938 & 0.917 \\ 
Max Coverage &  & 0.970 & 0.970 & 0.980 & 0.970 \\ 
Min Coverage &  & 0.910 & 0.890 & 0.900 & 0.820 \\ \hline
\multicolumn{6}{l}{\textbf{Panel B:} $n=500$} \\ 
Average Bias &  & 0.123 & 0.144 & 0.126 & 0.169 \\ 
Average SD &  & 0.146 & 0.160 & 0.146 & 0.177 \\ 
Average Coverage &  & 0.946 & 0.903 & 0.927 & 0.880 \\ 
Max Coverage &  & 0.980 & 0.960 & 0.970 & 0.950 \\ 
Min Coverage &  & 0.890 & 0.680 & 0.770 & 0.640 \\ \hline
\multicolumn{6}{l}{\textbf{Panel C:} $n=1000$} \\ 
Average Bias &  & 0.084 & 0.104 & 0.088 & 0.126 \\ 
Average SD &  & 0.097 & 0.105 & 0.097 & 0.115 \\ 
Average Coverage &  & 0.938 & 0.873 & 0.899 & 0.840 \\ 
Max Coverage &  & 0.980 & 0.960 & 0.990 & 0.940 \\ 
Min Coverage &  & 0.870 & 0.570 & 0.650 & 0.720 \\ \hline\hline
\end{tabular}%
\end{center}
\end{table}

To investigate the robustness of finite-sample performance of the locally
robust estimator, we design several different settings and report the
simulation results in Section \ref{appendix:tables} of the Appendix. In
Table \ref{tableA1}, we consider higher correlation of the errors, $\rho
=0.75$, and/or higher rate of censoring, $\Pr \left( D=0\right) =0.75$, to
increase the model complexity pertaining to sample selection. Panel A shows
that the estimators are fairly robust to different error correlation, while
Panels B and C show that higher rate of censoring has more serious impact on
the locally robust estimator than on the simpler Robinson estimator in terms
of the bias and standard deviation. Although large bias, the locally robust
estimator still performs well in the coverage probability. Tables \ref%
{tableA2} and \ref{tableA3} further verify the robustness of the locally
robust estimator, in terms of the coverage probability, to different error
distributions and different functional forms of $h\left( \cdot \right) $ in
the selection equation.

\section{Empirical Application}

\label{sec:appli}In this section, we apply the proposed estimator to \cite%
{honore2020selection}'s data on wages that are extracted from Current
Population Survey. In the survey, a considerable percentage of respondents
do not participate in work, whose market wages are latent and unobserved.
Specifically, the percentage working is only 67.7\% for man and 61.6\% for
women in \cite{honore2020selection}'s data, implying a potentially
substantial problem of sample selection. Summary statistics and more details
about the data and variables can be found in \cite{honore2020selection}. We
define the selection variable as whether or not to participate in work and
estimate the sample selection model on wages by using the same covariates as
in \cite{honore2020selection}. No exclusion restriction is imposed; namely,
all the covariates enter both the participation equation and the wage
equation. We estimate the model by the locally robust and Robinson
estimators, following the implementation details in Section \ref{sec:simul},
and report the estimated coefficients of the wage equation in the first two
data columns of Table \ref{table3}. The estimated bounds of \cite%
{honore2020selection} and \cite{lee2009training} in the last two data
columns in each panel of Table \ref{table3} are copied from \cite%
{honore2020selection}. The parameter of interest is the coefficient on being
third-generation Mexican-American as opposed on non-Hispanic white.

\begin{sidewaystable}[t]
\caption{Estimated wage regression}
\label{table3}
\medskip
\resizebox{\linewidth}{!}{
\begin{tabular}{ccccccccccc}
\hline\hline
&  & \multicolumn{4}{c}{Women} &  & \multicolumn{4}{c}{Men} \\ 
\cline{3-6}\cline{8-11}
&  & Locally & Robinson & Honor\'{e} and Hu & Lee (2009)'s &  & Locally & 
Robinson & Honor\'{e} and Hu & Lee (2009)'s \\ 
&  & Robust &  & (2020)'s Bound & Bound &  & Robust &  & (2020)'s Bound & 
Bound \\ \hline
Mexican-American &  & -0.141 & -0.114 & [-0.086, -0.080] & [-0.210, -0.041]
&  & -0.138 & -0.127 & [-0.109, -0.097] & [-0.249, -0.074] \\ 
&  & (0.010) & (0.012) &  &  &  & (0.009) & (0.012) &  &  \\ 
Age &  & 0.050 & 0.132 & [0.096, 0.106] &  &  & 0.047 & 0.082 & [0.077,
0.091] &  \\ 
&  & (0.012) & (0.015) &  &  &  & (0.011) & (0.013) &  &  \\ 
Age squared &  & 0.000 & 0.000 & [-0.000, -0.000] &  &  & 0.000 & 0.000 & 
[-0.001, -0.000] &  \\ 
&  & (0.000) & (0.000) &  &  &  & (0.000) & (0.000) &  &  \\ 
Experience &  & -0.057 & -0.112 & [-0.067, -0.062] &  &  & -0.026 & -0.046 & 
[-0.032, -0.023] &  \\ 
&  & (0.009) & (0.012) &  &  &  & (0.009) & (0.011) &  &  \\ 
Experience squared &  & 0.000 & 0.000 & [-0.000, -0.000] &  &  & -0.000 & 
0.000 & [-0.000, -0.000] &  \\ 
&  & (0.000) & (0.000) &  &  &  & (0.000) & (0.000) &  &  \\ 
Veteran &  & 0.115 & 0.022 & [0.029, 0.030] &  &  & 0.012 & 0.017 & [-0.001,
0.005] &  \\ 
&  & (0.018) & (0.023) &  &  &  & (0.010) & (0.013) &  &  \\ 
Married &  & 0.199 & 0.126 & [0.042, 0.052] &  &  & -0.008 & 0.050 & [0.133,
0.154] &  \\ 
&  & (0.009) & (0.010) &  &  &  & (0.008) & (0.010) &  &  \\ \hline
Year dummies &  & Yes & Yes & Yes & Yes &  & Yes & Yes & Yes & Yes \\ 
Education dummies &  & Yes & Yes & Yes & Yes &  & Yes & Yes & Yes & Yes \\ 
State dummies &  & Yes & Yes & Yes & Yes &  & Yes & Yes & Yes & Yes \\ 
No. observations &  & 127738 & 127738 & 127738 & 127738 &  & 118250 & 118250
& 118250 & 118250 \\ \hline\hline
\end{tabular}
}\\
\\
{\small {Note. (1) The locally robust and Robinson estimators are computed as in the
previous section of simulation. Their standard errors constructed from the
estimated asymptotic variance-covariance matrix are given in parentheses.
(2) \cite{honore2020selection}'s bound assumes the selection equation to
have a linear form, namely, assumes $h\left( \cdot \right) $ to be a linear
function, which is not imposed in the locally robust estimator, the Robinson
estimator, and \cite{lee2009training}'s bound. (3) \cite{lee2009training}'s
method only identifies the bound for the treatment effect (the coefficient of the
treatment variable). (4) Estimates of both bounds come directly from \cite%
{honore2020selection}.} }
\end{sidewaystable}

From Table \ref{table3}, \citeauthor{lee2009training}'s bound is long enough
to cover both \citeauthor{honore2020selection}'s bound and the implied
confidence intervals based on the two point-estimates as expected, because %
\citeauthor{lee2009training}'s assumption on the model is most relaxed and
uses the least amount of model information. The locally robust and Robinson
estimators are close to each other, especially for men, and they reveal a
significantly larger, in magnitude, effect of ethnicity on wages compared
with \citeauthor{honore2020selection}'s bound. Note that both for women and
men, the locally robust estimator is near the midpoint of %
\citeauthor{lee2009training}'s bound, and \citeauthor{honore2020selection}'s
bound is located at the very upper part of \citeauthor{lee2009training}'s
bound. This provides some support for the linear structure of the wage
equation assumed by both the locally robust estimator and %
\citeauthor{honore2020selection}'s bound, while casting some doubt on the
linear structure of the participation equation assumed by only %
\citeauthor{honore2020selection}'s bound.

\clearpage

\section{Conclusion}

\label{sec:concl}In this paper, we consider identification and estimation of
a semiparametric sample selection model without the exclusion restriction.
We establish identification of the model based on functional form.
Specifically, we assume nonlinearity on the selection equation and linearity
on the outcome equation, and utilize the nonlinear part of the selection
equation to provide excluded variation to control for selection and identify
coefficients of the outcome equation. We prove the identification by
representing the coefficients as known functions of conditional means of
observed variables.

According to the identification result, we propose to estimate the model by
a partially linear regression with a nonparametrically generated regressor.
To accommodate modern machine learning methods in generating the regressor,
we construct an orthogonalized moment by adding the first-step influence
function and develop a locally robust estimator by solving the cross-fitted
orthogonalized moment condition. Our derivation of the first-step influence
function differs from the current literature on double machine learning in
the three-step nature of the sample selection corrected estimation. Due to
the addition of an intermediate step, the first-step estimation bias has an
indirect influence along with the direct influence on the estimation of
target coefficients. We follow \cite{hahn2013asymptotic}'s method to address
this problem and show the first-order orthogonalization of the constructed
moment function with respect to the first-step estimation bias. Accordingly,
we then establish root-n-consistency and asymptotic normality of the
proposed locally robust estimator under mild regularity conditions.

By a simulation, we show that the locally robust estimator performs
desirably in terms of the coverage probability of an asymptotic 95\%
confidence interval, implying an effective correction for the size
distortion of hypothesis testing based on the naive Robinson-type estimator.
Moreover, the coverage probability of the locally robust estimator is quite
robust to various simulation designs. Finally, we provide an application to 
\cite{honore2020selection}' wage data to illustrate the usefulness of the
proposed method in sample selection models without exclusion restrictions.

\renewcommand{\theequation}{A.\arabic{equation}} \setcounter{equation}{0} %
\renewcommand{\thetheorem}{A.\arabic{theorem}} \setcounter{theorem}{0} %
\renewcommand{\thesubsection}{A.\arabic{subsection}} %
\setcounter{subsection}{0} \renewcommand{\thetable}{A.\arabic{table}} %
\setcounter{table}{0}

\bibliographystyle{Chicago}
\bibliography{0RefHeckman}
\addcontentsline{toc}{section}{References}\qquad

\newpage

\section*{Appendix}

\addcontentsline{toc}{section}{Appendix}

\subsection{Proof of Theorem \protect\ref{theorem:iden}}

\label{appendix:th1}Denote $m\left( x\right) =E\left[ Y\left\vert
X,D=1\right. \right] $ and $m_{0}\left( x^{C}\right) =m\left( x^{C},0\right)
=E\left[ Y\left\vert X^{C}=x^{C},X^{D}=0,D=1\right. \right] $. Note that $%
\pi \left( x\right) $ and $m\left( x\right) $, and thus $\pi _{0}\left(
x^{C}\right) $ and $m_{0}\left( x^{C}\right) $, are identified functions
because they are conditional expectations of observed variables. We first
consider identifying $\beta ^{C}$, the coefficients of continuous
covariates, from $\pi _{0}\left( x^{C}\right) $ and $m_{0}\left(
x^{C}\right) $. By equation (\ref{RF2}), we have%
\begin{equation}
m_{0}\left( x^{C}\right) =x^{C\prime }\beta ^{C}+g\left( \pi _{0}\left(
x^{C}\right) \right) .  \label{IDreg}
\end{equation}%
When $\dim \left( X^{C}\right) =1$, Assumption \ref{ass:ff}.(ii) implies that%
\begin{equation*}
m_{0}\left( x^{C}\right) -x^{C}\beta ^{C}=g\left( \pi _{0}\left(
x^{C}\right) \right) =g\left( \pi _{0}\left( \tilde{x}^{C}\right) \right)
=m_{0}\left( \tilde{x}^{C}\right) -\tilde{x}^{C}\beta ^{C}.
\end{equation*}%
Hence, $\beta ^{C}$ is identified by%
\begin{equation*}
\beta ^{C}=\frac{m_{0}\left( x^{C}\right) -m_{0}\left( \tilde{x}^{C}\right) 
}{x^{C}-\tilde{x}^{C}}.
\end{equation*}%
When $\dim \left( X^{C}\right) \geq 2$, for $k,j\in \left\{ 1,2,\cdots ,\dim
\left( X^{C}\right) \right\} $ satisfying Assumption \ref{ass:ff}.(i),
taking the partial derivatives of $m_{0}\left( x^{C}\right) $ with respect
to $x_{k}^{C}$ and $x_{j}^{C}$ yields that%
\begin{eqnarray*}
\partial _{k}m_{0}\left( x^{C}\right) &=&\beta _{k}^{C}+g^{\left( 1\right)
}\left( \pi _{0}\left( x^{C}\right) \right) \partial _{k}\pi _{0}\left(
x^{C}\right) , \\
\partial _{j}m_{0}\left( x^{C}\right) &=&\beta _{j}^{C}+g^{\left( 1\right)
}\left( \pi _{0}\left( x^{C}\right) \right) \partial _{j}\pi _{0}\left(
x^{C}\right) ,
\end{eqnarray*}%
where $g^{\left( 1\right) }$ denotes the derivative function of $g$. It
follows from $\partial _{k}\pi _{0}\left( x^{C}\right) \neq 0$ and $\partial
_{j}\pi _{0}\left( x^{C}\right) \neq 0$ in Assumption \ref{ass:ff}.(i) that%
\begin{equation*}
\frac{\partial _{k}m_{0}\left( x^{C}\right) -\beta _{k}^{C}}{\partial
_{k}\pi _{0}\left( x^{C}\right) }=g^{\left( 1\right) }\left( \pi _{0}\left(
x^{C}\right) \right) =\frac{\partial _{j}m_{0}\left( x^{C}\right) -\beta
_{j}^{C}}{\partial _{j}\pi _{0}\left( x^{C}\right) },
\end{equation*}%
so that%
\begin{equation}
\partial _{k}m_{0}\left( x^{C}\right) \partial _{j}\pi _{0}\left(
x^{C}\right) -\partial _{j}m_{0}\left( x^{C}\right) \partial _{k}\pi
_{0}\left( x^{C}\right) =\partial _{j}\pi _{0}\left( x^{C}\right) \beta
_{k}^{C}-\partial _{k}\pi _{0}\left( x^{C}\right) \beta _{j}^{C},  \label{11}
\end{equation}%
which is linear in $\beta _{k}^{C}$ and $\beta _{j}^{C}$. The same equation
is obtained if we evaluate the expression at another point $\tilde{x}^{C}$
that satisfies Assumption \ref{ass:ff}.(i), which gives%
\begin{equation}
\left( 
\begin{array}{c}
\partial _{k}m_{0}\left( x^{C}\right) \partial _{j}\pi _{0}\left(
x^{C}\right) -\partial _{j}m_{0}\left( x^{C}\right) \partial _{k}\pi
_{0}\left( x^{C}\right) \\ 
\partial _{k}m_{0}\left( \tilde{x}^{C}\right) \partial _{j}\pi _{0}\left( 
\tilde{x}^{C}\right) -\partial _{j}m_{0}\left( \tilde{x}^{C}\right) \partial
_{k}\pi _{0}\left( \tilde{x}^{C}\right)%
\end{array}%
\right) =\Upsilon \left( 
\begin{array}{c}
\beta _{k}^{C} \\ 
\beta _{j}^{C}%
\end{array}%
\right) ,  \label{betaC}
\end{equation}%
where%
\begin{equation*}
\Upsilon =\left( 
\begin{array}{cc}
\partial _{j}\pi _{0}\left( x^{C}\right) & -\partial _{k}\pi _{0}\left(
x^{C}\right) \\ 
\partial _{j}\pi _{0}\left( \tilde{x}^{C}\right) & -\partial _{k}\pi
_{0}\left( \tilde{x}^{C}\right)%
\end{array}%
\right) .
\end{equation*}%
The inequality of Assumption \ref{ass:ff}.(i) ensures that the determinant
of $\Upsilon $ is nonzero, which implies that $\Upsilon $ is nonsingular.
Therefore, equation (\ref{betaC}) can be solved for $\beta _{k}^{C}$ and $%
\beta _{j}^{C}$ by inverting $\Upsilon $, thereby identifying $\beta
_{k}^{C} $ and $\beta _{j}^{C}$. Given identification of $\beta _{k}^{C}$,
we can then identify all other coefficient $\beta _{l}^{C}$ in $\beta ^{C}$
by solving (\ref{11}) with the subscript $j$ replaced by $l$, which gives%
\begin{equation*}
\beta _{l}^{C}=\frac{\partial _{l}m_{0}\left( x^{C}\right) \partial _{k}\pi
_{0}\left( x^{C}\right) -\partial _{k}m_{0}\left( x^{C}\right) \partial
_{l}\pi _{0}\left( x^{C}\right) +\partial _{l}\pi _{0}\left( x^{C}\right)
\beta _{k}^{C}}{\partial _{k}\pi _{0}\left( x^{C}\right) }.
\end{equation*}%
Given the identification of $\beta ^{C}$, the function $g$ is identified on
the support of $\pi _{0}\left( X^{C}\right) $ by%
\begin{equation*}
g\left( p\right) =E\left[ \left. m_{0}\left( X^{C}\right) -X^{C\prime }\beta
^{C}\right\vert \pi _{0}\left( X^{C}\right) =p\right] .
\end{equation*}

Next, we consider identifying $\beta ^{D}$, the coefficients of discrete
covariates. For each $k\in \left\{ 1,2,\cdots ,\dim \left( X^{D}\right)
\right\} $, we have%
\begin{equation*}
m\left( x^{C},x^{Dk}\right) =x^{C\prime }\beta ^{C}+x_{k}^{D}\beta
_{k}^{D}+g\left( \pi \left( x^{C},x^{Dk}\right) \right)
\end{equation*}%
for any $x^{C}$ in the support of $X^{C}$. By Assumption \ref{ass:support},
there exists $x^{C}\left( k\right) $ in the support of $X^{C}$ such that $%
\pi \left( x^{C}\left( k\right) ,x^{Dk}\right) $ is in the support of $\pi
_{0}\left( X^{C}\right) $. It follows from the above identification result
that $g\left( \pi \left( x^{C}\left( k\right) ,x^{Dk}\right) \right) $ is
identified. Consequently, $\beta _{k}^{D}$ is identified by%
\begin{equation*}
\beta _{k}^{D}=\frac{m\left( x^{C}\left( k\right) ,x^{Dk}\right)
-x^{C}\left( k\right) ^{\prime }\beta ^{C}-g\left( \pi \left( x^{C}\left(
k\right) ,x^{Dk}\right) \right) }{x_{k}^{D}}.
\end{equation*}%
This argument holds for each $k\in \left\{ 1,2,\cdots ,\dim \left(
X^{D}\right) \right\} $, thereby identifying $\beta ^{D}$. Hence, $\beta
=\left( \beta ^{C},\beta ^{D}\right) $\ is identified. Additionally, given
the identification of $\beta $, it follows from equation (\ref{RF2}) that
the selectivity correction function $g$ is identified on the support of $\pi
\left( X\right) $ by%
\begin{equation*}
g\left( p\right) =E\left[ \left. m\left( X\right) -X^{\prime }\beta
\right\vert \pi \left( X\right) =p\right] .
\end{equation*}

\subsection{Proof of Theorem \protect\ref{theorem:AN}}

\label{appendix:th2}Denote%
\begin{equation*}
\hat{M}_{n}=\frac{1}{n}\sum_{\ell =1}^{L}\sum_{i\in I_{\ell }}D_{i}\hat{P}%
_{\ell i}\left( X_{i}-\hat{\mu}_{X\ell }\left( \hat{P}_{\ell i}\right)
\right) \left( X_{i}-\hat{\mu}_{X\ell }\left( \hat{P}_{\ell i}\right)
\right) ^{\prime }.
\end{equation*}%
Then we have%
\begin{eqnarray}
\sqrt{n}\left( \hat{\beta}^{LR}-\beta \right) &=&\hat{M}_{n}^{-1}\frac{1}{%
\sqrt{n}}\sum_{\ell =1}^{L}\sum_{i\in I_{\ell }}\left[ r\left( W_{i},\hat{\pi%
}_{\ell },\hat{\mu}_{\ell },\beta \right) +\hat{\alpha}_{\ell }\left(
X_{i}\right) \left( D_{i}-\hat{P}_{\ell i}\right) \right]  \notag \\
&=&\hat{M}_{n}^{-1}\frac{1}{\sqrt{n}}\sum_{\ell =1}^{L}\sum_{i\in I_{\ell }}%
\left[ \psi \left( W_{i},\hat{\pi}_{\ell },\hat{\mu}_{\ell },\alpha ,\beta
\right) +\left( \hat{\alpha}_{\ell }\left( X_{i}\right) -\alpha \left(
X_{i}\right) \right) \left( D_{i}-\hat{P}_{\ell i}\right) \right]  \notag \\
&=&\hat{M}_{n}^{-1}\frac{1}{\sqrt{n}}\sum_{\ell =1}^{L}\sum_{i\in I_{\ell }}%
\left[ \psi \left( W_{i},\pi ,\mu ,\alpha ,\beta \right) +\hat{R}_{1\ell i}+%
\hat{R}_{2\ell i}\right] ,  \label{ANexpansion}
\end{eqnarray}%
where%
\begin{eqnarray*}
\hat{R}_{1\ell i} &=&\psi \left( W_{i},\hat{\pi}_{\ell },\hat{\mu}_{\ell
},\alpha ,\beta \right) -\psi \left( W_{i},\pi ,\mu ,\alpha ,\beta \right) ,
\\
\hat{R}_{2\ell i} &=&\left( \hat{\alpha}_{\ell }\left( X_{i}\right) -\alpha
\left( X_{i}\right) \right) \left( D_{i}-\hat{P}_{\ell i}\right) .
\end{eqnarray*}

Let $\mathcal{W}_{\ell }^{c}=\left\{ W_{i}:i\notin I_{\ell }\right\} $
denote the observations not in $I_{\ell }$, so that $\hat{\pi}_{\ell }$, $%
\hat{\mu}_{\ell }$, and $\hat{\alpha}_{\ell }$ depend only on $\mathcal{W}%
_{\ell }^{c}$. Therefore, by equation (\ref{O_moment}) and Lemma \ref{lemma3}%
,%
\begin{eqnarray*}
\left\Vert E\left[ \left. \frac{1}{\sqrt{n}}\sum_{i\in I_{\ell }}\hat{R}%
_{1\ell i}\right\vert \mathcal{W}_{\ell }^{c}\right] \right\Vert &=&\frac{%
n_{\ell }}{\sqrt{n}}\left\Vert E\left[ \left. \psi \left( W_{i},\hat{\pi}%
_{\ell },\hat{\mu}_{\ell },\alpha ,\beta \right) \right\vert \mathcal{W}%
_{\ell }^{c}\right] \right\Vert \\
&\leq &\sqrt{n}\left[ 
\begin{array}{c}
C_{1}\left\Vert \hat{\pi}_{\ell }-\pi \right\Vert _{q,F}^{2}+C_{2}\left\Vert 
\hat{\mu}_{\ell }-\mu \right\Vert _{q,F}^{2} \\ 
+C_{3}\left\Vert \hat{\pi}_{\ell }-\pi \right\Vert _{q,F}\left\Vert \hat{\mu}%
_{\ell }-\mu \right\Vert _{q,F} \\ 
+C_{4}\left\Vert \hat{\pi}_{\ell }-\pi \right\Vert _{q,F}\left\Vert \hat{\mu}%
_{\ell }^{\left( 1\right) }-\mu ^{\left( 1\right) }\right\Vert _{q,F}%
\end{array}%
\right] .
\end{eqnarray*}%
It then follows from Assumption \ref{ass:rate} that for each $\ell =1,\cdots
,L$,%
\begin{equation}
E\left[ \left. \frac{1}{\sqrt{n}}\sum_{i\in I_{\ell }}\hat{R}_{1\ell
i}\right\vert \mathcal{W}_{\ell }^{c}\right] \overset{p}{\longrightarrow }0.
\label{ER1li}
\end{equation}%
Also, by observations in $I_{\ell }$ mutually independent conditional on $%
\mathcal{W}_{\ell }^{c}$,%
\begin{equation*}
Var\left( \left. \frac{1}{\sqrt{n}}\sum_{i\in I_{\ell }}\hat{R}_{1\ell
i}\right\vert \mathcal{W}_{\ell }^{c}\right) =\frac{n_{\ell }}{n}Var\left(
\left. \hat{R}_{1\ell i}\right\vert \mathcal{W}_{\ell }^{c}\right) \leq E%
\left[ \left. \hat{R}_{1\ell i}\hat{R}_{1\ell i}^{\prime }\right\vert 
\mathcal{W}_{\ell }^{c}\right] .
\end{equation*}%
It follows from Assumption \ref{ass:rate} and the continuity of $\psi \left(
W_{i},\pi ,\mu ,\alpha ,\beta \right) $ in $\pi $ and $\mu $ that%
\begin{equation*}
\left\Vert Var\left( \left. \frac{1}{\sqrt{n}}\sum_{i\in I_{\ell }}\hat{R}%
_{1\ell i}\right\vert \mathcal{W}_{\ell }^{c}\right) \right\Vert \leq E\left[
\left. \left\Vert \psi \left( W_{i},\hat{\pi}_{\ell },\hat{\mu}_{\ell
},\alpha ,\beta \right) -\psi \left( W_{i},\pi ,\mu ,\alpha ,\beta \right)
\right\Vert ^{2}\right\vert \mathcal{W}_{\ell }^{c}\right] \overset{p}{%
\longrightarrow }0.
\end{equation*}%
By the conditional Markov inequality,%
\begin{equation*}
\frac{1}{\sqrt{n}}\sum_{i\in I_{\ell }}\left( \hat{R}_{1\ell i}-E\left[
\left. \hat{R}_{1\ell i}\right\vert \mathcal{W}_{\ell }^{c}\right] \right) 
\overset{p}{\longrightarrow }0.
\end{equation*}%
It then follows from (\ref{ER1li}) that%
\begin{equation*}
\frac{1}{\sqrt{n}}\sum_{i\in I_{\ell }}\hat{R}_{1\ell i}\overset{p}{%
\longrightarrow }0
\end{equation*}%
for each $\ell =1,\cdots ,L$, and that%
\begin{equation}
\frac{1}{\sqrt{n}}\sum_{\ell =1}^{L}\sum_{i\in I_{\ell }}\hat{R}_{1\ell i}%
\overset{p}{\longrightarrow }0.  \label{R1li}
\end{equation}

For the second term $\hat{R}_{2\ell i}$, since by the law of iterated
expectations 
\begin{equation*}
E\left[ \left. \left( \hat{\alpha}_{\ell }\left( X_{i}\right) -\alpha \left(
X_{i}\right) \right) \left( D_{i}-P_{i}\right) \right\vert \mathcal{W}_{\ell
}^{c}\right] =E\left[ \left. \left( \hat{\alpha}_{\ell }\left( X_{i}\right)
-\alpha \left( X_{i}\right) \right) E\left[ \left. D_{i}-P_{i}\right\vert
X_{i}\right] \right\vert \mathcal{W}_{\ell }^{c}\right] =0,
\end{equation*}%
we have%
\begin{eqnarray*}
\left\Vert E\left[ \left. \frac{1}{\sqrt{n}}\sum_{i\in I_{\ell }}\hat{R}%
_{2\ell i}\right\vert \mathcal{W}_{\ell }^{c}\right] \right\Vert &=&\frac{%
n_{\ell }}{\sqrt{n}}\left\Vert E\left[ \left. \left( \hat{\alpha}_{\ell
}\left( X_{i}\right) -\alpha \left( X_{i}\right) \right) \left( \hat{P}%
_{\ell i}-P_{i}\right) \right\vert \mathcal{W}_{\ell }^{c}\right] \right\Vert
\\
&\leq &\sqrt{n}\left\Vert \hat{\alpha}_{\ell }-\alpha \right\Vert
_{F}\left\Vert \hat{\pi}_{\ell }-\pi \right\Vert _{F} \\
&\leq &\sqrt{n}\left\Vert \hat{\alpha}_{\ell }-\alpha \right\Vert
_{q,F}\left\Vert \hat{\pi}_{\ell }-\pi \right\Vert _{q,F} \\
&\overset{p}{\longrightarrow }&0,
\end{eqnarray*}%
for each $\ell =1,\cdots ,L$, where $\left\Vert \cdot \right\Vert _{F}$
denotes the $L_{2}\left( F\right) $-norm and the second inequality follows
from $q>2$ and H\"{o}lder's inequality. Similarly, for the variance, we have%
\begin{eqnarray*}
\left\Vert Var\left( \left. \frac{1}{\sqrt{n}}\sum_{i\in I_{\ell }}\hat{R}%
_{2\ell i}\right\vert \mathcal{W}_{\ell }^{c}\right) \right\Vert &\leq &%
\frac{n_{\ell }}{n}\left\Vert E\left[ \left. \hat{R}_{2\ell i}\hat{R}_{2\ell
i}^{\prime }\right\vert \mathcal{W}_{\ell }^{c}\right] \right\Vert \leq E%
\left[ \left. \left\Vert \hat{R}_{2\ell i}\right\Vert ^{2}\right\vert 
\mathcal{W}_{\ell }^{c}\right] \\
&=&E\left[ \left. \left\Vert \hat{\alpha}_{\ell }\left( X_{i}\right) -\alpha
\left( X_{i}\right) \right\Vert ^{2}\left( D_{i}-\hat{P}_{\ell i}\right)
^{2}\right\vert \mathcal{W}_{\ell }^{c}\right] \\
&\leq &E\left[ \left. \left\Vert \hat{\alpha}_{\ell }\left( X_{i}\right)
-\alpha \left( X_{i}\right) \right\Vert ^{2}\right\vert \mathcal{W}_{\ell
}^{c}\right] \\
&\leq &\left\Vert \hat{\alpha}_{\ell }-\alpha \right\Vert _{q,F}^{2}\overset{%
p}{\longrightarrow }0.
\end{eqnarray*}%
In combination, we have%
\begin{equation*}
\frac{1}{\sqrt{n}}\sum_{i\in I_{\ell }}\hat{R}_{2\ell i}\overset{p}{%
\longrightarrow }0
\end{equation*}%
for each $\ell =1,\cdots ,L$, and%
\begin{equation}
\frac{1}{\sqrt{n}}\sum_{\ell =1}^{L}\sum_{i\in I_{\ell }}\hat{R}_{2\ell i}%
\overset{p}{\longrightarrow }0.  \label{R2li}
\end{equation}%
Substituting (\ref{R1li}) and (\ref{R2li}) into (\ref{ANexpansion}) yields%
\begin{equation}
\sqrt{n}\left( \hat{\beta}^{LR}-\beta \right) =\hat{M}_{n}^{-1}\left[ \frac{1%
}{\sqrt{n}}\sum_{i=1}^{n}\psi \left( W_{i},\pi ,\mu ,\alpha ,\beta \right)
+o_{p}\left( 1\right) \right] .  \label{ANexpansion1}
\end{equation}

Next we prove $\hat{M}_{n}\overset{p}{\longrightarrow }M$, where $M$ is
defined in Assumption \ref{ass:pd}. To this end, we first show that%
\begin{equation}
\frac{1}{n}\sum_{\ell =1}^{L}\sum_{i\in I_{\ell }}\left\Vert \hat{\mu}%
_{X\ell }\left( \hat{P}_{\ell i}\right) -\mu _{X}\left( P_{i}\right)
\right\Vert ^{2}\overset{p}{\longrightarrow }0.  \label{muhatPhat}
\end{equation}%
Since%
\begin{eqnarray*}
\hat{\mu}_{X\ell }\left( \hat{P}_{\ell i}\right) -\mu _{X}\left(
P_{i}\right) &=&\hat{\mu}_{X\ell }\left( \hat{P}_{\ell i}\right) -\hat{\mu}%
_{X\ell }\left( P_{i}\right) +\hat{\mu}_{X\ell }\left( P_{i}\right) -\mu
_{X}\left( P_{i}\right) \\
&=&\hat{\mu}_{X\ell }^{\left( 1\right) }\left( \hat{P}_{\ell i}^{\ast
}\right) \left( \hat{P}_{\ell i}-P_{i}\right) +\hat{\mu}_{X\ell }\left(
P_{i}\right) -\mu _{X}\left( P_{i}\right) ,
\end{eqnarray*}%
we have%
\begin{equation*}
\left\Vert \hat{\mu}_{X\ell }\left( \hat{P}_{\ell i}\right) -\mu _{X}\left(
P_{i}\right) \right\Vert ^{2}\leq 2\left[ \left\Vert \bar{\mu}_{X}^{\left(
1\right) }\right\Vert ^{2}\left\vert \hat{P}_{\ell i}-P_{i}\right\vert
^{2}+\left\Vert \hat{\mu}_{X\ell }\left( P_{i}\right) -\mu _{X}\left(
P_{i}\right) \right\Vert ^{2}\right] .
\end{equation*}%
It follows that%
\begin{equation*}
E\left[ \left. \frac{1}{n}\sum_{i\in I_{\ell }}\left\Vert \hat{\mu}_{X\ell
}\left( \hat{P}_{\ell i}\right) -\mu _{X}\left( P_{i}\right) \right\Vert
^{2}\right\vert \mathcal{W}_{\ell }^{c}\right] \leq 2\left[ \left\Vert \bar{%
\mu}_{X}^{\left( 1\right) }\right\Vert ^{2}\left\Vert \hat{\pi}_{\ell }-\pi
\right\Vert _{F}^{2}+\left\Vert \hat{\mu}_{X\ell }-\mu _{X}\right\Vert
_{F}^{2}\right] \overset{p}{\longrightarrow }0
\end{equation*}%
and that%
\begin{equation*}
Var\left( \left. \frac{1}{n}\sum_{i\in I_{\ell }}\left\Vert \hat{\mu}_{X\ell
}\left( \hat{P}_{\ell i}\right) -\mu _{X}\left( P_{i}\right) \right\Vert
^{2}\right\vert \mathcal{W}_{\ell }^{c}\right) \leq \frac{1}{n}E\left[
\left. \left\Vert \hat{\mu}_{X\ell }\left( \hat{P}_{\ell i}\right) -\mu
_{X}\left( P_{i}\right) \right\Vert ^{4}\right\vert \mathcal{W}_{\ell }^{c}%
\right] \overset{p}{\longrightarrow }0
\end{equation*}%
for each $\ell =1,\cdots ,L$. Therefore, by the conditional Markov
inequality,%
\begin{equation*}
\frac{1}{n}\sum_{i\in I_{\ell }}\left\Vert \hat{\mu}_{X\ell }\left( \hat{P}%
_{\ell i}\right) -\mu _{X}\left( P_{i}\right) \right\Vert ^{2}\overset{p}{%
\longrightarrow }0
\end{equation*}%
for each $\ell =1,\cdots ,L$, and thus (\ref{muhatPhat}) holds.

Denote%
\begin{equation*}
M_{n}=\frac{1}{n}\sum_{i=1}^{n}D_{i}P_{i}\left( X_{i}-\mu _{X}\left(
P_{i}\right) \right) \left( X_{i}-\mu _{X}\left( P_{i}\right) \right)
^{\prime }.
\end{equation*}%
We have%
\begin{eqnarray*}
\left\Vert \hat{M}_{n}-M_{n}\right\Vert &\leq &\frac{1}{n}\sum_{\ell
=1}^{L}\sum_{i\in I_{\ell }}\left\Vert 
\begin{array}{c}
\hat{P}_{\ell i}\left( X_{i}-\hat{\mu}_{X\ell }\left( \hat{P}_{\ell
i}\right) \right) \left( X_{i}-\hat{\mu}_{X\ell }\left( \hat{P}_{\ell
i}\right) \right) ^{\prime } \\ 
-P_{i}\left( X_{i}-\mu _{X}\left( P_{i}\right) \right) \left( X_{i}-\mu
_{X}\left( P_{i}\right) \right) ^{\prime }%
\end{array}%
\right\Vert \\
&\leq &\frac{1}{n}\sum_{\ell =1}^{L}\sum_{i\in I_{\ell }}\left[ 
\begin{array}{c}
\left\vert \hat{P}_{\ell i}\right\vert \left\Vert \hat{\mu}_{X\ell }\left( 
\hat{P}_{\ell i}\right) -\mu _{X}\left( P_{i}\right) \right\Vert ^{2} \\ 
+2\left\vert \hat{P}_{\ell i}\right\vert \left\Vert \hat{\mu}_{X\ell }\left( 
\hat{P}_{\ell i}\right) -\mu _{X}\left( P_{i}\right) \right\Vert \left\Vert
X_{i}-\mu _{X}\left( P_{i}\right) \right\Vert \\ 
+\left\vert \hat{P}_{\ell i}-P_{i}\right\vert \left\Vert X_{i}-\mu
_{X}\left( P_{i}\right) \right\Vert ^{2}%
\end{array}%
\right] .
\end{eqnarray*}%
By $\left\vert \hat{P}_{\ell i}\right\vert \leq 1$ and (\ref{muhatPhat}), we
have%
\begin{eqnarray*}
\left\Vert \hat{M}_{n}-M_{n}\right\Vert &\leq &\frac{2}{n}\sum_{\ell
=1}^{L}\sum_{i\in I_{\ell }}\left\Vert \hat{\mu}_{X\ell }\left( \hat{P}%
_{\ell i}\right) -\mu _{X}\left( P_{i}\right) \right\Vert \left\Vert
X_{i}-\mu _{X}\left( P_{i}\right) \right\Vert \\
&&+\frac{1}{n}\sum_{\ell =1}^{L}\sum_{i\in I_{\ell }}\left\vert \hat{P}%
_{\ell i}-P_{i}\right\vert \left\Vert X_{i}-\mu _{X}\left( P_{i}\right)
\right\Vert ^{2}+o_{p}\left( 1\right) ,
\end{eqnarray*}%
For the first term of the right-hand side above, we have%
\begin{eqnarray*}
&&\frac{1}{n}\sum_{\ell =1}^{L}\sum_{i\in I_{\ell }}\left\Vert \hat{\mu}%
_{X\ell }\left( \hat{P}_{\ell i}\right) -\mu _{X}\left( P_{i}\right)
\right\Vert \left\Vert X_{i}-\mu _{X}\left( P_{i}\right) \right\Vert \\
&\leq &\sqrt{\frac{1}{n}\sum_{\ell =1}^{L}\sum_{i\in I_{\ell }}\left\Vert 
\hat{\mu}_{X\ell }\left( \hat{P}_{\ell i}\right) -\mu _{X}\left(
P_{i}\right) \right\Vert ^{2}}\sqrt{\frac{1}{n}\sum_{\ell =1}^{L}\sum_{i\in
I_{\ell }}\left\Vert X_{i}-\mu _{X}\left( P_{i}\right) \right\Vert ^{2}} \\
&\overset{p}{\longrightarrow }&0.
\end{eqnarray*}%
Similarly, we can show the convergence in probability of the second term to
zero. Consequently, $\hat{M}_{n}\overset{p}{\longrightarrow }M_{n}$. We also
have%
\begin{equation*}
M_{n}\overset{p}{\longrightarrow }E\left[ D_{i}P_{i}\left( X_{i}-\mu
_{X}\left( P_{i}\right) \right) \left( X_{i}-\mu _{X}\left( P_{i}\right)
\right) ^{\prime }\right] =M
\end{equation*}%
by Khintchine's law of large numbers, so $\hat{M}_{n}\overset{p}{%
\longrightarrow }M$ follows. By Assumption \ref{ass:pd}, we have $\hat{M}%
_{n}^{-1}\overset{p}{\longrightarrow }M^{-1}$, which in combination of (\ref%
{ANexpansion1}) yields%
\begin{equation*}
\sqrt{n}\left( \hat{\beta}^{LR}-\beta \right) =M^{-1}\frac{1}{\sqrt{n}}%
\sum_{i=1}^{n}\psi \left( W_{i},\pi ,\mu ,\alpha ,\beta \right) +o_{p}\left(
1\right) .
\end{equation*}%
The conclusion follows from the central limit theorem and $E\left[
\left\Vert \psi \left( W,\pi ,\mu ,\alpha ,\beta \right) \right\Vert ^{2}%
\right] <\infty $ in Assumption \ref{ass:pd}.

\subsection{Proofs of the Lemmas}

\label{appendix:lemmas}

\begin{proof}[\textbf{Proof of Lemma \ref{lemma1}}]
It follows from the law of iterated expectations that%
\begin{equation*}
E\left[ \tilde{\alpha}\left( X\right) \left( D-P\right) \right] =E\left[ 
\tilde{\alpha}\left( X\right) \left( E\left[ \left. D\right\vert X\right]
-P\right) \right] =0
\end{equation*}%
for any $\tilde{\alpha}$. Therefore,%
\begin{equation*}
E\left[ \psi \left( W,\pi ,\tilde{\mu},\tilde{\alpha},\tilde{\beta}\right) %
\right] =E\left[ r\left( W,\pi ,\tilde{\mu},\tilde{\beta}\right) \right] +E%
\left[ \tilde{\alpha}\left( X\right) \left( D-P\right) \right] =E\left[
r\left( W,\pi ,\tilde{\mu},\tilde{\beta}\right) \right] ,
\end{equation*}%
which is invariant with respect to $\tilde{\alpha}$\ for any $\tilde{\mu}$
and $\tilde{\beta}$.
\end{proof}

\begin{proof}[\textbf{Proof of Lemma \ref{lemma2}}]
\textbf{(i)} By definition of the function $\psi $, we have%
\begin{eqnarray*}
E\left[ \psi \left( W,\pi +t\delta ,\mu ,\alpha ,\beta \right) \right] &=&E%
\left[ 
\begin{array}{c}
\left( P+t\delta \left( X\right) \right) \left[ X-\mu _{X}\left( P+t\delta
\left( X\right) \right) \right] \\ 
\cdot \left[ Y-\mu _{Y}\left( P+t\delta \left( X\right) \right) -D\left(
X-\mu _{X}\left( P+t\delta \left( X\right) \right) \right) ^{\prime }\beta %
\right]%
\end{array}%
\right] \\
&&+E\left[ \alpha \left( X\right) \left( D-P-t\delta \left( X\right) \right) %
\right] .
\end{eqnarray*}%
Using the product rule of calculus results in that%
\begin{eqnarray*}
\left. \frac{\partial }{\partial t}E\left[ \psi \left( W,\pi +t\delta ,\mu
,\alpha ,\beta \right) \right] \right\vert _{t=0} &=&E\left[ \delta \left(
X\right) \left( X-\mu _{X}\left( P\right) \right) \left[ Y-\mu _{Y}\left(
P\right) -D\left( X-\mu _{X}\left( P\right) \right) ^{\prime }\beta \right] %
\right] \\
&&-E\left[ P\mu _{X}^{\left( 1\right) }\left( P\right) \delta \left(
X\right) \left[ Y-\mu _{Y}\left( P\right) -D\left( X-\mu _{X}\left( P\right)
\right) ^{\prime }\beta \right] \right] \\
&&-E\left[ P\left( X-\mu _{X}\left( P\right) \right) \mu _{Y}^{\left(
1\right) }\left( P\right) \delta \left( X\right) \right] \\
&&+E\left[ P\left( X-\mu _{X}\left( P\right) \right) D\mu _{X}^{\left(
1\right) }\left( P\right) ^{\prime }\beta \delta \left( X\right) \right] \\
&&-E\left[ \alpha \left( X\right) \delta \left( X\right) \right] .
\end{eqnarray*}%
It follows from the law of iterated expectations that the first and second
terms on the right-hand side will vanish, considering that by the equation (%
\ref{RF5}) 
\begin{equation*}
E\left[ \left. Y-\mu _{Y}\left( P\right) -D\left( X-\mu _{X}\left( P\right)
\right) ^{\prime }\beta \right\vert X\right] =E\left[ \left. Y-\mu
_{Y}\left( P\right) \right\vert X\right] -P\left( X-\mu _{X}\left( P\right)
\right) ^{\prime }\beta =0.
\end{equation*}%
Therefore,%
\begin{eqnarray*}
\left. \frac{\partial }{\partial t}E\left[ \psi \left( W,\pi +t\delta ,\mu
,\alpha ,\beta \right) \right] \right\vert _{t=0} &=&-E\left[ P\left( X-\mu
_{X}\left( P\right) \right) \left[ \mu _{Y}^{\left( 1\right) }\left(
P\right) -D\mu _{X}^{\left( 1\right) }\left( P\right) ^{\prime }\beta \right]
\delta \left( X\right) \right] \\
&&-E\left[ \alpha \left( X\right) \delta \left( X\right) \right] \\
&=&-E\left[ P\left( X-\mu _{X}\left( P\right) \right) \left[ \mu
_{Y}^{\left( 1\right) }\left( P\right) -P\mu _{X}^{\left( 1\right) }\left(
P\right) ^{\prime }\beta \right] \delta \left( X\right) \right] \\
&&-E\left[ \alpha \left( X\right) \delta \left( X\right) \right] \\
&=&0.
\end{eqnarray*}

\noindent \textbf{(ii)} Standard calculations yield that 
\begin{eqnarray*}
&&E\left[ r\left( W,\tilde{\pi},\mu ,\beta \right) \right] \\
&=&E\left[ \tilde{\pi}\left( X\right) \left( X-\mu _{X}\left( \tilde{\pi}%
\left( X\right) \right) \right) \left[ Y-\mu _{Y}\left( \tilde{\pi}\left(
X\right) \right) -D\left( X-\mu _{X}\left( \tilde{\pi}\left( X\right)
\right) \right) ^{\prime }\beta \right] \right] \\
&=&-E\left[ \tilde{\pi}\left( X\right) \left( X-\mu _{X}\left( \tilde{\pi}%
\left( X\right) \right) \right) \left[ \mu _{Y}\left( \tilde{\pi}\left(
X\right) \right) -\mu _{Y}\left( P\right) -P\left( \mu _{X}\left( \tilde{\pi}%
\left( X\right) \right) -\mu _{X}\left( P\right) \right) ^{\prime }\beta %
\right] \right] \\
&=&E\left[ \tilde{\pi}\left( X\right) \left( \mu _{X}\left( \tilde{\pi}%
\left( X\right) \right) -\mu _{X}\left( P\right) \right) \left[ \mu
_{Y}\left( \tilde{\pi}\left( X\right) \right) -\mu _{Y}\left( P\right)
-P\left( \mu _{X}\left( \tilde{\pi}\left( X\right) \right) -\mu _{X}\left(
P\right) \right) ^{\prime }\beta \right] \right] \\
&&-E\left[ \left( \tilde{\pi}\left( X\right) -P\right) \left( X-\mu
_{X}\left( P\right) \right) \left[ \mu _{Y}\left( \tilde{\pi}\left( X\right)
\right) -\mu _{Y}\left( P\right) -P\left( \mu _{X}\left( \tilde{\pi}\left(
X\right) \right) -\mu _{X}\left( P\right) \right) ^{\prime }\beta \right] %
\right] \\
&&-E\left[ P\left( X-\mu _{X}\left( P\right) \right) \left[ \mu _{Y}\left( 
\tilde{\pi}\left( X\right) \right) -\mu _{Y}\left( P\right) -P\left( \mu
_{X}\left( \tilde{\pi}\left( X\right) \right) -\mu _{X}\left( P\right)
\right) ^{\prime }\beta \right] \right] ,
\end{eqnarray*}%
where the second equality follows from the law of iterated expectations and (%
\ref{RF5}). Through Taylor expansion, we have%
\begin{eqnarray*}
&&E\left[ \psi \left( W,\tilde{\pi},\mu ,\alpha ,\beta \right) \right] =E%
\left[ r\left( W,\tilde{\pi},\mu ,\beta \right) \right] +E\left[ \alpha
\left( X\right) \left( D-\tilde{\pi}\left( X\right) \right) \right] \\
&=&E\left[ \tilde{\pi}\left( X\right) \left( \mu _{X}\left( \tilde{\pi}%
\left( X\right) \right) -\mu _{X}\left( P\right) \right) \left[ \mu
_{Y}\left( \tilde{\pi}\left( X\right) \right) -\mu _{Y}\left( P\right)
-P\left( \mu _{X}\left( \tilde{\pi}\left( X\right) \right) -\mu _{X}\left(
P\right) \right) ^{\prime }\beta \right] \right] \\
&&-E\left[ \left( \tilde{\pi}\left( X\right) -P\right) \left( X-\mu
_{X}\left( P\right) \right) \left[ \mu _{Y}\left( \tilde{\pi}\left( X\right)
\right) -\mu _{Y}\left( P\right) -P\left( \mu _{X}\left( \tilde{\pi}\left(
X\right) \right) -\mu _{X}\left( P\right) \right) ^{\prime }\beta \right] %
\right] \\
&&-E\left[ P\left( X-\mu _{X}\left( P\right) \right) \left[ 
\begin{array}{c}
\mu _{Y}\left( \tilde{\pi}\left( X\right) \right) -\mu _{Y}\left( P\right)
-\mu _{Y}^{\left( 1\right) }\left( P\right) \left( \tilde{\pi}\left(
X\right) -P\right) \\ 
-P\left[ \mu _{X}\left( \tilde{\pi}\left( X\right) \right) -\mu _{X}\left(
P\right) -\mu _{X}^{\left( 1\right) }\left( P\right) \left( \tilde{\pi}%
\left( X\right) -P\right) \right] ^{\prime }\beta%
\end{array}%
\right] \right] \\
&=&E\left[ \tilde{\pi}\left( X\right) \mu _{X}^{\left( 1\right) }\left( \pi
^{\ast }\left( X\right) \right) \left[ \mu _{Y}^{\left( 1\right) }\left( \pi
^{\ast }\left( X\right) \right) -P\mu _{X}^{\left( 1\right) }\left( \pi
^{\ast }\left( X\right) \right) ^{\prime }\beta \right] \left( \tilde{\pi}%
\left( X\right) -P\right) ^{2}\right] \\
&&-E\left[ \left( X-\mu _{X}\left( P\right) \right) \left[ \mu _{Y}^{\left(
1\right) }\left( \pi ^{\ast }\left( X\right) \right) -P\mu _{X}^{\left(
1\right) }\left( \pi ^{\ast }\left( X\right) \right) ^{\prime }\beta \right]
\left( \tilde{\pi}\left( X\right) -P\right) ^{2}\right] \\
&&-\frac{1}{2}E\left[ P\left( X-\mu _{X}\left( P\right) \right) \left[ \mu
_{Y}^{\left( 2\right) }\left( \pi ^{\ast }\left( X\right) \right) -P\mu
_{X}^{\left( 2\right) }\left( \pi ^{\ast }\left( X\right) \right) ^{\prime
}\beta \right] \left( \tilde{\pi}\left( X\right) -P\right) ^{2}\right] ,
\end{eqnarray*}%
where $\pi ^{\ast }$ represents an intermediate value between $\tilde{\pi}$
and $\pi $, which may have different values at each appearance. Therefore,%
\begin{eqnarray*}
\left\Vert E\left[ \psi \left( W,\tilde{\pi},\mu ,\alpha ,\beta \right) %
\right] \right\Vert &\leq &\left\Vert \bar{\mu}_{X}^{\left( 1\right)
}\right\Vert \left( \bar{\mu}_{Y}^{\left( 1\right) }+\left\Vert \bar{\mu}%
_{X}^{\left( 1\right) }\right\Vert \left\Vert \beta \right\Vert \right) E%
\left[ \left( \tilde{\pi}\left( X\right) -P\right) ^{2}\right] \\
&&+\left( \bar{\mu}_{Y}^{\left( 1\right) }+\left\Vert \bar{\mu}_{X}^{\left(
1\right) }\right\Vert \left\Vert \beta \right\Vert +\frac{1}{2}\bar{\mu}%
_{Y}^{\left( 2\right) }+\frac{1}{2}\left\Vert \bar{\mu}_{X}^{\left( 2\right)
}\right\Vert \left\Vert \beta \right\Vert \right) \\
&&\cdot E\left[ \left\Vert X-\mu _{X}\left( P\right) \right\Vert \left( 
\tilde{\pi}\left( X\right) -P\right) ^{2}\right] ,
\end{eqnarray*}%
where $\bar{\mu}_{X}^{\left( d\right) }$ and $\bar{\mu}_{Y}^{\left( d\right)
}$ denote the bounds of the $d$-th derivatives of $\mu _{X}$ and $\mu _{Y}$,
respectively. By H\"{o}lder's inequality,%
\begin{eqnarray*}
E\left[ \left\Vert X-\mu _{X}\left( P\right) \right\Vert \left( \tilde{\pi}%
\left( X\right) -P\right) ^{2}\right] &\leq &\left( E\left[ \left\Vert X-\mu
_{X}\left( P\right) \right\Vert ^{\left. q\right/ \left( q-2\right) }\right]
\right) ^{\left( q-2\right) \left/ q\right. }\left\Vert \tilde{\pi}-\pi
\right\Vert _{q,F}^{2} \\
E\left[ \left( \tilde{\pi}\left( X\right) -P\right) ^{2}\right] &\leq
&\left\Vert \tilde{\pi}-\pi \right\Vert _{q,F}^{2}
\end{eqnarray*}%
Since $E\left[ \left\Vert X\right\Vert ^{\left. q\right/ \left( q-2\right) }%
\right] =E\left[ \left\Vert X\right\Vert ^{s}\right] <\infty $, it follows
from the $C_{r}$-inequality \cite[p.80]{shao2003} that%
\begin{equation*}
E\left[ \left\Vert X-\mu _{X}\left( P\right) \right\Vert ^{\left. q\right/
\left( q-2\right) }\right] =E\left[ \left\Vert X-\mu _{X}\left( P\right)
\right\Vert ^{s}\right] \leq 2^{s-1}\left( E\left\Vert X\right\Vert
^{s}+\left\Vert \bar{\mu}_{X}\right\Vert ^{s}\right) <\infty .
\end{equation*}%
Consequently,%
\begin{equation*}
\left\Vert E\left[ \psi \left( W,\tilde{\pi},\mu ,\alpha ,\beta \right) %
\right] \right\Vert \leq C\left\Vert \tilde{\pi}-\pi \right\Vert _{q,F}^{2}
\end{equation*}%
where%
\begin{equation*}
C=\left\Vert \bar{\mu}_{X}^{\left( 1\right) }\right\Vert \left( \bar{\mu}%
_{Y}^{\left( 1\right) }+\left\Vert \bar{\mu}_{X}^{\left( 1\right)
}\right\Vert \left\Vert \beta \right\Vert \right) +\left[ \bar{\mu}%
_{Y}^{\left( 1\right) }+\frac{1}{2}\bar{\mu}_{Y}^{\left( 2\right) }+\frac{1}{%
2}\left( \left\Vert \bar{\mu}_{X}^{\left( 1\right) }\right\Vert +\left\Vert 
\bar{\mu}_{X}^{\left( 2\right) }\right\Vert \right) \left\Vert \beta
\right\Vert \right] \left\Vert X-\mu _{X}\left( P\right) \right\Vert _{s,F}
\end{equation*}
\end{proof}

\begin{proof}[\textbf{Proof of Lemma \ref{lemma3}}]
Since the bound for $\left\Vert E\left[ \psi \left( W,\tilde{\pi},\mu
,\alpha ,\beta \right) \right] \right\Vert $ has been developed in Lemma \ref%
{lemma2}.(ii), it is sufficient to establish the bound for $\left\Vert E%
\left[ \psi \left( W,\tilde{\pi},\tilde{\mu},\alpha ,\beta \right) \right] -E%
\left[ \psi \left( W,\tilde{\pi},\mu ,\alpha ,\beta \right) \right]
\right\Vert $. To this end, we write%
\begin{eqnarray}
&&E\left[ \psi \left( W,\tilde{\pi},\tilde{\mu},\alpha ,\beta \right) \right]
-E\left[ \psi \left( W,\tilde{\pi},\mu ,\alpha ,\beta \right) \right]  \notag
\\
&=&-E\left[ \tilde{\pi}\left( X\right) \left[ X-\tilde{\mu}_{X}\left( \tilde{%
\pi}\left( X\right) \right) \right] \left[ \tilde{\mu}_{Y}\left( \tilde{\pi}%
\left( X\right) \right) -\mu _{Y}\left( P\right) -P\left( \tilde{\mu}%
_{X}\left( \tilde{\pi}\left( X\right) \right) -\mu _{X}\left( P\right)
\right) ^{\prime }\beta \right] \right]  \notag \\
&&+E\left[ \tilde{\pi}\left( X\right) \left[ X-\mu _{X}\left( \tilde{\pi}%
\left( X\right) \right) \right] \left[ \mu _{Y}\left( \tilde{\pi}\left(
X\right) \right) -\mu _{Y}\left( P\right) -P\left( \mu _{X}\left( \tilde{\pi}%
\left( X\right) \right) -\mu _{X}\left( P\right) \right) ^{\prime }\beta %
\right] \right]  \notag \\
&=&E\left[ \tilde{\pi}\left( X\right) \left[ \tilde{\mu}_{X}\left( \tilde{\pi%
}\left( X\right) \right) -\mu _{X}\left( \tilde{\pi}\left( X\right) \right) %
\right] \left[ \tilde{\mu}_{Y}\left( \tilde{\pi}\left( X\right) \right) -\mu
_{Y}\left( P\right) -P\left( \tilde{\mu}_{X}\left( \tilde{\pi}\left(
X\right) \right) -\mu _{X}\left( P\right) \right) ^{\prime }\beta \right] %
\right]  \notag \\
&&+E\left[ \tilde{\pi}\left( X\right) \left[ \mu _{X}\left( \tilde{\pi}%
\left( X\right) \right) -\mu _{X}\left( P\right) \right] \left[ \tilde{\mu}%
_{Y}\left( \tilde{\pi}\left( X\right) \right) -\mu _{Y}\left( \tilde{\pi}%
\left( X\right) \right) -P\left( \tilde{\mu}_{X}\left( \tilde{\pi}\left(
X\right) \right) -\mu _{X}\left( \tilde{\pi}\left( X\right) \right) \right)
^{\prime }\beta \right] \right]  \notag \\
&&-E\left[ \tilde{\pi}\left( X\right) \left( X-\mu _{X}\left( P\right)
\right) \left[ \tilde{\mu}_{Y}\left( \tilde{\pi}\left( X\right) \right) -\mu
_{Y}\left( \tilde{\pi}\left( X\right) \right) -P\left( \tilde{\mu}_{X}\left( 
\tilde{\pi}\left( X\right) \right) -\mu _{X}\left( \tilde{\pi}\left(
X\right) \right) \right) ^{\prime }\beta \right] \right]  \notag \\
&\doteq &\left( \text{I}\right) +\left( \text{II}\right) -\left( \text{III}%
\right) .  \label{decompose}
\end{eqnarray}

By Taylor expansion, we have%
\begin{eqnarray}
\tilde{\mu}\left( \tilde{\pi}\left( X\right) \right) -\mu \left( P\right) &=&%
\left[ \tilde{\mu}\left( \tilde{\pi}\left( X\right) \right) -\tilde{\mu}%
\left( P\right) \right] +\left[ \tilde{\mu}\left( P\right) -\mu \left(
P\right) \right]  \notag \\
&=&\tilde{\mu}^{\left( 1\right) }\left( \pi ^{\ast }\left( X\right) \right)
\left( \tilde{\pi}\left( X\right) -P\right) +\left[ \tilde{\mu}\left(
P\right) -\mu \left( P\right) \right] ,  \label{111}
\end{eqnarray}%
and%
\begin{eqnarray}
&&\tilde{\mu}\left( \tilde{\pi}\left( X\right) \right) -\mu \left( \tilde{\pi%
}\left( X\right) \right)  \notag \\
&=&\left[ \tilde{\mu}\left( \tilde{\pi}\left( X\right) \right) -\tilde{\mu}%
\left( P\right) \right] -\left[ \mu \left( \tilde{\pi}\left( X\right)
\right) -\mu \left( P\right) \right] +\left[ \tilde{\mu}\left( P\right) -\mu
\left( P\right) \right]  \notag \\
&=&\left[ \tilde{\mu}^{\left( 1\right) }\left( \pi ^{\ast }\left( X\right)
\right) -\mu ^{\left( 1\right) }\left( \pi ^{\ast }\left( X\right) \right) %
\right] \left( \tilde{\pi}\left( X\right) -P\right) +\left[ \tilde{\mu}%
\left( P\right) -\mu \left( P\right) \right]  \label{222} \\
&=&\left[ \tilde{\mu}^{\left( 1\right) }\left( P\right) -\mu ^{\left(
1\right) }\left( P\right) \right] \left( \tilde{\pi}\left( X\right)
-P\right) +\frac{1}{2}\left[ \tilde{\mu}^{\left( 2\right) }\left( \pi ^{\ast
}\left( X\right) \right) -\mu ^{\left( 2\right) }\left( \pi ^{\ast }\left(
X\right) \right) \right] \left( \tilde{\pi}\left( X\right) -P\right) ^{2} 
\notag \\
&&+\left[ \tilde{\mu}\left( P\right) -\mu \left( P\right) \right] .
\label{333}
\end{eqnarray}%
where $\pi ^{\ast }$ represents an intermediate value between $\tilde{\pi}$
and $\pi $, which may have different values at each appearance.

Substituting (\ref{111}) and (\ref{222}) into the first term of (\ref%
{decompose}), we obtain%
\begin{eqnarray*}
\left( \text{I}\right) &=&E\left[ \tilde{\pi}\left( X\right) \left[ \tilde{%
\mu}_{X}^{\left( 1\right) }\left( \pi ^{\ast }\left( X\right) \right) -\mu
_{X}^{\left( 1\right) }\left( \pi ^{\ast }\left( X\right) \right) \right] %
\left[ \tilde{\mu}_{Y}^{\left( 1\right) }\left( \pi ^{\ast }\left( X\right)
\right) -P\tilde{\mu}_{X}^{\left( 1\right) }\left( \pi ^{\ast }\left(
X\right) \right) ^{\prime }\beta \right] \left( \tilde{\pi}\left( X\right)
-P\right) ^{2}\right] \\
&&+E\left[ \tilde{\pi}\left( X\right) \left[ \tilde{\mu}_{X}^{\left(
1\right) }\left( \pi ^{\ast }\left( X\right) \right) -\mu _{X}^{\left(
1\right) }\left( \pi ^{\ast }\left( X\right) \right) \right] \left( \tilde{%
\pi}\left( X\right) -P\right) \left[ \tilde{\mu}_{Y}\left( P\right) -\mu
_{Y}\left( P\right) \right] \right] \\
&&-E\left[ \tilde{\pi}\left( X\right) \left[ \tilde{\mu}_{X}^{\left(
1\right) }\left( \pi ^{\ast }\left( X\right) \right) -\mu _{X}^{\left(
1\right) }\left( \pi ^{\ast }\left( X\right) \right) \right] \left( \tilde{%
\pi}\left( X\right) -P\right) P\left[ \tilde{\mu}_{X}\left( P\right) -\mu
_{X}\left( P\right) \right] ^{\prime }\beta \right] \\
&&+E\left[ \left[ \tilde{\mu}_{Y}^{\left( 1\right) }\left( \pi ^{\ast
}\left( X\right) \right) -P\tilde{\mu}_{X}^{\left( 1\right) }\left( \pi
^{\ast }\left( X\right) \right) ^{\prime }\beta \right] \left( \tilde{\pi}%
\left( X\right) -P\right) \left[ \tilde{\mu}_{X}\left( P\right) -\mu
_{X}\left( P\right) \right] \right] \\
&&+E\left[ \left( \tilde{\mu}_{X}\left( P\right) -\mu _{X}\left( P\right)
\right) \left( \tilde{\mu}_{Y}\left( P\right) -\mu _{Y}\left( P\right)
\right) \right] -E\left[ \left( \tilde{\mu}_{X}\left( P\right) -\mu
_{X}\left( P\right) \right) P\left( \tilde{\mu}_{X}\left( P\right) -\mu
_{X}\left( P\right) \right) ^{\prime }\beta \right] .
\end{eqnarray*}%
It follows from H\"{o}lder's inequality that%
\begin{eqnarray}
\left\Vert \left( \text{I}\right) \right\Vert &\leq &\left( \left\Vert 
\overline{\tilde{\mu}}_{X}^{\left( 1\right) }\right\Vert +\left\Vert \bar{\mu%
}_{X}^{\left( 1\right) }\right\Vert \right) \left( \overline{\tilde{\mu}}%
_{Y}^{\left( 1\right) }+\left\Vert \overline{\tilde{\mu}}_{X}^{\left(
1\right) }\right\Vert \left\Vert \beta \right\Vert \right) \left\Vert \tilde{%
\pi}-\pi \right\Vert _{F}^{2}  \notag \\
&&+\left( \left\Vert \overline{\tilde{\mu}}_{X}^{\left( 1\right)
}\right\Vert +\left\Vert \bar{\mu}_{X}^{\left( 1\right) }\right\Vert \right)
\left\Vert \tilde{\pi}-\pi \right\Vert _{F}\left\Vert \tilde{\mu}_{Y}-\mu
_{Y}\right\Vert _{F}  \notag \\
&&+\left[ \left( \left\Vert \overline{\tilde{\mu}}_{X}^{\left( 1\right)
}\right\Vert +\left\Vert \bar{\mu}_{X}^{\left( 1\right) }\right\Vert \right)
\left\Vert \beta \right\Vert +\left( \overline{\tilde{\mu}}_{Y}^{\left(
1\right) }+\left\Vert \overline{\tilde{\mu}}_{X}^{\left( 1\right)
}\right\Vert \left\Vert \beta \right\Vert \right) \right] \left\Vert \tilde{%
\pi}-\pi \right\Vert _{F}\left\Vert \tilde{\mu}_{X}-\mu _{X}\right\Vert _{F}
\notag \\
&&+\left\Vert \tilde{\mu}_{X}-\mu _{X}\right\Vert _{F}\left\Vert \tilde{\mu}%
_{Y}-\mu _{Y}\right\Vert _{F}+\left\Vert \beta \right\Vert \left\Vert \tilde{%
\mu}_{X}-\mu _{X}\right\Vert _{F}^{2}  \label{(I)}
\end{eqnarray}%
where $\left\Vert \cdot \right\Vert _{F}$ denotes the $L_{2}\left( F\right) $%
-norm.

Substituting $\mu _{X}\left( \tilde{\pi}\left( X\right) \right) -\mu
_{X}\left( P\right) =\mu _{X}^{\left( 1\right) }\left( \pi ^{\ast }\left(
X\right) \right) \left( \tilde{\pi}\left( X\right) -P\right) $ and (\ref{222}%
) into the second term of (\ref{decompose}), we obtain%
\begin{equation*}
\left( \text{II}\right) =E\left[ \tilde{\pi}\left( X\right) \mu _{X}^{\left(
1\right) }\left( \pi ^{\ast }\left( X\right) \right) \left( \tilde{\pi}%
\left( X\right) -P\right) \left\{ 
\begin{array}{c}
\left[ 
\begin{array}{c}
\tilde{\mu}_{Y}^{\left( 1\right) }\left( \pi ^{\ast }\left( X\right) \right)
-\mu _{Y}^{\left( 1\right) }\left( \pi ^{\ast }\left( X\right) \right) \\ 
-P\left[ \tilde{\mu}_{X}^{\left( 1\right) }\left( \pi ^{\ast }\left(
X\right) \right) -\mu _{X}^{\left( 1\right) }\left( \pi ^{\ast }\left(
X\right) \right) \right] ^{\prime }\beta%
\end{array}%
\right] \left( \tilde{\pi}\left( X\right) -P\right) \\ 
+\left[ \tilde{\mu}_{Y}\left( P\right) -\mu _{Y}\left( P\right) \right] -P%
\left[ \tilde{\mu}_{X}\left( P\right) -\mu _{X}\left( P\right) \right]
^{\prime }\beta%
\end{array}%
\right\} \right] .
\end{equation*}%
It follows from H\"{o}lder's inequality that%
\begin{eqnarray}
\left\Vert \left( \text{II}\right) \right\Vert &\leq &\left\Vert \bar{\mu}%
_{X}^{\left( 1\right) }\right\Vert \left[ \overline{\tilde{\mu}}_{Y}^{\left(
1\right) }+\bar{\mu}_{Y}^{\left( 1\right) }+\left( \left\Vert \overline{%
\tilde{\mu}}_{X}^{\left( 1\right) }\right\Vert +\left\Vert \bar{\mu}%
_{X}^{\left( 1\right) }\right\Vert \right) \left\Vert \beta \right\Vert %
\right] \left\Vert \tilde{\pi}-\pi \right\Vert _{F}^{2}  \notag \\
&&+\left\Vert \bar{\mu}_{X}^{\left( 1\right) }\right\Vert \left\Vert \tilde{%
\pi}-\pi \right\Vert _{F}\left\Vert \tilde{\mu}_{Y}-\mu _{Y}\right\Vert _{F}
\notag \\
&&+\left\Vert \bar{\mu}_{X}^{\left( 1\right) }\right\Vert \left\Vert \beta
\right\Vert \left\Vert \tilde{\pi}-\pi \right\Vert _{F}\left\Vert \tilde{\mu}%
_{X}-\mu _{X}\right\Vert _{F}  \label{(II)}
\end{eqnarray}

Substituting (\ref{333}) into the third term of (\ref{decompose}), we obtain%
\begin{eqnarray*}
\left( \text{III}\right) &=&E\left[ \tilde{\pi}\left( X\right) \left( X-\mu
_{X}\left( P\right) \right) \left[ \tilde{\mu}_{Y}^{\left( 1\right) }\left(
P\right) -\mu _{Y}^{\left( 1\right) }\left( P\right) \right] \left( \tilde{%
\pi}\left( X\right) -P\right) \right] \\
&&-E\left[ \tilde{\pi}\left( X\right) \left( X-\mu _{X}\left( P\right)
\right) P\left[ \tilde{\mu}_{X}^{\left( 1\right) }\left( P\right) -\mu
_{X}^{\left( 1\right) }\left( P\right) \right] ^{\prime }\beta \left( \tilde{%
\pi}\left( X\right) -P\right) \right] \\
&&+\frac{1}{2}E\left[ \tilde{\pi}\left( X\right) \left( X-\mu _{X}\left(
P\right) \right) \left[ 
\begin{array}{c}
\tilde{\mu}_{Y}^{\left( 2\right) }\left( \pi ^{\ast }\left( X\right) \right)
-\mu _{Y}^{\left( 2\right) }\left( \pi ^{\ast }\left( X\right) \right) \\ 
-P\left[ \tilde{\mu}_{X}^{\left( 2\right) }\left( \pi ^{\ast }\left(
X\right) \right) -\mu _{X}^{\left( 2\right) }\left( \pi ^{\ast }\left(
X\right) \right) \right] ^{\prime }\beta%
\end{array}%
\right] \left( \tilde{\pi}\left( X\right) -P\right) ^{2}\right] \\
&&+E\left[ \left( \tilde{\pi}\left( X\right) -P\right) \left( X-\mu
_{X}\left( P\right) \right) \left[ \tilde{\mu}_{Y}\left( P\right) -\mu
_{Y}\left( P\right) -P\left( \tilde{\mu}_{X}\left( P\right) -\mu _{X}\left(
P\right) \right) ^{\prime }\beta \right] \right] \\
&&+E\left[ P\left( X-\mu _{X}\left( P\right) \right) \left[ \tilde{\mu}%
_{Y}\left( P\right) -\mu _{Y}\left( P\right) -P\left( \tilde{\mu}_{X}\left(
P\right) -\mu _{X}\left( P\right) \right) ^{\prime }\beta \right] \right] .
\end{eqnarray*}%
By the law of iterated expectations, we can show that the last term on the
right-hand side above equals to zero. Therefore, if we denote $c=\left\Vert
X-\mu _{X}\left( P\right) \right\Vert _{s,F}<\infty $, it will follow from H%
\"{o}lder's inequality that%
\begin{eqnarray}
\left\Vert \left( \text{III}\right) \right\Vert &\leq &c\left\Vert \tilde{\pi%
}-\pi \right\Vert _{q,F}\left\Vert \tilde{\mu}_{Y}^{\left( 1\right) }-\mu
_{Y}^{\left( 1\right) }\right\Vert _{q,F}+c\left\Vert \beta \right\Vert
\left\Vert \tilde{\pi}-\pi \right\Vert _{q,F}\left\Vert \tilde{\mu}%
_{X}^{\left( 1\right) }-\mu _{X}^{\left( 1\right) }\right\Vert _{q,F}  \notag
\\
&&+\frac{c}{2}\left[ \overline{\tilde{\mu}}_{Y}^{\left( 2\right) }+\bar{\mu}%
_{Y}^{\left( 2\right) }+\left( \left\Vert \overline{\tilde{\mu}}_{X}^{\left(
2\right) }\right\Vert +\left\Vert \bar{\mu}_{X}^{\left( 2\right)
}\right\Vert \right) \left\Vert \beta \right\Vert \right] \left\Vert \tilde{%
\pi}-\pi \right\Vert _{q,F}^{2}  \notag \\
&&+c\left\Vert \tilde{\pi}-\pi \right\Vert _{q,F}\left\Vert \tilde{\mu}%
_{Y}-\mu _{Y}\right\Vert _{q,F}+c\left\Vert \beta \right\Vert \left\Vert 
\tilde{\pi}-\pi \right\Vert _{q,F}\left\Vert \tilde{\mu}_{X}-\mu
_{X}\right\Vert _{q,F}  \label{(III)}
\end{eqnarray}

By the decomposition (\ref{decompose}), we have%
\begin{eqnarray*}
\left\Vert E\left[ \psi \left( W,\tilde{\pi},\tilde{\mu},\alpha ,\beta
\right) \right] \right\Vert &\leq &\left\Vert E\left[ \psi \left( W,\tilde{%
\pi},\tilde{\mu},\alpha ,\beta \right) \right] -E\left[ \psi \left( W,\tilde{%
\pi},\mu ,\alpha ,\beta \right) \right] \right\Vert +\left\Vert E\left[ \psi
\left( W,\tilde{\pi},\mu ,\alpha ,\beta \right) \right] \right\Vert \\
&\leq &\left\Vert \left( \text{I}\right) \right\Vert +\left\Vert \left( 
\text{II}\right) \right\Vert +\left\Vert \left( \text{III}\right)
\right\Vert +\left\Vert E\left[ \psi \left( W,\tilde{\pi},\mu ,\alpha ,\beta
\right) \right] \right\Vert .
\end{eqnarray*}%
Since by H\"{o}lder's inequality%
\begin{equation*}
\left\Vert \tilde{\pi}-\pi \right\Vert _{F}\leq \left\Vert \tilde{\pi}-\pi
\right\Vert _{q,F}\text{ , \ }\left\Vert \tilde{\mu}_{Y}-\mu _{Y}\right\Vert
_{F}\leq \left\Vert \tilde{\mu}_{Y}-\mu _{Y}\right\Vert _{q,F}\text{ ,}
\end{equation*}%
the desired result follows from combining (\ref{(I)}), (\ref{(II)}), (\ref%
{(III)}), and Lemma \ref{lemma2}.(ii).
\end{proof}

\begin{proof}[\textbf{Proof of Lemma \ref{lemma4}}]
We first prove $n^{1/4}\left\Vert \hat{\mu}_{\ell }-\mu \right\Vert _{q,F}%
\overset{p}{\longrightarrow }0$ for each $\ell =1,\cdots ,L$, which is implied by $\sup_{p}\left\vert 
\hat{\mu}_{Z\ell }\left( p\right) -\mu _{Z}\left( p\right) \right\vert
=o_{p}\left( n^{-1/4}\right) $ for $Z$ being $Y$ or any element of $X$.
Denote%
\begin{equation*}
\mu _{Z\ell n}\left( p\right) =\left. \left[ \sum_{j\in I_{\ell }^{c}}\frac{1%
}{h_{Z}}k\left( \frac{P_{j}-p}{h_{Z}}\right) Z_{j}\right] \right/ \left[
\sum_{j\in I_{\ell }^{c}}\frac{1}{h_{Z}}k\left( \frac{P_{j}-p}{h_{Z}}\right) %
\right] .
\end{equation*}%
It follows from \citet[Theorem 2.6]{li2007nonparametric} and Assumption \ref%
{ass:rate1} that%
\begin{equation}
\sup_{p}\left\vert \mu _{Z\ell n}\left( p\right) -\mu _{Z}\left( p\right)
\right\vert =O_{p}\left( \sqrt{\frac{\ln n}{nh_{Z}}}+h_{Z}^{2}\right)
=o_{p}\left( n^{-1/4}\right) .  \label{mu_Zn}
\end{equation}%
To prove $\sup_{p}\left\vert \hat{\mu}_{Z\ell }\left( p\right) -\mu _{Z\ell
n}\left( p\right) \right\vert =o_{p}\left( n^{-1/4}\right) $, we denote%
\begin{eqnarray*}
\hat{A}_{Z\ell }\left( p\right) &=&\frac{1}{n-n_{\ell }}\sum_{\ell ^{\prime
}\neq \ell }\sum_{j\in I_{\ell ^{\prime }}}\frac{1}{h_{Z}}k\left( \frac{\hat{%
P}_{\ell \ell ^{\prime }j}-p}{h_{Z}}\right) Z_{j} \\
A_{Z\ell n}\left( p\right) &=&\frac{1}{n-n_{\ell }}\sum_{j\in I_{\ell }^{c}}%
\frac{1}{h_{Z}}k\left( \frac{P_{j}-p}{h_{Z}}\right) Z_{j}
\end{eqnarray*}%
for $Z$ being $1$, $Y$, or any element of $X$. Then $\hat{\mu}_{Z\ell
}\left( p\right) =\left. \hat{A}_{Z\ell }\left( p\right) \right/ \hat{A}%
_{1\ell }\left( p\right) $, $\mu _{Z\ell n}\left( p\right) =\left. A_{Z\ell
n}\left( p\right) \right/ A_{1\ell n}\left( p\right) $, and%
\begin{equation}
\hat{\mu}_{Z\ell }-\mu _{Z\ell n}=\frac{\left( \hat{A}_{Z\ell }-A_{Z\ell
n}\right) -\left( \hat{A}_{1\ell }-A_{1\ell n}\right) \mu _{Z\ell n}}{\hat{A}%
_{1\ell }}.  \label{mu_Z}
\end{equation}%
So we need to show that $\sup_{p}\left\vert \hat{A}_{Z\ell }\left( p\right)
-A_{Z\ell n}\left( p\right) \right\vert =o_{p}\left( n^{-1/4}\right) $ for $%
Z $ being $1$, $Y$, or any element of $X$.

By Taylor expansion,%
\begin{eqnarray}
\hat{A}_{Z\ell }\left( p\right) -A_{Z\ell n}\left( p\right) &=&\frac{1}{%
n-n_{\ell }}\sum_{\ell ^{\prime }\neq \ell }\sum_{j\in I_{\ell ^{\prime }}}%
\frac{1}{h_{Z}^{2}}k^{\left( 1\right) }\left( \frac{P_{j}-p}{h_{Z}}\right)
Z_{j}\left( \hat{P}_{\ell \ell ^{\prime }j}-P_{j}\right)  \notag \\
&+&\frac{1}{2\left( n-n_{\ell }\right) }\sum_{\ell ^{\prime }\neq \ell
}\sum_{j\in I_{\ell ^{\prime }}}\frac{1}{h_{Z}^{3}}k^{\left( 2\right)
}\left( \frac{P_{\ell \ell ^{\prime }j}^{\ast }-p}{h_{Z}}\right) Z_{j}\left( 
\hat{P}_{\ell \ell ^{\prime }j}-P_{j}\right) ^{2},  \label{AZ}
\end{eqnarray}%
where $P_{\ell \ell ^{\prime }j}^{\ast }$ is an intermediate value between $%
\hat{P}_{\ell \ell ^{\prime }j}$ and $P_{j}$. Let $\mathcal{W}_{\ell \ell
^{\prime }}^{c}=\left\{ W_{i}:i\notin I_{\ell },i\notin I_{\ell ^{\prime
}}\right\} $ denote the observations not in $I_{\ell }$ and not in $I_{\ell
^{\prime }}$. For the first term, by H\"{o}lder's inequality and standard
arguments in the kernel estimation,%
\begin{eqnarray*}
&&\left\vert E\left[ \left. \frac{1}{n-n_{\ell }}\sum_{j\in I_{\ell ^{\prime
}}}\frac{1}{h_{Z}^{2}}k^{\left( 1\right) }\left( \frac{P_{j}-p}{h_{Z}}%
\right) Z_{j}\left( \hat{P}_{\ell \ell ^{\prime }j}-P_{j}\right) \right\vert 
\mathcal{W}_{\ell \ell ^{\prime }}^{c}\right] \right\vert \\
&=&\frac{n_{\ell ^{\prime }}}{\left( n-n_{\ell }\right) h_{Z}^{2}}\left\vert
E\left[ \left. k^{\left( 1\right) }\left( \frac{P_{j}-p}{h_{Z}}\right)
Z_{j}\left( \hat{P}_{\ell \ell ^{\prime }j}-P_{j}\right) \right\vert 
\mathcal{W}_{\ell \ell ^{\prime }}^{c}\right] \right\vert \\
&\leq &\frac{1}{h_{Z}^{2}}\left( E\left[ \left\vert k^{\left( 1\right)
}\left( \frac{P_{j}-p}{h_{Z}}\right) Z_{j}\right\vert ^{q\left/ \left(
q-1\right) \right. }\right] \right) ^{1-1/q}\left\Vert \hat{\pi}_{\ell \ell
^{\prime }}-\pi \right\Vert _{q,F} \\
&=&\frac{1}{h_{Z}^{2}}\left( E\left[ \left\vert k^{\left( 1\right) }\left( 
\frac{P_{j}-p}{h_{Z}}\right) \right\vert ^{\tilde{s}}\sigma _{\tilde{s}%
}\left( P_{j}\right) \right] \right) ^{1/\tilde{s}}\left\Vert \hat{\pi}%
_{\ell \ell ^{\prime }}-\pi \right\Vert _{q,F} \\
&=&\frac{1}{h_{Z}^{2}}\left( \int_{0}^{1}\left\vert k^{\left( 1\right)
}\left( \frac{p_{j}-p}{h_{Z}}\right) \right\vert ^{\tilde{s}}\sigma _{\tilde{%
s}}\left( p_{j}\right) f_{P}\left( p_{j}\right) dp_{j}\right) ^{1/\tilde{s}%
}\left\Vert \hat{\pi}_{\ell \ell ^{\prime }}-\pi \right\Vert _{q,F} \\
&=&\frac{1}{h_{Z}^{2}}\left( h_{Z}\int_{0}^{1}\left\vert k^{\left( 1\right)
}\left( v\right) \right\vert ^{\tilde{s}}\sigma _{\tilde{s}}\left(
p+vh_{Z}\right) f_{P}\left( p+vh_{Z}\right) dv\right) ^{1/\tilde{s}%
}\left\Vert \hat{\pi}_{\ell \ell ^{\prime }}-\pi \right\Vert _{q,F} \\
&=&\frac{1}{h_{Z}^{1+1/q}}\left[ \sigma _{\tilde{s}}\left( p\right)
f_{P}\left( p\right) \int_{0}^{1}\left\vert k^{\left( 1\right) }\left(
v\right) \right\vert ^{\tilde{s}}dv+O\left( h_{Z}\right) \right] ^{1/\tilde{s%
}}\left\Vert \hat{\pi}_{\ell \ell ^{\prime }}-\pi \right\Vert _{q,F}
\end{eqnarray*}%
where $\tilde{s}=q\left/ \left( q-1\right) \right. $ and $\sigma _{\tilde{s}%
}\left( p\right) =E\left[ \left. \left\vert Z\right\vert ^{\tilde{s}%
}\right\vert P=p\right] $. It follows from Assumption \ref{ass:rate1}.(iii),
(iv), and (v) that $f_{P}\left( p\right) $ and $\sigma _{\tilde{s}}\left(
p\right) $ are bounded over $p\in \left( 0,1\right) $, and $%
\int_{0}^{1}\left\vert k^{\left( 1\right) }\left( v\right) \right\vert ^{%
\tilde{s}}dv\leq \left( \overline{k^{\left( 1\right) }}\right) ^{\tilde{s}%
}<\infty $. Therefore,%
\begin{equation*}
\sup_{p}\left\vert E\left[ \left. \frac{1}{n-n_{\ell }}\sum_{j\in I_{\ell
^{\prime }}}\frac{1}{h_{Z}^{2}}k^{\left( 1\right) }\left( \frac{P_{j}-p}{%
h_{Z}}\right) Z_{j}\left( \hat{P}_{\ell \ell ^{\prime }j}-P_{j}\right)
\right\vert \mathcal{W}_{\ell \ell ^{\prime }}^{c}\right] \right\vert
=O\left( \frac{\left\Vert \hat{\pi}_{\ell \ell ^{\prime }}-\pi \right\Vert
_{q,F}}{h_{Z}^{1+1/q}}\right) =o_{p}\left( n^{-1/4}\right) ,
\end{equation*}%
where the second equality follows from Assumption \ref{ass:rate1}.(ii). On
the other hand,%
\begin{eqnarray*}
&&Var\left( \left. \frac{1}{n-n_{\ell }}\sum_{j\in I_{\ell ^{\prime }}}\frac{%
1}{h_{Z}^{2}}k^{\left( 1\right) }\left( \frac{P_{j}-p}{h_{Z}}\right)
Z_{j}\left( \hat{P}_{\ell \ell ^{\prime }j}-P_{j}\right) \right\vert 
\mathcal{W}_{\ell \ell ^{\prime }}^{c}\right) \\
&\leq &\frac{1}{n-n_{\ell }}E\left[ \left. \frac{1}{h_{Z}^{4}}k^{\left(
1\right) }\left( \frac{P_{j}-p}{h_{Z}}\right) ^{2}Z_{j}^{2}\left( \hat{P}%
_{\ell \ell ^{\prime }j}-P_{j}\right) ^{2}\right\vert \mathcal{W}_{\ell \ell
^{\prime }}^{c}\right] \\
&\leq &\frac{\left( \overline{k^{\left( 1\right) }}\right) ^{2}}{\left(
n-n_{\ell }\right) h_{Z}^{4}}\left( E\left[ \left\vert Z_{j}\right\vert ^{2s}%
\right] \right) ^{1/s}\left\Vert \hat{\pi}_{\ell \ell ^{\prime }}-\pi
\right\Vert _{q,F}^{2} \\
&=&O\left( \frac{\left\Vert \hat{\pi}_{\ell \ell ^{\prime }}-\pi \right\Vert
_{q,F}^{2}}{nh_{Z}^{4}}\right) =o_{p}\left( \frac{h_{Z}^{2/q}}{%
n^{3/2}h_{Z}^{2}}\right) =o_{p}\left( n^{-1/2}\right)
\end{eqnarray*}%
uniformly over $p\in \left( 0,1\right) $, where the last equality follows
from $h_{Z}\rightarrow 0$ and $nh_{Z}^{2}\rightarrow \infty $. In
combination, we have%
\begin{equation}
\sup_{p}\left\vert \frac{1}{n-n_{\ell }}\sum_{j\in I_{\ell ^{\prime }}}\frac{%
1}{h_{Z}^{2}}k^{\left( 1\right) }\left( \frac{P_{j}-p}{h_{Z}}\right)
Z_{j}\left( \hat{P}_{\ell \ell ^{\prime }j}-P_{j}\right) \right\vert
=o_{p}\left( n^{-1/4}\right)  \label{B1}
\end{equation}%
for each $\ell =1,\cdots ,L$ and $\ell ^{\prime }\neq \ell $.

For the second term of $\hat{A}_{Z\ell }\left( p\right) -A_{Z\ell n}\left(
p\right) $, it follows from Assumption \ref{ass:rate1}.(v) and (vi) that%
\begin{eqnarray*}
\sup_{p}\left\vert E\left[ \left. \frac{1}{n-n_{\ell }}\sum_{j\in I_{\ell
^{\prime }}}\frac{1}{h_{Z}^{3}}k^{\left( 2\right) }\left( \frac{P_{\ell \ell
^{\prime }j}^{\ast }-p}{h_{Z}}\right) Z_{j}\left( \hat{P}_{\ell \ell
^{\prime }j}-P_{j}\right) ^{2}\right\vert \mathcal{W}_{\ell \ell ^{\prime
}}^{c}\right] \right\vert &\leq &\frac{\overline{k^{\left( 2\right) }}}{%
h_{Z}^{3}}E\left[ \left. \left\vert Z_{j}\right\vert \left( \hat{P}_{\ell
\ell ^{\prime }j}-P_{j}\right) ^{2}\right\vert \mathcal{W}_{\ell \ell
^{\prime }}^{c}\right] \\
&\leq &\frac{\overline{k^{\left( 2\right) }}}{h_{Z}^{3}}\left( E\left\vert
Z_{j}\right\vert ^{s}\right) ^{1/s}\left\Vert \hat{\pi}_{\ell \ell ^{\prime
}}-\pi \right\Vert _{q,F}^{2} \\
&=&O\left( \frac{\left\Vert \hat{\pi}_{\ell \ell ^{\prime }}-\pi \right\Vert
_{q,F}^{2}}{h_{Z}^{3}}\right) \\
&=&o_{p}\left( \frac{1}{n^{1/2}h_{Z}^{1-2/q}}\right) \\
&=&o_{p}\left( n^{-1/4}\right) .
\end{eqnarray*}%
On the other hand,%
\begin{eqnarray*}
&&Var\left( \left. \frac{1}{n-n_{\ell }}\sum_{j\in I_{\ell ^{\prime }}}\frac{%
1}{h_{Z}^{3}}k^{\left( 2\right) }\left( \frac{P_{\ell \ell ^{\prime
}j}^{\ast }-p}{h_{Z}}\right) Z_{j}\left( \hat{P}_{\ell \ell ^{\prime
}j}-P_{j}\right) ^{2}\right\vert \mathcal{W}_{\ell \ell ^{\prime
}}^{c}\right) \\
&\leq &\frac{1}{n-n_{\ell }}E\left[ \left. \frac{1}{h_{Z}^{6}}k^{\left(
2\right) }\left( \frac{P_{\ell \ell ^{\prime }j}^{\ast }-p}{h_{Z}}\right)
^{2}Z_{j}^{2}\left( \hat{P}_{\ell \ell ^{\prime }j}-P_{j}\right)
^{4}\right\vert \mathcal{W}_{\ell \ell ^{\prime }}^{c}\right] \\
&\leq &\frac{\overline{k^{\left( 2\right) }}}{\left( n-n_{\ell }\right)
h_{Z}^{6}}\left( E\left[ \left\vert Z_{j}\right\vert ^{2s\left/ \left(
2-s\right) \right. }\right] \right) ^{\left. \left( 2-s\right) \right/
s}\left\Vert \hat{\pi}_{\ell \ell ^{\prime }}-\pi \right\Vert _{q,F}^{4} \\
&=&O\left( \frac{\left\Vert \hat{\pi}_{\ell \ell ^{\prime }}-\pi \right\Vert
_{q,F}^{4}}{nh_{Z}^{6}}\right) =o_{p}\left( \frac{h_{Z}^{4/q}}{n^{2}h_{Z}^{2}%
}\right) =o_{p}\left( \frac{1}{n}\right)
\end{eqnarray*}%
uniformly over $p\in \left( 0,1\right) $. In combination, we have%
\begin{equation}
\sup_{p}\left\vert \frac{1}{n-n_{\ell }}\sum_{j\in I_{\ell ^{\prime }}}\frac{%
1}{h_{Z}^{3}}k^{\left( 2\right) }\left( \frac{P_{\ell \ell ^{\prime
}j}^{\ast }-p}{h_{Z}}\right) Z_{j}\left( \hat{P}_{\ell \ell ^{\prime
}j}-P_{j}\right) ^{2}\right\vert =o_{p}\left( n^{-1/4}\right)  \label{B2}
\end{equation}%
for each $\ell =1,\cdots ,L$ and $\ell ^{\prime }\neq \ell $.

By (\ref{AZ}), (\ref{B1}), (\ref{B2}), and the triangle inequality,%
\begin{equation*}
\sup_{p}\left\vert \hat{A}_{Z\ell }\left( p\right) -A_{Z\ell n}\left(
p\right) \right\vert =o_{p}\left( n^{-1/4}\right)
\end{equation*}%
for $Z$ being $1$, $Y$, or any element of $X$. Substituting into (\ref{mu_Z}%
) yields%
\begin{eqnarray*}
\sup_{p}\left\vert \hat{\mu}_{Z\ell }\left( p\right) -\mu _{Z\ell n}\left(
p\right) \right\vert &=&\frac{\sup_{p}\left\vert \hat{A}_{Z\ell }\left(
p\right) -A_{Z\ell n}\left( p\right) \right\vert +\sup_{p}\left\vert \hat{A}%
_{1\ell }\left( p\right) -A_{1\ell n}\left( p\right) \right\vert
\sup_{p}\left\vert \mu _{Z\ell n}\left( p\right) \right\vert }{%
\inf_{p}\left\vert \hat{A}_{1\ell }\left( p\right) \right\vert } \\
&=&o_{p}\left( n^{-1/4}\right) \frac{1+\sup_{p}\left\vert \mu _{Z\ell
n}\left( p\right) \right\vert }{\inf_{p}\left\vert \hat{A}_{1\ell }\left(
p\right) \right\vert }.
\end{eqnarray*}%
Note that%
\begin{equation*}
\sup_{p}\left\vert \mu _{Z\ell n}\left( p\right) \right\vert \leq
\sup_{p}\left\vert \mu _{Z}\left( p\right) \right\vert +\sup_{p}\left\vert
\mu _{Z\ell n}\left( p\right) -\mu _{Z}\left( p\right) \right\vert =O\left(
1\right) +o_{p}\left( n^{-1/4}\right) =O_{p}\left( 1\right)
\end{equation*}%
and%
\begin{equation*}
\inf_{p}\left\vert \hat{A}_{1\ell }\left( p\right) \right\vert \geq
\inf_{p}\left\vert f_{P}\left( p\right) \right\vert -\sup_{p}\left\vert
A_{1\ell n}\left( p\right) -f_{P}\left( p\right) \right\vert
-\sup_{p}\left\vert \hat{A}_{1\ell }\left( p\right) -A_{1\ell n}\left(
p\right) \right\vert .
\end{equation*}%
It follows from Assumption \ref{ass:rate1}.(iii) that $\inf_{p}\left\vert
f_{P}\left( p\right) \right\vert \geq \epsilon $ for an $\epsilon >0$, and
from \citet[Theorem 1.4]{li2007nonparametric} that%
\begin{equation*}
\sup_{p}\left\vert A_{1\ell n}\left( p\right) -f_{P}\left( p\right)
\right\vert =O_{p}\left( \sqrt{\frac{\ln n}{nh_{Z}}}+h_{Z}^{2}\right)
=o_{p}\left( n^{-1/4}\right) ,
\end{equation*}%
so that $\inf_{p}\left\vert \hat{A}_{1\ell }\left( p\right) \right\vert \geq 
\tilde{\epsilon}>0$ holds with probability approaching to one for any $%
\tilde{\epsilon}<\epsilon $. Hence, we have%
\begin{equation*}
\sup_{p}\left\vert \hat{\mu}_{Z\ell }\left( p\right) -\mu _{Z\ell n}\left(
p\right) \right\vert =o_{p}\left( n^{-1/4}\right) \cdot O_{p}\left( 1\right)
=o_{p}\left( n^{-1/4}\right)
\end{equation*}%
for each $\ell =1,\cdots ,L$ and each $Z$ being $Y$ or any element of $X$,
which in combination with (\ref{mu_Zn}) implies $\sup_{p}\left\vert \hat{\mu}%
_{Z\ell }\left( p\right) -\mu _{Z}\left( p\right) \right\vert =o_{p}\left(
n^{-1/4}\right) $ and thus $n^{1/4}\left\Vert \hat{\mu}_{\ell }-\mu
\right\Vert _{q,F}\overset{p}{\longrightarrow }0$. Analogously, we can show
that $\sup_{p}\left\vert \hat{\mu}_{Z\ell }^{\left( 1\right) }\left(
p\right) -\mu _{Z}^{\left( 1\right) }\left( p\right) \right\vert
=o_{p}\left( n^{-1/4}\right) $ and $n^{1/4}\left\Vert \hat{\mu}_{\ell
}^{\left( 1\right) }-\mu ^{\left( 1\right) }\right\Vert _{q,F}\overset{p}{%
\longrightarrow }0$.

It remains to show that $n^{1/4}\left\Vert \hat{\alpha}_{\ell }-\alpha
\right\Vert _{q,F}\overset{p}{\longrightarrow }0$. By definition of $\hat{%
\alpha}_{\ell }$ in (\ref{alphahat}), it is sufficient to show that $%
n^{1/4}\left\Vert \hat{\mu}_{\ell }\left( \hat{\pi}_{\ell }\right) -\mu
\left( \pi \right) \right\Vert _{q,F}\overset{p}{\longrightarrow }0$, $%
n^{1/4}\left\Vert \hat{\mu}_{\ell }^{\left( 1\right) }\left( \hat{\pi}_{\ell
}\right) -\mu ^{\left( 1\right) }\left( \pi \right) \right\Vert _{q,F}%
\overset{p}{\longrightarrow }0$, and $n^{1/4}\left\Vert \hat{\beta}_{\ell
}-\beta \right\Vert \overset{p}{\longrightarrow }0$. Since%
\begin{eqnarray*}
\hat{\mu}_{Z\ell }\left( \hat{\pi}_{\ell }\left( x\right) \right) -\mu
_{Z}\left( \pi \left( x\right) \right)  &=&\hat{\mu}_{Z\ell }\left( \hat{\pi}%
_{\ell }\left( x\right) \right) -\mu _{Z}\left( \hat{\pi}_{\ell }\left(
x\right) \right) +\mu _{Z}^{\left( 1\right) }\left( \pi _{\ell }^{\ast
}\left( x\right) \right) \left[ \hat{\pi}_{\ell }\left( x\right) -\pi \left(
x\right) \right]  \\
\hat{\mu}_{Z\ell }^{\left( 1\right) }\left( \hat{\pi}_{\ell }\left( x\right)
\right) -\mu _{Z}^{\left( 1\right) }\left( \pi \left( x\right) \right)  &=&%
\hat{\mu}_{Z\ell }^{\left( 1\right) }\left( \hat{\pi}_{\ell }\left( x\right)
\right) -\mu _{Z}^{\left( 1\right) }\left( \hat{\pi}_{\ell }\left( x\right)
\right) +\mu _{Z}^{\left( 2\right) }\left( \pi _{\ell }^{\ast }\left(
x\right) \right) \left[ \hat{\pi}_{\ell }\left( x\right) -\pi \left(
x\right) \right] 
\end{eqnarray*}%
where $\pi _{\ell }^{\ast }$ represents an intermediate value between $\hat{%
\pi}_{\ell }$ and $\pi $ which may have different values at each appearance,
we have%
\begin{flalign}
\left\Vert \hat{\mu}_{Z\ell }\left( \hat{\pi}_{\ell }\right) -\mu _{Z}\left(
\pi \right) \right\Vert _{q,F} &\leq \sup_{p}\left\vert \hat{\mu}_{Z\ell
}\left( p\right) -\mu _{Z}\left( p\right) \right\vert +\overline{\mu
_{Z}^{\left( 1\right) }}\left\Vert \hat{\pi}_{\ell }-\pi \right\Vert
_{q,F}=o_{p}\left( n^{-1/4}\right)  & \label{31} \\
\left\Vert \hat{\mu}_{Z\ell }^{\left( 1\right) }\left( \hat{\pi}_{\ell
}\right) -\mu _{Z}^{\left( 1\right) }\left( \pi \right) \right\Vert _{q,F}
&\leq \sup_{p}\left\vert \hat{\mu}_{Z\ell }^{\left( 1\right) }\left(
p\right) -\mu _{Z}^{\left( 1\right) }\left( p\right) \right\vert +\overline{%
\mu _{Z}^{\left( 2\right) }}\left\Vert \hat{\pi}_{\ell }-\pi \right\Vert
_{q,F}=o_{p}\left( n^{-1/4}\right)  &  \label{32}
\end{flalign}for $Z$ being $Y$ or any element of $X$. As to $\hat{\beta}%
_{\ell }$, we denote%
\begin{gather*}
\hat{M}_{\ell }=\frac{1}{n-n_{\ell }}\sum_{\ell ^{\prime }\neq \ell
}\sum_{j\in I_{\ell ^{\prime }}}\hat{P}_{\ell \ell ^{\prime }j}^{2}\left(
X_{j}-\hat{\mu}_{X\ell \ell ^{\prime }}\left( \hat{P}_{\ell \ell ^{\prime
}j}\right) \right) \left( X_{j}-\hat{\mu}_{X\ell \ell ^{\prime }}\left( \hat{%
P}_{\ell \ell ^{\prime }j}\right) \right) ^{\prime }, \\
r_{0}\left( W,\pi ,\mu ,\beta \right) =P\left( X-\mu _{X}\left( P\right)
\right) \left[ Y-\mu _{Y}\left( P\right) -P\left( X-\mu _{X}\left( P\right)
\right) ^{\prime }\beta \right] ,
\end{gather*}%
then%
\begin{equation*}
\hat{\beta}_{\ell }-\beta =\hat{M}_{\ell }^{-1}\frac{1}{n-n_{\ell }}%
\sum_{\ell ^{\prime }\neq \ell }\sum_{j\in I_{\ell ^{\prime }}}r_{0}\left(
W_{j},\hat{\pi}_{\ell \ell ^{\prime }},\hat{\mu}_{\ell \ell ^{\prime
}},\beta \right) .
\end{equation*}%
As in the proof of Theorem \ref{theorem:AN}, we can show that%
\begin{equation*}
\hat{M}_{\ell }\overset{p}{\longrightarrow }E\left[ P_{j}^{2}\left(
X_{j}-\mu _{X}\left( P_{j}\right) \right) \left( X_{j}-\mu _{X}\left(
P_{j}\right) \right) ^{\prime }\right] =M
\end{equation*}%
for each $\ell =1,\cdots ,L$. For the term $r_{0}$, we have%
\begin{equation}
r_{0}\left( W_{j},\hat{\pi}_{\ell \ell ^{\prime }},\hat{\mu}_{\ell \ell
^{\prime }},\beta \right) -r_{0}\left( W_{j},\pi ,\mu ,\beta \right) =\hat{Q}%
_{1\ell \ell ^{\prime }j}+\hat{Q}_{2\ell \ell ^{\prime }j}+\hat{Q}_{3\ell
\ell ^{\prime }j}+\hat{Q}_{4\ell \ell ^{\prime }j},  \label{35}
\end{equation}%
where%
\begin{eqnarray*}
\hat{Q}_{1\ell \ell ^{\prime }j} &=&\hat{P}_{\ell \ell ^{\prime }j}\left( 
\hat{\mu}_{X\ell \ell ^{\prime }}\left( \hat{P}_{\ell \ell ^{\prime
}j}\right) -\mu _{X}\left( P_{j}\right) \right) \left[ 
\begin{array}{c}
\hat{\mu}_{Y\ell \ell ^{\prime }}\left( \hat{P}_{\ell \ell ^{\prime
}j}\right) -\mu _{Y}\left( P_{j}\right) -\hat{P}_{\ell \ell ^{\prime
}j}\left( \hat{\mu}_{X\ell \ell ^{\prime }}\left( \hat{P}_{\ell \ell
^{\prime }j}\right) -\mu _{X}\left( P_{j}\right) \right) ^{\prime }\beta  \\ 
+\left( \hat{P}_{\ell \ell ^{\prime }j}-P_{j}\right) \left( X_{j}-\mu
_{X}\left( P_{j}\right) \right) ^{\prime }\beta 
\end{array}%
\right] , \\
\hat{Q}_{2\ell \ell ^{\prime }j} &=&-\hat{P}_{\ell \ell ^{\prime }j}\left(
X_{j}-\mu _{X}\left( P_{j}\right) \right) \left[ 
\begin{array}{c}
\hat{\mu}_{Y\ell \ell ^{\prime }}\left( \hat{P}_{\ell \ell ^{\prime
}j}\right) -\mu _{Y}\left( P_{j}\right) -\hat{P}_{\ell \ell ^{\prime
}j}\left( \hat{\mu}_{X\ell \ell ^{\prime }}\left( \hat{P}_{\ell \ell
^{\prime }j}\right) -\mu _{X}\left( P_{j}\right) \right) ^{\prime }\beta  \\ 
+\left( \hat{P}_{\ell \ell ^{\prime }j}-P_{j}\right) \left( X_{j}-\mu
_{X}\left( P_{j}\right) \right) ^{\prime }\beta 
\end{array}%
\right] , \\
\hat{Q}_{3\ell \ell ^{\prime }j} &=&-\hat{P}_{\ell \ell ^{\prime }j}\left( 
\hat{\mu}_{X\ell \ell ^{\prime }}\left( \hat{P}_{\ell \ell ^{\prime
}j}\right) -\mu _{X}\left( P_{j}\right) \right) \left[ Y_{j}-\mu _{Y}\left(
P_{j}\right) -P_{j}\left( X_{j}-\mu _{X}\left( P_{j}\right) \right) ^{\prime
}\beta \right] , \\
\hat{Q}_{2\ell \ell ^{\prime }j} &=&\left( \hat{P}_{\ell \ell ^{\prime
}j}-P_{j}\right) \left( X_{j}-\mu _{X}\left( P_{j}\right) \right) \left[
Y_{j}-\mu _{Y}\left( P_{j}\right) -P_{j}\left( X_{j}-\mu _{X}\left(
P_{j}\right) \right) ^{\prime }\beta \right] .
\end{eqnarray*}%
Similar to (\ref{31}) and (\ref{32}), it is ready to show that $\left\Vert 
\hat{\mu}_{\ell \ell ^{\prime }}\left( \hat{\pi}_{\ell \ell ^{\prime
}}\right) -\mu \left( \pi \right) \right\Vert _{q,F}=o_{p}\left(
n^{-1/4}\right) $ for each $\ell =1,\cdots ,L$ and $\ell ^{\prime }\neq \ell 
$. It follows that%
\begin{eqnarray*}
\left\Vert E\left[ \left. \frac{1}{n-n_{\ell }}\sum_{j\in I_{\ell ^{\prime
}}}\hat{Q}_{1\ell \ell ^{\prime }j}\right\vert \mathcal{W}_{\ell \ell
^{\prime }}^{c}\right] \right\Vert  &\leq &\left\Vert \hat{\mu}_{X\ell \ell
^{\prime }}\left( \hat{\pi}_{\ell \ell ^{\prime }}\right) -\mu _{X}\left(
\pi \right) \right\Vert _{F}\cdot \left\Vert \hat{\mu}_{Y\ell \ell ^{\prime
}}\left( \hat{\pi}_{\ell \ell ^{\prime }}\right) -\mu _{Y}\left( \pi \right)
\right\Vert _{F} \\
&&+\left\Vert \hat{\mu}_{X\ell \ell ^{\prime }}\left( \hat{\pi}_{\ell \ell
^{\prime }}\right) -\mu _{X}\left( \pi \right) \right\Vert _{F}^{2}\cdot
\left\Vert \beta \right\Vert +\left\Vert \hat{\mu}_{X\ell \ell ^{\prime
}}\left( \hat{\pi}_{\ell \ell ^{\prime }}\right) -\mu _{X}\left( \pi \right)
\right\Vert _{q,F} \\
&&\cdot \left\Vert \hat{\pi}_{\ell \ell ^{\prime }}-\pi \right\Vert
_{q,F}\cdot \left( E\left\vert X_{j}-\mu _{X}\left( P_{j}\right) \right\vert
^{s}\right) ^{1/s}\cdot \left\Vert \beta \right\Vert  \\
&=&o_{p}\left( n^{-1/2}\right) ,
\end{eqnarray*}%
\begin{eqnarray*}
\left\Vert E\left[ \left. \frac{1}{n-n_{\ell }}\sum_{j\in I_{\ell ^{\prime
}}}\hat{Q}_{2\ell \ell ^{\prime }j}\right\vert \mathcal{W}_{\ell \ell
^{\prime }}^{c}\right] \right\Vert  &\leq &\left( E\left\vert X_{j}-\mu
_{X}\left( P_{j}\right) \right\vert ^{\tilde{s}}\right) ^{1/\tilde{s}}\left[ 
\begin{array}{c}
\left\Vert \hat{\mu}_{Y\ell \ell ^{\prime }}\left( \hat{\pi}_{\ell \ell
^{\prime }}\right) -\mu _{Y}\left( \pi \right) \right\Vert _{q,F} \\ 
+\left\Vert \hat{\mu}_{X\ell \ell ^{\prime }}\left( \hat{\pi}_{\ell \ell
^{\prime }}\right) -\mu _{X}\left( \pi \right) \right\Vert _{q,F}\left\Vert
\beta \right\Vert 
\end{array}%
\right]  \\
&&+\left( E\left\vert X_{j}-\mu _{X}\left( P_{j}\right) \right\vert ^{2%
\tilde{s}}\right) ^{\left. 1\right/ \left( 2\tilde{s}\right) }\left\Vert 
\hat{\pi}_{\ell \ell ^{\prime }}-\pi \right\Vert _{q,F}\left\Vert \beta
\right\Vert  \\
&=&o_{p}\left( n^{-1/4}\right) , \\
\left\Vert E\left[ \left. \frac{1}{n-n_{\ell }}\sum_{j\in I_{\ell ^{\prime
}}}\hat{Q}_{3\ell \ell ^{\prime }j}\right\vert \mathcal{W}_{\ell \ell
^{\prime }}^{c}\right] \right\Vert  &\leq &O\left( \left\Vert \hat{\mu}%
_{X\ell \ell ^{\prime }}\left( \hat{\pi}_{\ell \ell ^{\prime }}\right) -\mu
_{X}\left( \pi \right) \right\Vert _{q,F}\right) =o_{p}\left(
n^{-1/4}\right) , \\
\left\Vert E\left[ \left. \frac{1}{n-n_{\ell }}\sum_{j\in I_{\ell ^{\prime
}}}\hat{Q}_{4\ell \ell ^{\prime }j}\right\vert \mathcal{W}_{\ell \ell
^{\prime }}^{c}\right] \right\Vert  &\leq &O\left( \left\Vert \hat{\pi}%
_{\ell \ell ^{\prime }}-\pi \right\Vert _{q,F}\right) =o_{p}\left(
n^{-1/4}\right) .
\end{eqnarray*}%
On the other hand, we can show that%
\begin{equation*}
\left\Vert Var\left( \left. \frac{1}{n-n_{\ell }}\sum_{j\in I_{\ell ^{\prime
}}}\hat{Q}_{k\ell \ell ^{\prime }j}\right\vert \mathcal{W}_{\ell \ell
^{\prime }}^{c}\right) \right\Vert =o_{p}\left( n^{-1/2}\right) 
\end{equation*}%
for each $k=1,2,3,4$. In consequence,%
\begin{equation}
\frac{n^{1/4}}{n-n_{\ell }}\sum_{j\in I_{\ell ^{\prime }}}\hat{Q}_{k\ell
\ell ^{\prime }j}\overset{p}{\longrightarrow }E\left[ \left. \frac{n^{1/4}}{%
n-n_{\ell }}\sum_{j\in I_{\ell ^{\prime }}}\hat{Q}_{k\ell \ell ^{\prime
}j}\right\vert \mathcal{W}_{\ell \ell ^{\prime }}^{c}\right] \overset{p}{%
\longrightarrow }0  \label{36}
\end{equation}%
for each $k=1,2,3,4$. By (\ref{35}) and (\ref{36}), we have%
\begin{eqnarray*}
\frac{1}{n-n_{\ell }}\sum_{\ell ^{\prime }\neq \ell }\sum_{j\in I_{\ell
^{\prime }}}r_{0}\left( W_{j},\hat{\pi}_{\ell \ell ^{\prime }},\hat{\mu}%
_{\ell \ell ^{\prime }},\beta \right)  &=&\frac{1}{n-n_{\ell }}\sum_{j\notin
I_{\ell }}r_{0}\left( W_{j},\pi ,\mu ,\beta \right)
+\sum_{k=1}^{4}\sum_{\ell ^{\prime }\neq \ell }\left( \frac{1}{n-n_{\ell }}%
\sum_{j\in I_{\ell ^{\prime }}}\hat{Q}_{k\ell \ell ^{\prime }j}\right)  \\
&=&O_{p}\left( n^{-1/2}\right) +o_{p}\left( n^{-1/4}\right) =o_{p}\left(
n^{-1/4}\right) .
\end{eqnarray*}%
Therefore,%
\begin{eqnarray}
\hat{\beta}_{\ell }-\beta  &=&\left( M^{-1}+o_{p}\left( 1\right) \right) 
\frac{1}{n-n_{\ell }}\sum_{\ell ^{\prime }\neq \ell }\sum_{j\in I_{\ell
^{\prime }}}r_{0}\left( W_{j},\hat{\pi}_{\ell \ell ^{\prime }},\hat{\mu}%
_{\ell \ell ^{\prime }},\beta \right)   \notag \\
&=&O_{p}\left( 1\right) \cdot o_{p}\left( n^{-1/4}\right) =o_{p}\left(
n^{-1/4}\right) .  \label{33}
\end{eqnarray}%
The conclusion $n^{1/4}\left\Vert \hat{\alpha}_{\ell }-\alpha \right\Vert
_{q,F}\overset{p}{\longrightarrow }0$ follows from (\ref{31}), (\ref{32}), (%
\ref{33}), and arguments similar to Lemmas \ref{lemma2}.(ii) and \ref{lemma3}%
, which completes the proof.
\end{proof}

\subsection{Derivation of the First-Step Influence Function}

\label{appendix:derivation}Denote the \citeauthor{robinson1988root}-type
moment function in (\ref{R_moment}) as%
\begin{equation*}
r_{0}\left( \pi _{1},\pi _{2},\pi _{3},\pi _{4},\pi _{5},\beta \right) =\pi
_{1}\left( X\right) \left[ X-\mu _{X1}\left( \pi _{2}\left( X\right) \right) %
\right] \left\{ Y-\mu _{Y}\left( \pi _{3}\left( X\right) \right) -\pi
_{4}\left( X\right) \left[ X-\mu _{X2}\left( \pi _{5}\left( X\right) \right) %
\right] ^{\prime }\beta \right\} ,
\end{equation*}%
where the dependences on the data $W=\left( Y,D,X\right) $ and the
second-step estimand $\mu =\left( \mu _{X1},\mu _{X2},\mu _{Y}\right) $ are
supressed for simplicity. The notations $\pi _{j}$ ($j=1,\cdots ,5$) and $%
\mu _{Xj}$ ($j=1,2$) are just an expositional device, since $\pi _{j}=\pi $
and $\mu _{Xj}=\mu _{X}$.

Note that estimation of $\pi _{1}$ and $\pi _{4}$ has only direct effect on $%
r_{0}$, which is%
\begin{eqnarray*}
\alpha _{1}\left( \pi ,\beta \right) &=&E\left[ \left. \frac{\partial
r_{0}\left( \pi ,\beta \right) }{\partial \pi _{1}}\right\vert X\right] \\
&=&E\left[ \left. \left( X-\mu _{X}\left( P\right) \right) \left( Y-\mu
_{Y}\left( P\right) -P\left( X-\mu _{X}\left( P\right) \right) ^{\prime
}\beta \right) \right\vert X\right] \\
&=&\left( X-\mu _{X}\left( P\right) \right) \left[ E\left( \left.
Y\right\vert X\right) -\mu _{Y}\left( P\right) -P\left( X-\mu _{X}\left(
P\right) \right) ^{\prime }\beta \right] \\
&=&0, \\
\alpha _{4}\left( \pi ,\beta \right) &=&E\left[ \left. \frac{\partial
r_{0}\left( \pi ,\beta \right) }{\partial \pi _{4}}\right\vert X\right] \\
&=&E\left[ \left. -P\left( X-\mu _{X}\left( P\right) \right) \left( X-\mu
_{X}\left( P\right) \right) ^{\prime }\beta \right\vert X\right] \\
&=&-P\left( X-\mu _{X}\left( P\right) \right) \left( X-\mu _{X}\left(
P\right) \right) ^{\prime }\beta ,
\end{eqnarray*}%
where the fourth equality follows from (\ref{RF5}). By Proposition 4 of \cite%
{newey1994asymptotic}, the influence function corresponding to the direct
effect is%
\begin{equation*}
\phi _{D}\left( \beta \right) =\alpha _{4}\left( \pi ,\beta \right) \cdot
\left( D-P\right) =-P\left( X-\mu _{X}\left( P\right) \right) \left( X-\mu
_{X}\left( P\right) \right) ^{\prime }\beta \left( D-P\right) .
\end{equation*}

In comparison, estimation of $\pi _{2}$, $\pi _{3}$, and $\pi _{5}$ affects $%
r_{0}$ indirectly through $\mu $. Due to the influence function of $\mu $ is
zero, the indirect effect of first-step estimation is merely the naive
derivative of $r_{0}$ that only accounts for first-step estimation as an
argument \cite[Remark 3]{hahn2013asymptotic}. Specifically, it follows from %
\citet[Theorem 5]{hahn2013asymptotic} that the indirect effect is%
\begin{eqnarray*}
\alpha _{2}\left( \pi ,\beta \right) &=&E\left[ \left. \frac{\partial r_{0}}{%
\partial \mu _{X1}}\frac{d\mu _{X1}}{d\pi _{2}}\right\vert X\right] \\
&=&E\left[ \left. -P\left( Y-\mu _{Y}\left( P\right) -P\left( X-\mu
_{X}\left( P\right) \right) ^{\prime }\beta \right) \mu _{X}^{\left(
1\right) }\left( P\right) \right\vert X\right] \\
&=&0, \\
\alpha _{3}\left( \pi \right) &=&E\left[ \left. \frac{\partial r_{0}}{%
\partial \mu _{Y}}\frac{d\mu _{Y}}{d\pi _{3}}\right\vert X\right] =-P\left(
X-\mu _{X}\left( P\right) \right) \mu _{Y}^{\left( 1\right) }\left( P\right)
, \\
\alpha _{5}\left( \pi ,\beta \right) &=&E\left[ \left. \frac{\partial r_{0}}{%
\partial \mu _{X2}}\frac{d\mu _{X2}}{d\pi _{5}}\right\vert X\right]
=P^{2}\left( X-\mu _{X}\left( P\right) \right) \mu _{X}^{\left( 1\right)
}\left( P\right) ^{\prime }\beta ,
\end{eqnarray*}%
and the corresponding influence function is%
\begin{equation*}
\phi _{I}\left( \beta \right) =\left[ \alpha _{3}\left( \pi \right) +\alpha
_{5}\left( \pi ,\beta \right) \right] \cdot \left( D-P\right) =-P\left(
X-\mu _{X}\left( P\right) \right) \left[ \mu _{Y}^{\left( 1\right) }\left(
P\right) -P\mu _{X}^{\left( 1\right) }\left( P\right) ^{\prime }\beta \right]
\left( D-P\right) .
\end{equation*}

In summary, the first-step influence function is%
\begin{equation*}
\phi _{D}\left( \beta \right) +\phi _{I}\left( \beta \right) =-P\left( X-\mu
_{X}\left( P\right) \right) \left[ \mu _{Y}^{\left( 1\right) }\left(
P\right) -P\mu _{X}^{\left( 1\right) }\left( P\right) ^{\prime }\beta
+\left( X-\mu _{X}\left( P\right) \right) ^{\prime }\beta \right] \left(
D-P\right) ,
\end{equation*}%
and the orthogonalized moment function is constructed as%
\begin{equation*}
\psi \left( \beta \right) =r_{0}\left( \pi ,\beta \right) +\phi _{D}\left(
\beta \right) +\phi _{I}\left( \beta \right) .
\end{equation*}%
Since $E\left[ \left. D-P\right\vert X\right] =0$, it follows from the law
of iterated expectations that $E\left[ \phi _{D}\left( \tilde{\beta}\right)
+\phi _{I}\left( \tilde{\beta}\right) \right] =0$ for any $\tilde{\beta}$.
As a consequence, we have four ways of formulating a consistent estimator
for $\beta $ according to the orthogonalized moment function $\psi \left(
\beta \right) $.

\begin{formulation}
\label{form1}Regard $\beta $ as a solution to the moment condition $E\left[
\psi \left( \beta \right) \right] =0:$%
\begin{eqnarray*}
\beta &=&\left\{ E\left[ P\left( X-\mu _{X}\left( P\right) \right) \left[
D\left( X-\mu _{X}\left( P\right) \right) -P\left( D-P\right) \mu
_{X}^{\left( 1\right) }\left( P\right) \right] ^{\prime }\right] \right\}
^{-1} \\
&&\cdot E\left[ P\left( X-\mu _{X}\left( P\right) \right) \left( Y-\mu
_{Y}\left( P\right) -\left( D-P\right) \mu _{Y}^{\left( 1\right) }\left(
P\right) \right) \right] ,
\end{eqnarray*}%
and substitute the unknown functions with their estimates and the
expectations with their sample analogs.
\end{formulation}

\begin{formulation}
\label{form2}Regard $\beta $ as a solution to the moment condition $E\left[
r_{0}\left( \beta \right) +\phi _{D}\left( \tilde{\beta}\right) +\phi
_{I}\left( \beta \right) \right] =0$ for a predetermined $\tilde{\beta}:$%
\begin{eqnarray*}
\beta &=&\left\{ E\left[ P^{2}\left( X-\mu _{X}\left( P\right) \right)
\left( X-\mu _{X}\left( P\right) -\left( D-P\right) \mu _{X}^{\left(
1\right) }\left( P\right) \right) ^{\prime }\right] \right\} ^{-1} \\
&&\cdot E\left[ P\left( X-\mu _{X}\left( P\right) \right) \left[ Y-\mu
_{Y}\left( P\right) -\left( D-P\right) \left( \mu _{Y}^{\left( 1\right)
}\left( P\right) +\left( X-\mu _{X}\left( P\right) \right) ^{\prime }\tilde{%
\beta}\right) \right] \right] ,
\end{eqnarray*}%
and substitute the unknowns with their estimates.
\end{formulation}

\begin{formulation}
\label{form3}Regard $\beta $ as a solution to the moment condition $E\left[
r_{0}\left( \beta \right) +\phi _{D}\left( \beta \right) +\phi _{I}\left( 
\tilde{\beta}\right) \right] =0$ for a predetermined $\tilde{\beta}:$%
\begin{eqnarray*}
\beta &=&\left\{ E\left[ PD\left( X-\mu _{X}\left( P\right) \right) \left(
X-\mu _{X}\left( P\right) \right) ^{\prime }\right] \right\} ^{-1} \\
&&\cdot E\left[ P\left( X-\mu _{X}\left( P\right) \right) \left[ Y-\mu
_{Y}\left( P\right) -\left( D-P\right) \left( \mu _{Y}^{\left( 1\right)
}\left( P\right) -P\mu _{X}^{\left( 1\right) }\left( P\right) ^{\prime }%
\tilde{\beta}\right) \right] \right] ,
\end{eqnarray*}%
and substitute the unknowns with their estimates.
\end{formulation}

\begin{formulation}
\label{form4}Regard $\beta $ as a solution to the moment condition $E\left[
r_{0}\left( \beta \right) +\phi _{D}\left( \tilde{\beta}\right) +\phi
_{I}\left( \tilde{\beta}\right) \right] =0$ for a predetermined $\tilde{\beta%
}:$%
\begin{eqnarray*}
\beta &=&\left\{ E\left[ P^{2}\left( X-\mu _{X}\left( P\right) \right)
\left( X-\mu _{X}\left( P\right) \right) ^{\prime }\right] \right\} ^{-1} \\
&&\cdot E\left[ P\left( X-\mu _{X}\left( P\right) \right) \left[ Y-\mu
_{Y}\left( P\right) -\left( D-P\right) \left( \mu _{Y}^{\left( 1\right)
}\left( P\right) +\left( X-\mu _{X}\left( P\right) -P\mu _{X}^{\left(
1\right) }\left( P\right) \right) ^{\prime }\tilde{\beta}\right) \right] %
\right] ,
\end{eqnarray*}%
and substitute the unknowns with their estimates.
\end{formulation}

The estimators derived from the above four formulations have identical
asymptotics if the predetermined $\tilde{\beta}$ is a well-defined
consistent estimate of $\beta $. However, these estimators may differ
substantially in finite-sample performance. For example, the estimators\ in
Formulations \ref{form1} and \ref{form2} are expected to behave badly in
finite samples, because the \textquotedblleft Jacobian\textquotedblright\
matrices are not symmetric and thus difficult to find the inverse. On the
other side, Formulation \ref{form4} relies heavily on the predetermined $%
\tilde{\beta}$, which may induce more bias in the estimation of $\beta $
than necessary. Overall, we think of Formulation \ref{form3} as the best
compromise and adopt it when constructing the locally robust estimator for $%
\beta $ in Section \ref{sec:estim}, where we set $r=r_{0}+\phi _{D}$ and $%
\alpha =\alpha _{3}+\alpha _{5}$.

\subsection{Additional Results of the Simulation}

\label{appendix:tables}

\begin{table}[htb]
\caption{Simulation results for higher error correlation and censoring rate}
\label{tableA1}
\begin{center}
\begin{tabular}{cccccc}
\hline\hline
$n=1000$ &  & (1) & (2) & (3) & (4) \\ 
&  & \ \ Locally \ \  & \ Robinson \  & Robinson with & Robinson with \\ 
&  & Robust &  & Orthogonalization & Cross-fitting \\ \hline
\multicolumn{6}{l}{\textbf{Panel A:} $Cov\left( U,\varepsilon \right) =0.75,$
$\Pr \left( D=0\right) =0.5$} \\ 
Average Bias &  & 0.108 & 0.114 & 0.074 & 0.254 \\ 
Average SD &  & 0.103 & 0.097 & 0.083 & 0.160 \\ 
Average Coverage &  & 0.938 & 0.773 & 0.900 & 0.549 \\ 
Max Coverage &  & 0.990 & 0.960 & 0.980 & 0.940 \\ 
Min Coverage &  & 0.820 & 0.340 & 0.770 & 0.020 \\ \hline
\multicolumn{6}{l}{\textbf{Panel B:} $Cov\left( U,\varepsilon \right) =0.5,$ 
$\Pr \left( D=0\right) =0.75$} \\ 
Average Bias &  & 0.229 & 0.149 & 0.082 & 0.469 \\ 
Average SD &  & 0.232 & 0.129 & 0.099 & 0.324 \\ 
Average Coverage &  & 0.945 & 0.797 & 0.921 & 0.627 \\ 
Max Coverage &  & 0.990 & 0.950 & 0.970 & 0.960 \\ 
Min Coverage &  & 0.830 & 0.590 & 0.870 & 0.110 \\ \hline
\multicolumn{6}{l}{\textbf{Panel C:} $Cov\left( U,\varepsilon \right) =0.75,$
$\Pr \left( D=0\right) =0.75$} \\ 
Average Bias &  & 0.217 & 0.159 & 0.100 & 0.460 \\ 
Average SD &  & 0.224 & 0.150 & 0.120 & 0.326 \\ 
Average Coverage &  & 0.957 & 0.827 & 0.933 & 0.640 \\ 
Max Coverage &  & 1.000 & 0.960 & 0.960 & 0.940 \\ 
Min Coverage &  & 0.870 & 0.660 & 0.880 & 0.130 \\ \hline\hline
\end{tabular}%
\end{center}
\end{table}

\begin{table}[htb]
\caption{Simulation results for different error distributions}
\label{tableA2}
\begin{center}
\begin{tabular}{cccccc}
\hline\hline
$n=1000$ &  & (1) & (2) & (3) & (4) \\ 
&  & \ \ Locally \ \  & \ Robinson \  & Robinson with & Robinson with \\ 
&  & Robust &  & Orthogonalization & Cross-fitting \\ \hline
\multicolumn{6}{l}{\textbf{Panel A:} $\varepsilon \sim Logistic\left(
0,1\right) $} \\ 
Average Bias &  & 0.135 & 0.140 & 0.103 & 0.274 \\ 
Average SD &  & 0.133 & 0.130 & 0.111 & 0.190 \\ 
Average Coverage &  & 0.926 & 0.832 & 0.880 & 0.618 \\ 
Max Coverage &  & 0.990 & 0.960 & 0.960 & 0.940 \\ 
Min Coverage &  & 0.770 & 0.620 & 0.650 & 0.090 \\ \hline
\multicolumn{6}{l}{\textbf{Panel B:} $\varepsilon \sim t\left( 3\right) $}
\\ 
Average Bias &  & 0.147 & 0.147 & 0.112 & 0.283 \\ 
Average SD &  & 0.215 & 0.190 & 0.169 & 0.254 \\ 
Average Coverage &  & 0.934 & 0.843 & 0.898 & 0.590 \\ 
Max Coverage &  & 0.980 & 0.960 & 0.970 & 0.910 \\ 
Min Coverage &  & 0.850 & 0.660 & 0.730 & 0.090 \\ \hline
\multicolumn{6}{l}{\textbf{Panel C:} $\varepsilon \sim t\left( 2\right) $}
\\ 
Average Bias &  & 0.171 & 0.192 & 0.166 & 0.286 \\ 
Average SD &  & 0.191 & 0.231 & 0.209 & 0.226 \\ 
Average Coverage &  & 0.935 & 0.899 & 0.917 & 0.695 \\ 
Max Coverage &  & 0.980 & 0.950 & 0.970 & 0.980 \\ 
Min Coverage &  & 0.850 & 0.760 & 0.780 & 0.090 \\ \hline\hline
\end{tabular}%
\end{center}
\end{table}

\begin{table}[htb]
\caption{Simulation results for different selection mechanisms}
\label{tableA3}
\begin{center}
\begin{tabular}{cccccc}
\hline\hline
$n=1000$ &  & (1) & (2) & (3) & (4) \\ 
&  & \ \ Locally \ \  & \ Robinson \  & Robinson with & Robinson with \\ 
&  & Robust &  & Orthogonalization & Cross-fitting \\ \hline
\multicolumn{6}{l}{\textbf{Panel A:} $h\left( x\right) =x_{1}+\log \left(
x_{1}^{2}\right) -x_{2}-\log \left( x_{2}^{2}\right) +x_{3}x_{4}-x_{5}x_{6}$}
\\ 
Average Bias &  & 0.090 & 0.102 & 0.059 & 0.211 \\ 
Average SD &  & 0.092 & 0.086 & 0.074 & 0.147 \\ 
Average Coverage &  & 0.944 & 0.787 & 0.939 & 0.568 \\ 
Max Coverage &  & 0.990 & 0.960 & 0.960 & 0.950 \\ 
Min Coverage &  & 0.870 & 0.070 & 0.910 & 0.060 \\ \hline
\multicolumn{6}{l}{\textbf{Panel B:} $h\left( x\right) =x_{1}+\exp \left(
x_{1}\right) -x_{2}-\exp \left( x_{2}\right) +x_{3}x_{4}-x_{5}x_{6}$} \\ 
Average Bias &  & 0.093 & 0.103 & 0.064 & 0.178 \\ 
Average SD &  & 0.101 & 0.091 & 0.079 & 0.150 \\ 
Average Coverage &  & 0.946 & 0.773 & 0.923 & 0.688 \\ 
Max Coverage &  & 0.980 & 0.940 & 0.950 & 0.900 \\ 
Min Coverage &  & 0.830 & 0.250 & 0.890 & 0.110 \\ \hline
\multicolumn{6}{l}{\textbf{Panel C:} $h\left( x\right) =x_{1}+\exp \left(
x_{1}\right) +\log \left( x_{1}^{2}\right) -x_{2}-\exp \left( x_{2}\right)
-\log \left( x_{2}^{2}\right) +x_{3}x_{4}-x_{5}x_{6}$} \\ 
Average Bias &  & 0.079 & 0.089 & 0.061 & 0.154 \\ 
Average SD &  & 0.093 & 0.084 & 0.077 & 0.138 \\ 
Average Coverage &  & 0.954 & 0.849 & 0.946 & 0.720 \\ 
Max Coverage &  & 0.980 & 0.970 & 0.980 & 0.940 \\ 
Min Coverage &  & 0.920 & 0.390 & 0.920 & 0.090 \\ \hline\hline
\end{tabular}%
\end{center}
\end{table}

\end{document}